\documentclass[review]{elsarticle}

\usepackage[utf8]{inputenc}
\usepackage{hyperref}
\usepackage{tabularx}
\usepackage{graphicx}
\usepackage{subfigure}
\usepackage{amsmath}
\usepackage{amsfonts}
\usepackage{amssymb} 
\usepackage{rawfonts} 
\usepackage{latexsym}   
\usepackage{textcomp}
\usepackage{subfigmat}
\usepackage{gensymb}
\usepackage{multirow}
\usepackage[multiuser,draft]{fixme}
\usepackage{vmargin}
\usepackage{placeins}

\setpapersize{A4}
\setmarginsrb{30mm}{20mm}{30mm}{20mm}{12pt}{10mm}{12pt}{10mm}

\newcommand{\mat}[1]{\mbox{\boldmath{$#1$}}}

\newcommand{\Rey}{\mathit{Re}}
\FXRegisterAuthor{bsc}{absc}{BSC}
\fxusetheme{color}
\fxloadtargetlayouts{colorcb}

\begin{document}
\begin{frontmatter}

\title{Sensitivity of the least stable modes to passive control for a flow around an elastically-mounted circular cylinder}

\author[PME-EPUSP]{Daiane Iglesia Dolci$^1$}
\author[PME-EPUSP]{Bruno Souza Carmo$^2$}
\address{$^1$daia.dolci@gmail.com, $^2$bruno.carmo@usp.br}

\address[PME-EPUSP]{Department of Mechanical Engineering, Escola 
Politécnica, University of São Paulo. Av. Prof. Mello Moraes, 2231, 
São Paulo, SP, 05508-030, Brazil.}

\begin{abstract}

In this paper, a methodology to calculate the sensitivity of the least stable modes of fluid-structure interaction systems with respect to local forces is presented. We make use of the adjoint equations of the flow-structure coupled system to calculate the gradients, and the algorithms were implemented using the spectral/\emph{hp} element method for the spatial discretization. The methodology was applied to two-dimensional incompressible laminar steady flows around an elastically-mounted circular cylinder, and we obtained the gradients of the real and imaginary parts of the least stable eigenvalues with respect to forces located at arbitrary points in the flow domain. Selected values of mass ratio and reduced velocity were considered in the simulations, and the results were compared to those obtained for a fixed cylinder at the same Reynolds number. We noticed that the sensitivity fields of the fluid-structure interaction system can be very different from its fixed structure counterpart, and amongst the cases, with an elastic structure, the fields vary greatly according to the reduced velocity. Finally, the sensitivity results were verified against linear and nonlinear simulations of flows with small control cylinders placed at locations selected according to the sensitivity fields. The agreement between the predictions made with the sensitivity analyses and the linear and nonlinear results of the forced flows was excellent. In some cases, it was possible to completely suppress the cylinder vibration.
\end{abstract}
\end{frontmatter}

\section{Introduction} 

Linear stability analysis has been used for decades to study fluid flows. However, its use in fluid-structure interaction (FSI) problems has only been introduced in the last few years.
The flow around an elastically-mounted bluff body has been commonly used as the model problem, and, in general, the  investigations focus on the characterisation of the least stable modes close to the first instability. For an elastically-mounted circular cylinder, \cite{Cossu2000, Meliga2011,Zhang2015,Mittal2016, Yao2017,Kou2017} showed that the first instability can occur for Reynolds number, $\Rey$, less than 47, which is the critical Reynolds number of the first bifurcation in the flow around a fixed circular cylinder. We remind that the Reynolds number for this flow is defined as $\Rey = U_\infty D/\nu$, where $U_\infty$ is the free stream speed, $D$ is the cylinder diameter and $\nu$ is the fluid kinematic viscosity. We also remind that the first bifurcation in the flow around a fixed circular cylinder is that at which the originally steady flow becomes time-dependent, giving rise to the Bérnard-von Kármán vortex wake.

\citet{Cossu2000} were the first to apply linear stability analysis to the flow around an elastically-mounted circular cylinder.
They calculated the threshold for the primary instability to be $\Rey =23.5$ for mass ratios (mass of the moving body divided by the mass of displaced fluid) $m^\ast=7$ and $m^\ast=70$. \citet{Meliga2011} introduced a map of $\Rey$ and $m^{\ast}$ showing stable and unstable regions for the elastically-mounted circular cylinder free to oscillate in both transverse and in-line directions. They showed that the critical Reynolds number drops as the mass ratio decreases. On the other hand, when the mass ratio increases the critical Reynolds tends to that of the flow around a fixed cylinder. Like \citet{Cossu2000} and \citet{Meliga2011}, \cite{Zhang2015} and \citet{Mittal2016} also employed linear stability to study this flow and considered the two least stable modes in the analyses. \cite{Mittal2016} described these modes as coupled and decoupled modes. They presented a map  of $\Rey$ and $m^{\ast}$ showing regions where the modes were coupled and decoupled. The decoupled modes had marked distinctions between them, i.e., one mode has eigen-frequency closer to the natural frequency of the structure, and the other mode has an eigen-frequency closer to the eigen-frequency of the flow around a fixed cylinder. For the coupled modes, this distinction could not be made. In addition, for the flow around an elastically-mounted cylinder, \cite{Dolci2019} investigated the nonlinear character of the first instability for reduced velocities ($V_r = U_\infty/(f_n D)$, with $f_n = \sqrt{k/m}$ the natural frequency of the structure in vacuum, $k$ the stiffness of the structure, and $m$ the moving mass of the structure) inside and  outside of the lock-in range. They verified a subcritical character for reduced velocities inside of the lock-in range, and a supercritical character for $V_r$ outside the lock-in region.

Regarding the mathematical approach used to linearize an FSI system, \citet{Fernandez2002, Fernandez2003} introduced a formulation to obtain a linear FSI system independent of the mesh displacement. The algebra consisted of writing the spatial coordinates as a first-order Taylor expansion around the coordinates of the non-deformed, fixed mesh, $\mathbf{x}_0$. Next, the nonlinear FSI system was rewritten by considering the spatial coordinates described as Taylor expansions, and by supposing that the FSI state vector is given by the steady base flow plus a perturbation. Such assumptions led to the steady Navier-Stokes system that governs the base flow and the FSI system which governs the perturbation. Transpiration boundary conditions at the structure wall and additional stresses in the governing structure equation appear in the final linearized system. This methodology has been the base for recent works of linear stability analyses for different FSI problems. \cite{Pfister2019, Pfister2020} investigated flutter instabilities for FSI problems with flexible structures, among them, for flow around a circular cylinder with an attached elastic plate. \cite{Prabal2020} employed global linear analysis to investigate the symmetry breaking in the flow around a rotating cylinder with an attached splitter plate. Besides that, they presented structural sensitivity results considering the least stable mode for an elastically-mounted circular cylinder. These results showed that the structural sensitivity can be very different from the sensitivity of the flow around a fixed cylinder at the same Reynolds number.

In global linear analyses, the study of sensitivity can provide regions more sensitive to external forcing, for instance, passive control. For the flow around a fixed circular cylinder with Reynolds number closer to the critical value of the first stability,  \citet{Giannetti2007} and \citet{Marquet2008} applied computations of sensitivity and receptivity based on the least stable direct and adjoint modes. \citet{Giannetti2007} identified regions of the domain where the flow is more receptive to the presence of momentum forcing and mass injection in the perturbation field. Based on the spatial distribution of the product of the direct and adjoint least stable eigen-vectors/modes, they identified the regions more sensitive to a structural change. \cite{Marquet2008} also presented results of sensitivity analyses considering the hypothesis that the presence of a control cylinder also modifies the base flow and can change its dynamics. Based on that, they evaluated the regions at which the imposition of a steady forcing in the base flow would lead to a larger variation of the eigenvalues. \cite{Pralitis:2010} carried out sensitivity analyses by considering the sum of the structural and base flow sensitivities to perform the sensitivity analyses in the two-dimensional flow around a rotating circular cylinder at $\Rey = 100$. Later, \cite{Carini:2014} employed the same approach of sensitivity calculation to suppress the vortex shedding using passive control in the flow around two side-by-side circular cylinders. \cite{Fani:2012} used the sum of sensitivities to apply a passive control to suppress vortex shedding in the backward-facing step flow. 

Fluid-structure interaction is ubiquitous in nature and engineering devices. In many cases, it is necessary to be able to predict the dynamical behaviour of systems with this type of interaction and, if possible, to control it, in order to prevent failures or to obtain some benefit out of it, like in energy harvesting, for instance. In this work, we propose a method to calculate the sensitivity of eigen-modes to passive control in a FSI system. We use this method to perform sensitivity analyses of low Reynolds number flows around an elastically-mounted cylinder, and validate the predictions using non-linear FSI calculations employing small cylinders as passive flow control. The linearization of the FSI system is based on the methodology introduced by \citet{Fernandez_thesis2001, Fernandez2002, Fernandez2003}. The sensitivity computations are carried out based on the approach presented by \cite{Pralitis:2010,Fani:2012,Carini:2014}, i.e., the structural and base flow sensitivities are considered in the analyses. The passive control is considered as a localised external forcing proportional to the local velocity, inserted in the domain at an arbitrary point $(x_p, y_p)$. The critical Reynolds number of the primary bifurcation in the flow around a fixed circular cylinder is chosen to be used in every case. The sensitivity fields of the fixed cylinder are used as a benchmark to perform comparisons with the fields of sensitivity of the flows around the elastically-mounted cylinder.
To the best of the authors' knowledge, no effective method of sensitivity calculation that could be employed for passive control of FSI phenomena has been presented before. Thus, this piece of research is an essential contribution towards the control of flow-induced vibrations and flow-induced motion.

The remainder of the paper is organised as follows.
In section~\ref{math}, we present the equations which govern the nonlinear FSI system and the mathematical approach to perform the linearisation of the FSI system. These constitute the starting point to obtain the algebraic expressions of sensitivity and the adjoint FSI systems. Next, section~\ref{Num_Method} gives some general information about the numerical method employed to discretize the equations, the numerical approaches adopted to simulate the FSI system and to compute the modes and their respective eigenvalues. The results and analyses are presented in section~\ref{Results}. In section~\ref{Conclusions}, we draw final comments about the results, and highlight the main contributions of this work. The algebraic development to obtain the adjoint equations, the numerical convergence analyses and the verification of the adjoint modes which is performed by comparing their eigenvalues to those of the direct modes are presented in the appendices.

\section{Mathematical fundamentals} \label{math}
\subsection{Non-linear system}

The two-dimensional flow around an elastically-mounted circular cylinder free to oscillated in the cross-flow direction is governed by the Navier-Stokes equations and the mass-spring-damper system equation, described in nondimensional form by:
\begin{eqnarray}\label{NonLFSI}
\frac{\mathrm{D} \mathbf{u}}{\mathrm{D} t} - \frac{1}{\Rey}\nabla^2 \mathbf{u} + \nabla p&=&  \mathbf{0},\label{momento_FSI}\\
\nabla \cdot \mathbf{u} &=& 0,\label{massaFSI}\\
M^{\ast} \ddot{\gamma}(t)  +  C^{\ast}\dot{\gamma}(t) + K^{\ast} \gamma(t) &=& F (\mathbf{u}, p),\label{structure}\\
\mathbf{u} &=&\dot{\gamma}(t)   \quad \forall \mathbf{x} \in \partial \mat{\Omega}_w.
\end{eqnarray}
The vector $\mathbf{u}(\mathbf{x}, t)$ represents the velocity field, $p(\mathbf{x}, t)$ is the pressure and the variable $\gamma(t)$ is the cylinder displacement. We assume that the free stream is in the $x$ direction, so the cylinder displacement will occur only in the $y$ direction. The variables $\dot{\gamma}(t) = \dfrac{\mathrm{d}\gamma}{\mathrm{d} t}$ and $\ddot{\gamma}(t) = \dfrac{\mathrm{d}^2\gamma}{\mathrm{d} t^2}$  are the velocity and acceleration of the structure. For a circular cylinder, $K^{\ast} = \dfrac{\pi^{3} m^{\ast}}{V_r^2}$, $M^{\ast} = \dfrac{\pi m^{\ast}}{4} $ and $C^{\ast} = \dfrac{\pi^{2} \zeta m^{\ast}}{V_r}$. The coefficient $\zeta$ is the  structural  damping, $F (\mathbf{u}, p)$ is the instantaneous nondimensional fluid force acting on the cylinder in the cross-flow direction, given by:
\begin{equation}\label{force_FSI}
   F (\mathbf{u}, p) = \int_{\partial \mat{\Omega}_w} \mathbf{n}_y\cdot \sigma (\mathbf{u}, p) \text{d}S_{w} = \int_{\partial \mat{\Omega}_w} \mathbf{n}_y \cdot\left[ -p \mathbf{I} + Re^{-1} (\nabla \mathbf{u} + (\nabla\mathbf{u})^{T})\right] \text{d}S_{w},
\end{equation}
where $\partial \mat{\Omega}_w$ refers to the cylinder surface. 

\subsection{Linear system}\label{linear_sys}

We linearize this fluid-structure interaction (FSI) problem around a steady flow field and zero structural displacement and velocity. So, the state vector is written as the sum of the steady base field $\mathbf{Q}=[\mathbf{U}(\mathbf{x})_f, P(\mathbf{x})_f]^{\text{T}}$ and the perturbation $\mathbf{q}' = [\mathbf{u}', p', \gamma(t)', \dot{\gamma}(t)']^{\text{T}}$. In this case, the steady base field $\mathbf{Q}$ is solved in the domain $\mat{\Omega}_0$ of steady coordinates $\mathbf{x}_0$, while the $\mathbf{q}'$ is defined in a domain $\mat{\Omega}$ which is time dependent, i.e, the coordinates $\mathbf{x}'(t)\in \mat{\Omega}$. 
To deal with this kind of problem, we adopt a strategy based on the approach previously introduced by \cite{Fernandez_thesis2001, Fernandez2002, Fernandez2003}. In what follows, we present the formulation for the flow equations (Navier-Stokes system) and structure equations (mass-spring-damper system).

\subsubsection{Flow equations} 
For the linear analysis, the state vector $\mathbf{q}$ is expressed as a first order Taylor expansion around the fixed coordinates $\mathbf{x}_{0}$ (equilibrium points):
\begin{align*}
    \mathbf{u} =\mathbf{u}(\mathbf{x},t) &= \mathbf{u}(\mathbf{x}_0 + \mathbf{x}'(t),t) = \mathbf{u}(\mathbf{x}_{0},t) +  \nabla_0 \mathbf{u}(\mathbf{x}_{0},t)\cdot \mathbf{x}'(t), \\
    p = p (\mathbf{x},t)  &=  p (\mathbf{x}_0 + \mathbf{x}'(t),t) = p(\mathbf{x}_{0},t) +  \nabla_0 p(\mathbf{x}_{0},t)\cdot\mathbf{x}'(t),
\end{align*}
where $\nabla_0 = \left[\dfrac{\partial}{\partial x_0},  \dfrac{\partial}{\partial y_0}\right]^{T}$ set in the domain $\mat{\Omega}_0\subset \mathbb{R}^2$.
Imposing $\mathbf{q} = \mathbf{Q} + \mathbf{q}'$, velocity and pressure fields read:
\begin{align*}
    \mathbf{u} &= \mathbf{U}(\mathbf{x}_{0}) + \nabla_0 \mathbf{U}(\mathbf{x}_{0})\cdot\mathbf{x}'(t) + \mathbf{u}'(\mathbf{x}_{0}, t) +  \nabla_0 \mathbf{u}'(\mathbf{x}_{0},t)\cdot\mathbf{x}'(t),\\
    p &= P(\mathbf{x}_{0}) +  \nabla_0 P(\mathbf{x}_{0})\cdot\mathbf{x}'(t) + p'(\mathbf{x}_{0}, t) +   \nabla_0 p'(\mathbf{x}_{0},t)\cdot\mathbf{x}'(t).
\end{align*}
Using the hypothesis that the perturbation variables are infinitesimal, the second-order terms are neglected and the variables are finally described by: 
\begin{align}
    \mathbf{u} &= \mathbf{U}(\mathbf{x}_{0}) + \nabla_0 \mathbf{U}(\mathbf{x}_{0})\cdot\mathbf{x}'(t) + \mathbf{u}'(\mathbf{x}_{0}, t) ,\label{velocity_x0}\\
    p &= P(\mathbf{x}_{0}) +  \nabla_0 P(\mathbf{x}_{0})\cdot\mathbf{x}'(t) + p'(\mathbf{x}_{0}, t) .\label{pressure_x0}
\end{align}

Based on the strategy introduced by \cite{Fernandez_thesis2001, Fernandez2002, Fernandez2003}, the linear analysis can be carried out by working with a FSI system defined in the fixed system of coordinates $\mathbf{x}_0 \in \Omega_0$. To do that, we first consider the coordinates translation given by the application of a transformation matrix $\mathbf{T}$ in $\mathbf{x}_{0}$, i. e., $\mathbf{x} = \mathbf{T}\mathbf{x}_0$. The transformation matrix is written as $\mathbf{T} = \mathbf{I} + \mathbf{R}'$, where $\mathbf{I}$ is the identity matrix and for translation displacement, $\mathbf{R}'$ is a diagonal matrix such that $||\mathbf{R}'||<<1$. Therefore, to rewrite the flow equations on the system of coordinates $\mathbf{x}_0 \in \Omega_0$, we start by working with the total acceleration that is given by (recall that $\mathbf{U}$ is steady):
\[
\frac{\mathrm{D}\mathbf{u}}{\mathrm{D}t} = \frac{\mathrm{D}}{\mathrm{D}t}\left(\mathbf{U} + \nabla_0 \mathbf{U}\cdot\mathbf{x}' + \mathbf{u}'\right) = 
\frac{\partial \mathbf{u}'}{\partial t} + \frac{\mathrm{d} \mathbf{x}'}{\mathrm{d} t}\cdot\nabla_0 \mathbf{U} + \frac{\mathrm{d} \mathbf{x}_0}{\mathrm{d} t}\cdot\nabla_0\left(\mathbf{U} + \nabla_0 \mathbf{U}\cdot\mathbf{x}' + \mathbf{u}'\right)
\]
Using the fact that $\mathbf{x}_0 = \mathbf{x} - \mathbf{x}'$, so that
\[ \frac{\mathrm{d} \mathbf{x}_0}{\mathrm{d} t} = \frac{\mathrm{d}}{\mathrm{d} t}(\mathbf{x} - \mathbf{x}') = \mathbf{u} - \frac{\mathrm{d} \mathbf{x}'}{\mathrm{d} t} = \left(\mathbf{U} + \nabla_0 \mathbf{U}\cdot\mathbf{x}' + \mathbf{u}'\right) - \frac{\mathrm{d} \mathbf{x}'}{\mathrm{d} t}. \]
For linear analysis, second-order terms are neglected and the final expression of the total acceleration is
\begin{align}
\frac{\mathrm{D}\mathbf{u}}{\mathrm{D}t} =\frac{\partial \mathbf{u}'}{\partial t} 
+ \mathbf{U}\cdot\nabla_0 \mathbf{U}
+ \mathbf{U}\cdot\nabla_0 \mathbf{u}'
+ \mathbf{u}'\cdot\nabla_0 \mathbf{U}
+ \mathbf{x}'\cdot\nabla_0(\mathbf{U}\cdot\nabla_0 \mathbf{U})
+ \mathbf{U}\cdot\nabla_0 \mathbf{U} \nabla_0 \mathbf{x}',
\label{eq:totalacc}    
\end{align}

The next step is to rewrite the spatial derivatives by making $\nabla (\,\,) = \nabla_0 (\,\,) \dfrac{\partial \mathbf{x}_0}{\partial \mathbf{x}}$. This way, the Laplacian term, the pressure gradient, and the velocity divergent from eqs.~(\ref{momento_FSI}) -- (\ref{massaFSI}) are respectively given by:
\begin{align}
    \nabla^2 \mathbf{u} &= \left[\nabla_0^2 \mathbf{U} + \nabla_0^2 \mathbf{u}' +  \nabla_0^2(\nabla_0 \mathbf{U}\cdot\mathbf{x}') \right]\left(\dfrac{\partial \mathbf{x}_0}{\partial \mathbf{x}}\right)^2\label{lap_u}\\
    \nabla \mathbf{p} &= \left(\nabla_0 P + \nabla_0 p' + \nabla_0(\nabla_0 P\cdot\mathbf{x}')\right)\dfrac{\partial \mathbf{x}_0}{\partial \mathbf{x}} \\
    \nabla \cdot \mathbf{u} &= \left(\nabla_0 \cdot \mathbf{U} + \nabla_0 \cdot \mathbf{u}' + \nabla_0\cdot (\nabla_0 \mathbf{U}\cdot\mathbf{x}')\right)\dfrac{\partial \mathbf{x}_0}{\partial \mathbf{x}} \label{div_u}. 
\end{align}
The derivative $\dfrac{\partial \mathbf{x}_0}{\partial \mathbf{x}}$ can be rewritten as:
\[\dfrac{\partial \mathbf{x}_0}{\partial \mathbf{x}} =\left(\dfrac{\partial \mathbf{x}_0}{\partial \mathbf{x}}\right)^{-1} = \left(\dfrac{\partial }{\partial \mathbf{x}_0}[(\mathbf{I} + \mathbf{R}')\mathbf{x}_0]\right)^{-1} = \left(\mathbf{I} + \mathbf{R}'\right)^{-1}.\]
For translation displacement, the matrix $(\mathbf{I} + \mathbf{R}')$ is diagonal, and therefore, it is invertible. Also, let be $r'_{i,j}<<1$ the elements of $\mathbf{R}'$ with $i$ lines and $j$ columns, we have $(\mathbf{I} + \mathbf{R}')^{-1}$ diagonal matrix with the elements $[(\mathbf{I} + \mathbf{R}')_{i,i}]^{-1} = \dfrac{1}{1+r'_{i,i}}$. On rewriting the $[(\mathbf{I} + \mathbf{R}')_{i,i}]^{-1}$ diagonal elements as Taylor series and neglecting the non-linear terms, we reach $\left(\mathbf{I} + \mathbf{R}'\right)^{-1} \approx \mathbf{I} - \mathbf{R}'$. Therefore, from eqs.~(\ref{lap_u})-(\ref{div_u}), it is arrived at:
\begin{align}
    \nabla^2 \mathbf{u} &= \nabla_0^2 \mathbf{U} (\mathbf{I} - \mathbf{R}')^{2} + \nabla_0^2 \mathbf{u}' + \mathbf{x}' \cdot \nabla_0 (\nabla_0^2 \mathbf{U}) \label{final_lap_u}\\
    \nabla \mathbf{p} &= \nabla_0 P(\mathbf{I} - \mathbf{R}') + \nabla_0 p' + \mathbf{x}'\cdot\nabla_0(\nabla_0 P)\\
    \nabla \cdot \mathbf{u} &= \nabla_0 \cdot \mathbf{U} (\mathbf{I} - \mathbf{R}') + \nabla_0 \cdot \mathbf{u}' + \mathbf{x}' \cdot\nabla_0 (\nabla_0 \mathbf{U}). \label{final_div_u}.
\end{align}
Once again, the non-linear terms were discarded. 

Finally, substituting eqs.~(\ref{eq:totalacc}) and (\ref{final_lap_u})--(\ref{final_div_u}) into eqs.~(\ref{momento_FSI}) -- (\ref{massaFSI}), the momentum and mass conservation equations read:
\begin{eqnarray*}
\left.\frac{\partial \mathbf{u}'}{\partial t}\right|_{\mathbf{x}_0} +  \mathbf{U} \cdot \nabla_0 \mathbf{u}'  + \mathbf{u}' \cdot \nabla_0 \mathbf{U}  - \frac{1}{Re}\nabla^2_0 \mathbf{u}' + \nabla_0 p' + \nonumber\\
+ \mathbf{x}' \cdot  \nabla_0 \left[\nabla_0 \mathbf{U} \cdot \mathbf{U}  - \frac{1}{Re}\nabla^2_0 \mathbf{U} + \nabla_0  P \right] +   \left[\mathbf{U} \cdot\nabla_0 \mathbf{U}  - \frac{1}{Re}\nabla^2_0 \mathbf{U} + \nabla_0  P \right] &=& \mathbf{0}, \\[6pt]
(\nabla_0 \cdot \mathbf{U} + \nabla_0 \cdot \mathbf{u}') +  \mathbf{x}' \cdot  \nabla_0 (\nabla_0 \cdot \mathbf{U})  = 0.
\end{eqnarray*}
The algebraic development to achieve the linearized flow equations is presented in the current work in brief way; more details can be found in \cite{Prabal2020}.

Since the steady base flow is governed by the system:
\begin{equation} \label{NS}
    \mathbb{N} (\mathbf{Q}) = \left\{
    \begin{array}{rcl}
        \nabla_0 \mathbf{U} \cdot \mathbf{U}  - \dfrac{1}{Re}\nabla^2_0 \mathbf{U} + \nabla_0  P   &=& \mathbf{0},\\
        \nabla_0 \cdot \mathbf{U}&=& 0,  
    \end{array}
    \right.
\end{equation}
the linearized Navier-Stokes system can be simplified to
\begin{align*}
     \left.\frac{\partial \mathbf{u}'}{\partial t}\right|_{\mathbf{x}_0} +  \mathbf{U} \cdot \nabla_0 \mathbf{u}'  + \mathbf{u}' \cdot \nabla_0 \mathbf{U} - \frac{1}{Re}\nabla^2_0 \mathbf{u}' + \nabla_0 p'  &= \mathbf{0},\\
    \nabla_0 \cdot \mathbf{u}' &= 0.
\end{align*}
It is worth remembering that $\mathbf{U} = \mathbf{U}(\mathbf{x}_{0})$ and $\mathbf{u}' = \mathbf{u}'(\mathbf{x}_{0}, t)$. 

\subsubsection{Structure equations}

Following \cite{Fernandez2002}, the structure displacement $\gamma(t)$ is here written around the equilibrium configuration as a linear combination of a finite number of vibration modes. For the linear analysis, we consider a small displacement around the equilibrium configuration. So the structure displacement is written as a linear combination of the first order $\gamma(t) = \gamma_0 + \gamma'(t)$, where $\gamma_0$ represents the displacement for the equilibrium configuration which is achieved when the structure is steady. Therefore, $\gamma_0 = 0$ is a proper assumption.

The fluid force $F(\mathbf{u}(\mathbf{x}, t), p(\mathbf{x}, t))$ is rewritten to be computed in the domain $\mat{\Omega}_0$. To do so, the variables $\mathbf{u}(\mathbf{x}, t)$ and $p(\mathbf{x}, t)$ are written as the Taylor expansions (\ref{velocity_x0}) -- (\ref{pressure_x0}), and spatial gradients and the spatial integral are rewritten in the domain $\mat{\Omega}_0$. Therefore, according to these assumptions, the fluid force acting on the structure is rewritten as:
\begin{align*}
    F (\mathbf{u}(\mathbf{x}, t), p(\mathbf{x}, t))  &= \int_{\partial \mat{\Omega}_w} \mathbf{n}_y\cdot \sigma (\mathbf{u}(\mathbf{x}, t), p(\mathbf{x}, t)) \text{d}S_{w}\\
    &= \int_{\partial \mat{\Omega}_{w,0}} \mathbf{n}_y\cdot \sigma (\mathbf{U}(\mathbf{x}_0), P(\mathbf{x}_0)) \left|\dfrac{\partial \mathbf{x}}{\partial \mathbf{x}_0}\right| \text{d}S_{w,0} + \\
    &+  \int_{\partial \mat{\Omega}_{w,0}} \mathbf{n}_y\cdot \sigma (\mathbf{u}'(\mathbf{x}_0,t), p'(\mathbf{x}_0,t)) \left|\dfrac{\partial \mathbf{x}}{\partial \mathbf{x}_0}\right| \text{d}S_{w,0} +\\
    &+ \int_{\partial \mat{\Omega}_{w,0}} \mathbf{n}_y\cdot \left[\nabla\sigma (\mathbf{U}(\mathbf{x}_0), P(\mathbf{x}_0))\cdot\mathbf{x}'\right] \left|\dfrac{\partial \mathbf{x}}{\partial \mathbf{x}_0}\right| \text{d}S_{w,0}, 
\end{align*}
where $\partial \mat{\Omega}_{w, 0}$ refers to the structure surface in the domain $\mat{\Omega}_0$. The expression $ \left|\dfrac{\partial \mathbf{x}}{\partial \mathbf{x}_0}\right|$ is the determinant of the Jacobian matrix $\dfrac{\partial \mathbf{x}}{\partial \mathbf{x}_0} = \mathbf{I} + \mathbf{R}'$. This way, by discarding the nonlinear terms evolving the perturbation field, the force in the cross flow direction is  written as:
\begin{align*}
    F (\mathbf{u}(\mathbf{x}, t), p(\mathbf{x}, t))  &=\int_{\partial \mat{\Omega}_{w,0}} \mathbf{n}_y\cdot \sigma (\mathbf{U}(\mathbf{x}_0), P(\mathbf{x}_0)) \left|\mathbf{I} + \mathbf{R}'\right| \text{d}S_{w,0} + \\
    &+  \int_{\partial \mat{\Omega}_{w,0}} \mathbf{n}_y\cdot \sigma (\mathbf{u}'(\mathbf{x}_0,t), p'(\mathbf{x}_0,t))  \text{d}S_{w,0} +\\
    &+ \int_{\partial \mat{\Omega}_{w,0}} \mathbf{n}_y\cdot \left[\nabla\sigma (\mathbf{U}(\mathbf{x}_0), P(\mathbf{x}_0))\cdot \mathbf{x}'\right] \text{d}S_{w,0}, 
\end{align*}
For a steady base flow around a circular cylinder with zero lift, also by considering the cylinder displacement in the cross flow direction, we obtain
\[\int_{\partial \mat{\Omega}_{w,0}} \mathbf{n}_y\cdot \sigma (\mathbf{U}(\mathbf{x}_0), P(\mathbf{x}_0)) \left|\mathbf{I} + \mathbf{R}'\right| \text{d}S_{w,0} = 0.\]
Besides that, the term
\[ \int_{\partial \mat{\Omega}_{w,0}} \mathbf{n}_y\cdot \left[\nabla\sigma (\mathbf{U}(\mathbf{x}_0), P(\mathbf{x}_0))\cdot\mathbf{x}'\right]  \text{d}S_{w,0},
\]
which was referred by 'added stiffness' \cite{Fernandez2003}, is neglected in this work. As showed by \citep{Prabal2020}, for the flow around rigid bodies (including rigid circular cylinder) 'added stiffness' has a smaller order than that of the forces of fluid perturbations. Therefore, the fluid force is reduced to:
\begin{align*}
    F (\mathbf{u}(\mathbf{x}, t), p(\mathbf{x}, t))  &= \int_{\partial \mat{\Omega}_{w,0}} \mathbf{n}_y\cdot \sigma (\mathbf{u}'(\mathbf{x}_0, t), p'(\mathbf{x}_0, t)) \text{d}S_{w,0}.
\end{align*}
In conclusion, for steady base flow and for displacement only on the $y$-axis, the mass-spring-damper system that governs the displacement of the structure due to a perturbation is given by:
\begin{equation}\label{MSD_eq}
    M^{\ast} \ddot{\gamma'} + C^{\ast} \dot{\gamma}' + K^{\ast} \gamma' = \int_{\partial \mat{\Omega}_{w, 0}} \mathbf{n}_y \cdot\, \sigma (\mathbf{u}'(\mathbf{x}_0, t), p'(\mathbf{x}_0, t))  \text{d}S_{w, 0}.
\end{equation}

Again, it is worth emphasise that this structure equation for the perturbation is valid for this particular case. As discussed in \cite{Pfister2020}, the 'added stiffness' has a non-negligible effect in the linear analysis for a flexible splitter plate interacting with a circular cylinder flow. Therefore, discarding the added stiffness must not be understood as a step to be taken in all cases. 

\subsubsection{Boundary conditions}
By applying the first-order Taylor expansion around an equilibrium point $\mathbf{x}_0$ and by using $\mathbf{u} = \mathbf{U} + \mathbf{u}'$, the boundary conditions imposed on the linearized FSI system are set as follows:
\begin{itemize}
    \item \underline{Inlet} : $\mathbf{u} = \mathbf{U}_c + \nabla \mathbf{U}_c \cdot \mathbf{x}' + \mathbf{u}'(\mathbf{x}_{0}, t)$. 
    By imposing an uniform velocity at inlet ($\mathbf{U}_c$), the boundary condition of the perturbation velocity is :
    \begin{equation}\label{inlet}
        \mathbf{u}'(\mathbf{x}_{0}, t) = \mathbf{0}.
    \end{equation}
    
    \item \underline{Structure wall}: At this boundary $\mathbf{u} = \mathbf{U} + \nabla \mathbf{U} \cdot \mathbf{x}' + \mathbf{u}'(\mathbf{x}_{0}, t) = [0,\dot{\gamma}] = [0, \dot{\gamma}_0+\dot{\gamma}']^{T} = [0, \dot{\gamma}']^{T}$, in which $\mathbf{U}=0$. Therefore,
    \begin{align*}
        \mathbf{u} &= \left[0, \dot{\gamma}'\right]^{T} =  \left[\frac{\partial U}{\partial x_0}x' + \frac{\partial U}{\partial y_0}  y' + u'(\mathbf{x}_{0}, t), \frac{\partial V}{\partial x_0}x' + \frac{\partial V}{\partial y_0}  y' + v'(\mathbf{x}_{0}, t)\right]^{T} \\
        &= \left[\frac{\partial U}{\partial y_0} \gamma' + u'(\mathbf{x}_{0}, t),  \frac{\partial V}{\partial y_0}  \gamma' + v'(\mathbf{x}_{0}, t)\right]^{T}  \quad \Rightarrow 
    \end{align*}
    \begin{equation}\label{wall}
        \Rightarrow \quad \mathbf{u}'(\mathbf{x}_{0}, t) =\left [u'(\mathbf{x}_{0}, t), v'(\mathbf{x}_{0}, t)\right]^{T} =  \left[-\frac{\partial U}{\partial y_0} \gamma', \dot{\gamma}'(t) - \frac{\partial V}{\partial y_0}  \gamma' \right]^{T}
    \end{equation}
     
    \item \underline{Outlet}: To impose the boundary condition at the outlet we consider that this boundary is very far from cylinder. Therefore, a Neumann boundary condition $\nabla \mathbf{u}\cdot \mathbf{n} = 0$ is applied at this boundary, implying  $\nabla_0 \mathbf{U}\cdot \mathbf{n} = 0$ and 
    \begin{equation}\label{outlet}
        \nabla_0 \mathbf{u}'\cdot \mathbf{n} = 0.   
    \end{equation}
\end{itemize}

\subsubsection{Final systems/Modal analysis}
In conclusion, the steady base flow is computed by solving the system (\ref{NS}) which satisfies the following boundary conditions:
\begin{align}
    (U, V) &= (1,0) \quad \forall \mathbf{x}_0 \in \partial \mat{\Omega}_{i, 0},  \\
    \mathbf{U} &= \mathbf{0}  \quad \forall \mathbf{x}_0 \in \partial \mat{\Omega}_{w, 0}, \\
    \nabla_0 \mathbf{U}\cdot \mathbf{n} &= \mathbf{0} \quad \forall \mathbf{x}_0 \in \partial \mat{\Omega}_{o, 0} .
\end{align} 
The inlet and outlet boundaries are represented by $\partial \mat{\Omega}_{i, 0}$ and $\partial \mat{\Omega}_{o, 0}$, respectively.

For a rigid structure free to oscillate in the cross flow direction, the linearized FSI system that governs the perturbation is given by:
\begin{align}
    \dfrac{\partial \mathbf{u}'}{\partial t} +  \mathbf{U} \cdot \nabla_0 \mathbf{u}'  + \nabla_0 \mathbf{U} \cdot \mathbf{u}' - \frac{1}{Re}\nabla_0^2 \mathbf{u}' + \nabla  p'  &= 0,\label{FSI1} \\  
    \nabla_0 \cdot \mathbf{u}' &= 0, \\
    \dot{\gamma}' &= \gamma_1',\\
    M^{\ast} \dot{\gamma'}_1 + C^{\ast} \gamma_1' + K^{\ast} \gamma' &= \int_{\partial \mat{\Omega}_{w, 0}} \mathbf{n}_y\cdot\sigma (\mathbf{u}'(\mathbf{x}_0, t), p'(\mathbf{x}_0, t))  \text{d}S_{w, 0} \label{FSI3},
\end{align}
satisfying the boundary conditions defined in previous section, eqs.~(\ref{inlet})--(\ref{outlet}).

Since this is a linear system, for modal analysis, the solution of the perturbation is:
\[\mathbf{q}'(\mathbf{x}_0, t) = \widehat{\mathbf{q}} \text{exp}(\lambda t) = \left[ \widehat{\mathbf{u}}(\mathbf{x}_0) \text{exp}(\lambda t),\quad \widehat{p}(\mathbf{x}_0) \text{exp}(\lambda t),\quad \widehat{\gamma} \text{exp}(\lambda t), \quad \dot{\widehat{\gamma}} \text{exp}(\lambda t) \right]^{T}.\] 
Therefore, the linearized FSI system can be rewritten as:
\begin{equation} \label{PAG}
    (\mat{\lambda} \mathbb{B} - \mathbb{L}) \widehat{\mathbf{q}} = \left\{
    \begin{array}{rcl}
        \lambda \widehat{\mathbf{u}}  +  \mathbf{U} \cdot \nabla_0 \widehat{\mathbf{u}}  + \nabla_0 \mathbf{U} \cdot \widehat{\mathbf{u}} - \dfrac{1}{Re}\nabla^2_0 \widehat{\mathbf{u}} + \nabla_0  \widehat{p} &=& 0,  \\
        \nabla_0 \cdot \widehat{\mathbf{u}} &=& 0, \\
        \lambda \widehat{\gamma} - \widehat{\gamma}_1 &=& 0, \\
        \lambda \dot{\widehat{\gamma}}_{1}  +  \dfrac{C^{\ast}}{M^{\ast}} \widehat{\gamma}_{1} + \dfrac{K^{\ast}}{M^{\ast}} \widehat{\gamma} -\dfrac{1}{M^{\ast}}F(\widehat{\mathbf{u}}, \widehat{p}) &=&  0. 
    \end{array}
    \right.
\end{equation}

\citet{Fernandez2002, Pfister2019, Prabal2020} presented the formulation to linearize the FSI system using the Arbitrary Lagrangian Eulerian method (ALE). In the end, they showed the steady base flow equations and the linear FSI system are independent of the arbitrary velocity of the mesh displacement. The transforms to reach this independence of mesh motion generate the transpiration boundary conditions at the structure wall, as shown in eq.~(\ref{wall}). \citet{Pettit2000, Gao2005, Kirshman2006, Bekka2015} used the transpiration approach in the inviscid flow around structures with small displacements. In those works, the authors did not employ any methodology (like ALE, non-inertial frame of reference) to deal with the mesh motion. Likewise, in this work we did not choose a methodology to capture the mesh displacement. Nonetheless, we obtained a linearized FSI system that is the same as that presented by \cite{Prabal2020}. They employed the same transform $\mathbf{x}(t) = \mathbf{x}_0+\mathbf{x}'(t)$ such that $\mathbf{x}'(t) = \mathbf{R}'\mathbf{x}_0$, and the structure was rigid as well. 
\subsection{Sensitivity to external forcing}\label{Sens_SteadyForceFSI}
To obtain the sensitivity field to an external forcing, this work follows the approach previously introduced by \cite{Pralitis:2010, Fani:2012, Carini:2014}. In those works, an external forcing proportional to the velocity $\mathbf{u}$ was considered. Following the methodology presented by them, the forcing can be written as
\begin{equation*} 
    \mathbf{G} = -G \delta(x-x_p, y-y_p) \mathbf{u}(\mathbf{x}, t).
\end{equation*}
The local forcing is here represented by a Gaussian function $f(\mathbf{x})$, where $G \delta(x-x_p, y-y_p) = G f(\mathbf{x})$. By writing $f(\mathbf{x})$ as a first order Taylor expansion around an equilibrium point $\mathbf{x}_0$, we have $f(\mathbf{x}) = f(\mathbf{x}_0) + \nabla_0 f(\mathbf{x}_0) \cdot \mathbf{x}'$. We assume $\mathbf{G}$ is a local forcing, such that $\mathbf{G}\neq 0$ only for $|\mathbf{x}_0-\mathbf{x}_p| \ll 1$. As $\nabla_0 f(\mathbf{x}_0)$ has a coefficient given by $\mathbf{x}_0-\mathbf{x}_p$, the term $\nabla_0 f(\mathbf{x}_0) \cdot \mathbf{x}'$ has smaller order than the perturbation and can be neglected. Therefore, the Gaussian function may be written as $f(\mathbf{x}) \approx f(\mathbf{x}_0)$ and the external forcing is then rewritten in the following way:
\begin{equation} \label{forcing}
    \mathbf{G} = -G \delta(x_0-x_p, y_0-y_p) \mathbf{u}(\mathbf{x}, t).
\end{equation}
We substitute $\mathbf{u}(\mathbf{x}, t)$ by the expression (\ref{velocity_x0}), arriving at
\begin{equation} \label{forcing_final}
     \mathbf{G} = - G \delta(x_0-x_p, y_0-y_p) \left[\mathbf{U}(\mathbf{x}_{0}) + \nabla_0 \mathbf{U}(\mathbf{x}_{0},t)\cdot\mathbf{x}' + \mathbf{u}'(\mathbf{x}_{0}, t)\right].
\end{equation}
By adopting the approach presented in subsection \ref{linear_sys}, the linearized FSI system with a forcing added in the momentum equation can be written as:
\begin{eqnarray*}
    \left.\frac{\partial \mathbf{u}'}{\partial t}\right|_{\mathbf{x}_0} +  \mathbf{U} \cdot \nabla_0 \mathbf{u}'  + \mathbf{u}' \cdot \nabla_0 \mathbf{U}  - \frac{1}{Re}\nabla^2_0 \mathbf{u}' + \nabla_0 p' +  \alpha \mathbf{u}'(\mathbf{x}_{0}) +& &\nonumber\\
    + \mathbf{x}' \cdot  \nabla_0 \left[\nabla_0 \mathbf{U} \cdot \mathbf{U}  - \frac{1}{Re}\nabla^2_0 \mathbf{U} + \nabla_0  P + \alpha \mathbf{U}(\mathbf{x}_{0}) \right] + & &\nonumber \\
    \left[\nabla_0 \mathbf{U} \cdot \mathbf{U}  - \frac{1}{Re}\nabla^2_0 \mathbf{U} + \nabla_0  P +  \alpha \mathbf{U}(\mathbf{x}_{0}) \right] &=& \mathbf{0}, \\
    (\nabla_0 \cdot \mathbf{U} + \nabla_0 \cdot \mathbf{u}') + \mathbf{x}' \cdot \nabla_0 (\nabla_0 \cdot \mathbf{U})  &=& 0,
\end{eqnarray*}
where $\alpha=G \delta(x_0-x_p, y_0-y_p)$. The mathematical expression of sensitivity is here obtained by using the Lagrangian functional. According to what was shown in the previous section, the linearized FSI system can be written as a generalized eigenvalue problem. Therefore, the Lagrangian functional reads:
\begin{align}\label{augmented_funct}
    \mathcal{L}(\mathbf{Q},\alpha, \widehat{\mathbf{q}}, \lambda,  \mathbf{Q}^{\dagger}, \widehat{\mathbf{q}}^{\dagger}) = \lambda - 
    \int_{\mat{\Omega}_0} \widehat{\mathbf{q}}^{\dagger} \cdot \left[(\lambda \mathbb{B} - \mathbb{L}) \widehat{\mathbf{q}} + \alpha\widehat{\mathbf{u}}\right] \text{d}\mathcal{V}_0  +
    \nonumber \\
    -\int_{\mat{\Omega}_0} \mathbf{Q}^{\dagger}\cdot\left\{(1+\mathbf{x}'\cdot \nabla_0) \left[ \mathbb{N}(\mathbf{Q}) + \alpha \mathbf{U}\right]\right\}\text{d}\mathcal{V}_0. 
\end{align}

The variables $\widehat{\mathbf{q}}^{\dagger}$ and $\mathbf{Q}^{\dagger}$ are the Lagrange multipliers defined in the domain $\mat{\Omega}_0$. To reach an optimum value, the gradient of the Lagrangian functional with respect to any variable must be zero. To reach such objective, the gradient with respect to any variable $s$ is computed here using the Gateaux derivative
\begin{equation} \label{Gateaux}
    \frac{\partial \mathcal{L}}{\partial s} = \lim_{\epsilon \rightarrow 0} \frac{\mathcal{L}(s + \epsilon \delta s) - \mathcal{L}(s)}{\epsilon} .
\end{equation}
Therefore, computing the derivative $\dfrac{\partial \mathcal{L}}{\partial \mathbf{Q}^{\dagger}} \delta \mathbf{Q}^{\dagger}$ and making it equal to zero, we obtain:
\begin{align*}
    &\dfrac{\partial \mathcal{L}}{\partial \mathbf{Q}^{\dagger}} \delta \mathbf{Q}^{\dagger} =\\
    &=\lim_{\epsilon \rightarrow 0} \frac{ \int_{\mat{\Omega}_0} (\mathbf{Q}^{\dagger} +  \epsilon \delta \mathbf{Q}^{\dagger})\cdot (1+\mathbf{x}'\cdot \nabla_0) \left[ \mathbb{N}(\mathbf{Q}) + \alpha \mathbf{U}\right] - \mathbf{Q}^{\dagger} \cdot (1+\mathbf{x}'\cdot \nabla_0)  \left[ \mathbb{N}(\mathbf{Q}) + \alpha \mathbf{U}\right] \, \text{d}\mathcal{V}_0}{\epsilon}  \\
    &= \lim_{\rightarrow 0}  \int_{\mat{\Omega}_0} \epsilon\delta \mathbf{Q}^{\dagger} \cdot (1+\mathbf{x}'\cdot \nabla_0)\left[ \mathbb{N}(\mathbf{Q}) + \alpha \mathbf{U}\right]  \text{d}\mathcal{V}_0  = 0.
\end{align*}
This expression is true if $\mathbb{N} \mathbf{Q} + \alpha \mathbf{U} = \mathbf{0}$ for all $\mathbf{x}_0 \in \mat{\Omega}_0$. Similarly, the gradient $\dfrac{\partial \mathcal{L}}{\partial \widehat{\mathbf{q}}^{\dagger}} \delta \widehat{\mathbf{q}}^{\dagger}=0$ requires the constraint $(\mat{\lambda} \mathbb{B} - \mathbb{L}) \cdot \widehat{\mathbf{q}} + \alpha\widehat{\mathbf{u}} = \mathbf{0}$ be satisfied for $\mathbf{x}_0 \in \mat{\Omega}_0$.

\subsubsection{Sensitivity}
In the sensitivity analysis, we evaluate the variation of the eigenvalue $\lambda$ as a response to the external forcing (\ref{forcing}), which is given by the gradient:
\begin{equation*} 
    \delta \lambda = S1+S2 = \nabla_{\alpha} \lambda  = -\int_{\mat{\Omega_0}} \mathbf{U}^{\dagger}\cdot (1+\mathbf{x}'\cdot \nabla_0) (\alpha \mathbf{U}) \, \text{d} \mathcal{V}_0 - \int_{\mat{\Omega_0}} \widehat{\mathbf{u}}^{\dagger}\cdot (\alpha \widehat{\mathbf{u}}) \, \text{d} \mathcal{V}_0.
\end{equation*}
By supposing $\alpha$ a local and small amplitude actuation, we can neglect the term $\mathbf{x}'\cdot \nabla_0 (\alpha \mathbf{U})$. An equal assumption cannot be adopted for the $S2$ term due to the normalisation given by eq.~($\ref{normalization}$) presented in the appendix, which was necessary to obtain the adjoint system. Therefore, the final sensitivity expression is given by:
\begin{equation} \label{eig_variation}
    \delta \lambda = S1+S2 = \nabla_{\alpha} \lambda  = -\int_{\mat{\Omega_0}} \alpha \mathbf{U}^{\dagger}\cdot \mathbf{U} \, \text{d} \mathcal{V}_0 - \int_{\mat{\Omega_0}} \alpha \widehat{\mathbf{u}}^{\dagger}\cdot \widehat{\mathbf{u}} \, \text{d} \mathcal{V}_0.
\end{equation}

The first integral on the right side will be referred to as the base flow sensitivity to a steady forcing (S1), and the second term is referred to as structural sensitivity (S2). These names are based on the works by \cite{Marquet2008} and \cite{Giannetti2007}, respectively.

\subsubsection{Adjoint systems}

The sensitivity computations depend on the Lagrange multipliers $\mathbf{Q}^{\dagger}$ and $\hat{\mathbf{q}}^{\dagger}$. These variables are the solution of the adjoint systems that are obtained by considering the gradient of the augmented functional, eq.~(\ref{augmented_funct}) with respect to the state variables $\mathbf{Q}$ and $\widehat{\mathbf{q}}$, respectively. If we make the algebraic development of theses gradients, we conclude that the Lagrange multiplier/adjoint variable $\mathbf{U}^{\dagger}$ is solution of the adjoint system:
\begin{eqnarray}
    \nabla  \mathbf{U} \cdot \mathbf{U}^{\dagger} -  \nabla \mathbf{U}^{\dagger} \cdot \mathbf{U} - Re^{-1} \nabla^2 \mathbf{U}^{\dagger} - \nabla P^{\dagger} &=& -(\nabla  \widehat{\mathbf{u}} \cdot \widehat{\mathbf{u}}^{\dagger} -  \nabla \widehat{\mathbf{u}}^{\dagger}  \cdot  \widehat{\mathbf{u}}), \label{steadyforce_adj1} \\
    \nabla \cdot \mathbf{U}^{\dagger} &=& 0, \label{steadyforce_adj2}\\
    \mathbf{U}^{\dagger}(\mathbf{x}_0, t)  &=& \mathbf{0}  \quad \forall\mathbf{x}_0 \in \partial\mat{\Omega}_{i,0}, \nonumber\\
     \mathbf{U}^{\dagger}(\mathbf{x}_0, t)  &=& \mathbf{0} \quad \forall \mathbf{x}_0 \in \partial\mat{\Omega}_{w,0}, \nonumber\\
    (\hat{\mathbf{u}} \cdot \mathbf{n})\mathbf{u}^{\dagger} + (\mathbf{U} \cdot \mathbf{n})\mathbf{U}^{\dagger} + P^{\dagger} + Re^{-1} \nabla \mathbf{U}^{\dagger} &=& \mathbf{0} \quad \forall \mathbf{x}_0 \in \partial\mat{\Omega}_{o,0}. \nonumber
\end{eqnarray}
The adjoint system (\ref{steadyforce_adj1})--(\ref{steadyforce_adj2}) is equal to the adjoint system introduced by \citet{Marquet2008}. However, here the direct ($\widehat{\mathbf{u}}$) and adjoint ($\widehat{\mathbf{u}}^{\dagger}$) modes are solutions of the generalized eigenvalue problem of the FSI system, whereas in \citet{Marquet2008} the direct and adjoint modes were related to the generalized eigenvalue problem of a fluid flow system.

The Lagrange multiplier $\widehat{\mathbf{q}}^{\dagger}$ is computed by solving the adjoint generalized eigenvalue problem given by:
\begin{align}
    -\lambda \widehat{\mathbf{u}}^{\dagger}  - \mathbf{U} \cdot \nabla_0 \widehat{\mathbf{u}}^{\dagger} + \nabla_0 \mathbf{U}\cdot \widehat{\mathbf{u}}^{\dagger}  - Re^{-1}\nabla^{2} \widehat{\mathbf{u}}^{\dagger}  - \nabla_0 p^{\dagger} &= \mathbf{0}, \label{GEP_adj1}\\
    \nabla_0 \cdot \widehat{\mathbf{u}}^{\dagger} &= 0,\\
    \lambda\widehat{\gamma}^{\dagger}  + K^{\ast}_1\widehat{\gamma}_{1}^{\dagger}  - \int_{\partial \mat{\Omega}_{w, 0}}\left[
    \dfrac{\partial V}{\partial y_0} \sigma(\widehat{\mathbf{u}}^{\dagger}, - p^{\dagger})\right] \cdot \mathbf{n}_y\, \text{d}S_{w, 0}&= \mathbf{0},\\
  - M^{\ast} \lambda\dot{\widehat{\gamma}}_{1}^{\dagger} + C^{\ast} \widehat{\gamma}_{1}^{\dagger} - \widehat{\gamma}^{\dagger} + \int_{\partial \mat{\Omega}_{w, 0}} \sigma(\widehat{\mathbf{u}}^{\dagger}, - p^{\dagger}) \cdot \mathbf{n}_y\, \text{d}S_{w, 0} &= \mathbf{0}, \label{GEP_adj4}\\
  \widehat{\mathbf{u}}^{\dagger} &= \mathbf{0} \quad \forall\mathbf{x}_0 \in \partial\mat{\Omega}_i, \nonumber\\
  (\widehat{u}^{\dagger}, \widehat{v}^{\dagger})  &= -(0, \widehat{\gamma}_{1}^{\dagger} ) \quad \forall \mathbf{x}_0 \in \partial\mat{\Omega}_w \nonumber\\
  (\mathbf{U} \cdot \mathbf{n})\widehat{\mathbf{u}}^{\dagger} + Re^{-1} \nabla_0 \widehat{\mathbf{u}}^{\dagger}\cdot \mathbf{n} &= p^{\dagger} = \mathbf{0} \quad \forall \mathbf{x}_0  \in \partial\mat{\Omega}_o \nonumber
\end{align}
Details of the algebraic development to obtain the adjoint systems are described in appendix~\ref{app:adjoint}.

\section{Numerical methodology} \label{Num_Method} 

Numerical results were computed using the Nektar++ software, which is based on the Spectral/\textit{hp} Element Method \citep{Karn/Sher05}. A second-order stiffly stable time-stepping scheme \citep{Karn/Isra/Orsz91} was employed to advance the Navier-Stokes equations in time. In this work, the direct and adjoint modes were computed by considering a steady base flow that was obtained by solving the time dependent Navier-Stokes equations for a sufficiently large time to reach the steady state. In the base flow calculations, the cylinder was fixed. For the flow around an elastically-mounted cylinder, the non-linear FSI system was solved using a coordinate mapping technique which is used for the treatment of the moving wall \citep{Serson:2016}. In this case, the simulations were carried out in a time interval long enough for the structure displacement to reach a constant amplitude of oscillation. The mass-spring-damper system that provides the dynamic response of the cylinder was calculated using a Newmark-$\beta$ solver \citep{Newm59}. The generalised eigenvalue problems were solved using a modified Arnoldi algorithm \citep{Saad92}. All FSI results presented in this work considered an elastically-mounted cylinder free to oscillate only in the cross flow direction. 

The structure considered was a circular cylinder of diameter $D = 1$. This solid body was immersed in a uniform flow of magnitude $U = 1$ parallel to the $x$-axis, pointing to the $x+$ direction. The origin of the coordinate system was at the centre of the cylinder, and the domain had dimensions: $x+ = 50D$ downstream, $x- = -50D$ upstream and $y\pm = 50D$ cross-stream. In order to assure low numerical error, we have performed a convergence analysis of the polynomial degree. After performing this convergence test, we decided to adopt seventh-degree polynomial order for all simulations. Every FSI computation performed adopted structural damping $\zeta=0$.

\section{Results}\label{Results} 

\begin{figure}
    \centering
    \subfigure[\label{Amp_Re46_m20}]{\includegraphics[scale=0.155]{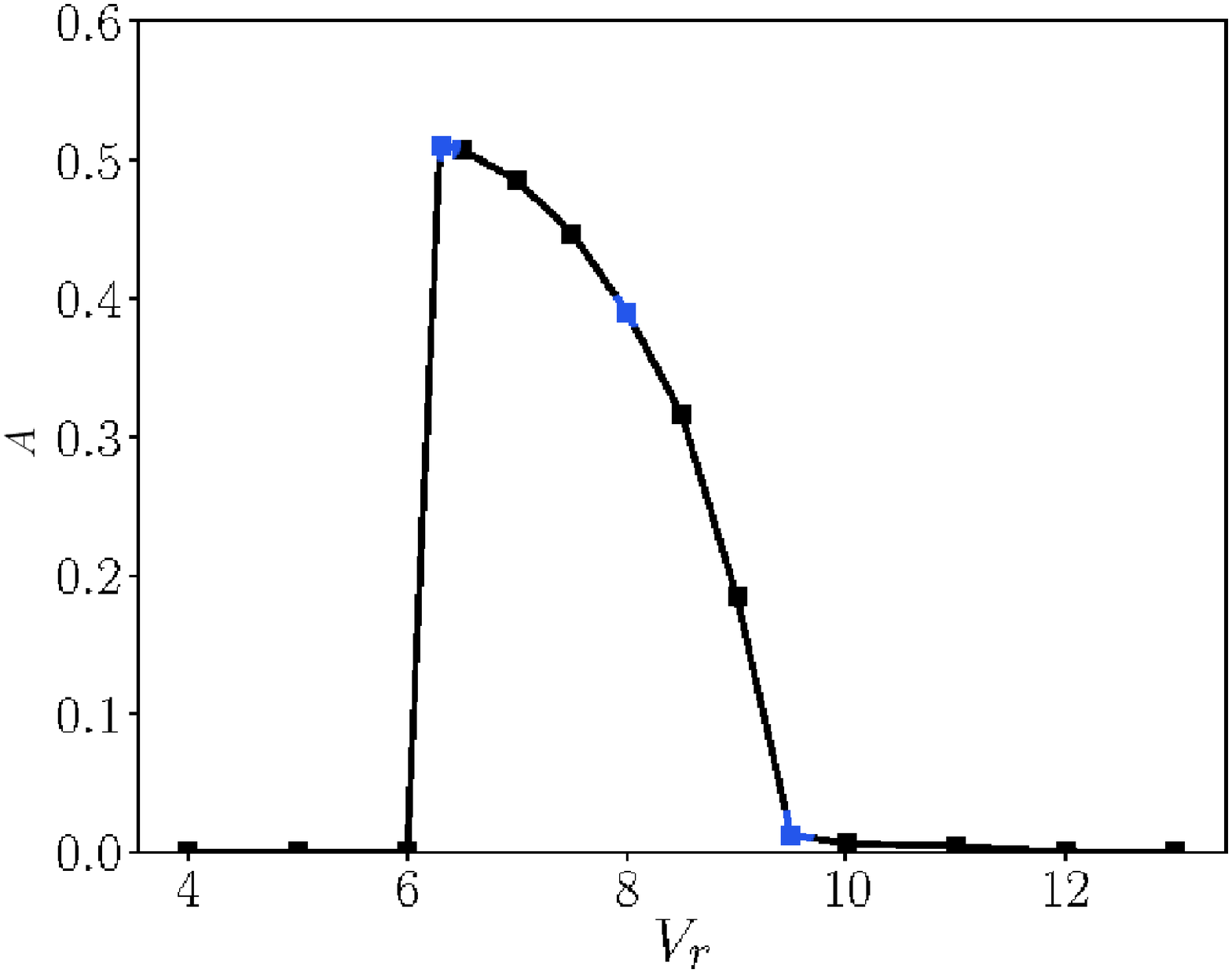}}
    \subfigure[\label{eigen_Re46_m20}]{\includegraphics[scale=0.3]{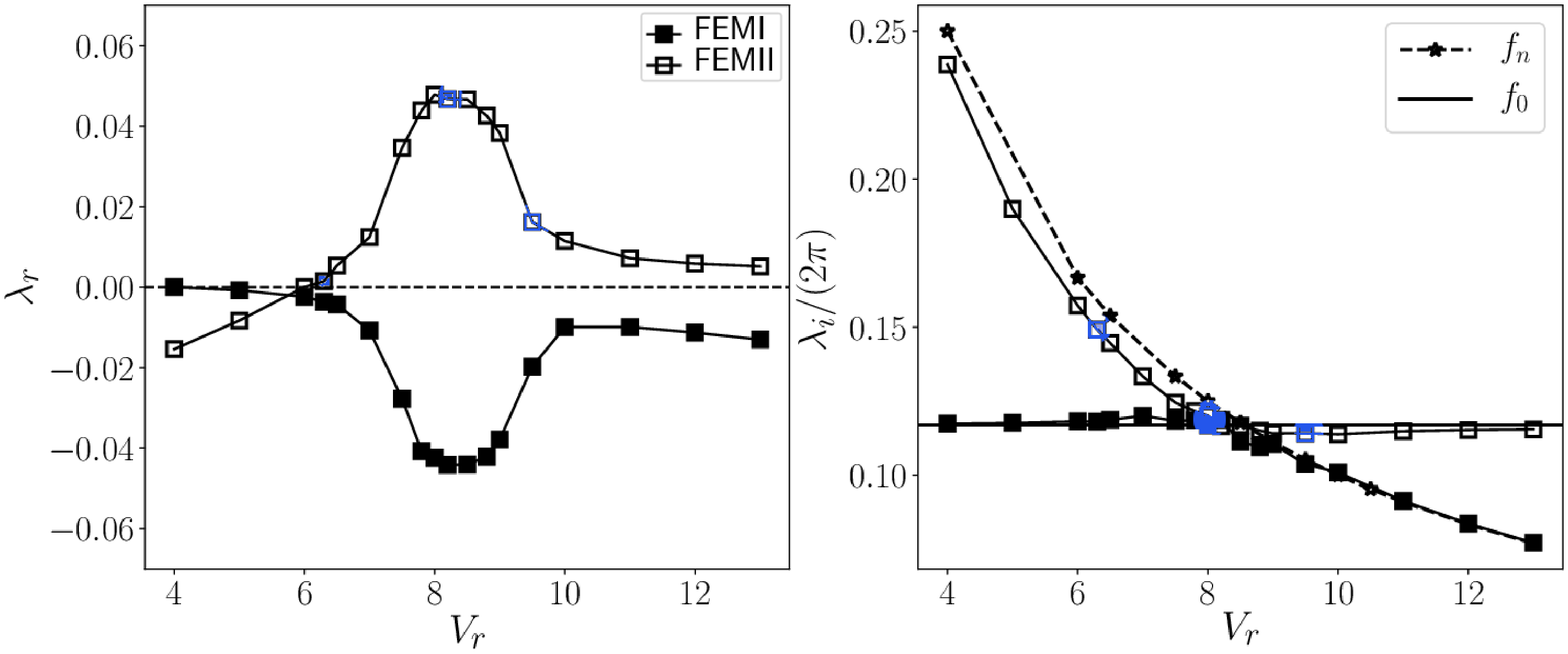}}
     \subfigure[\label{Amp_Re46_m7}]{\includegraphics[scale=0.165]{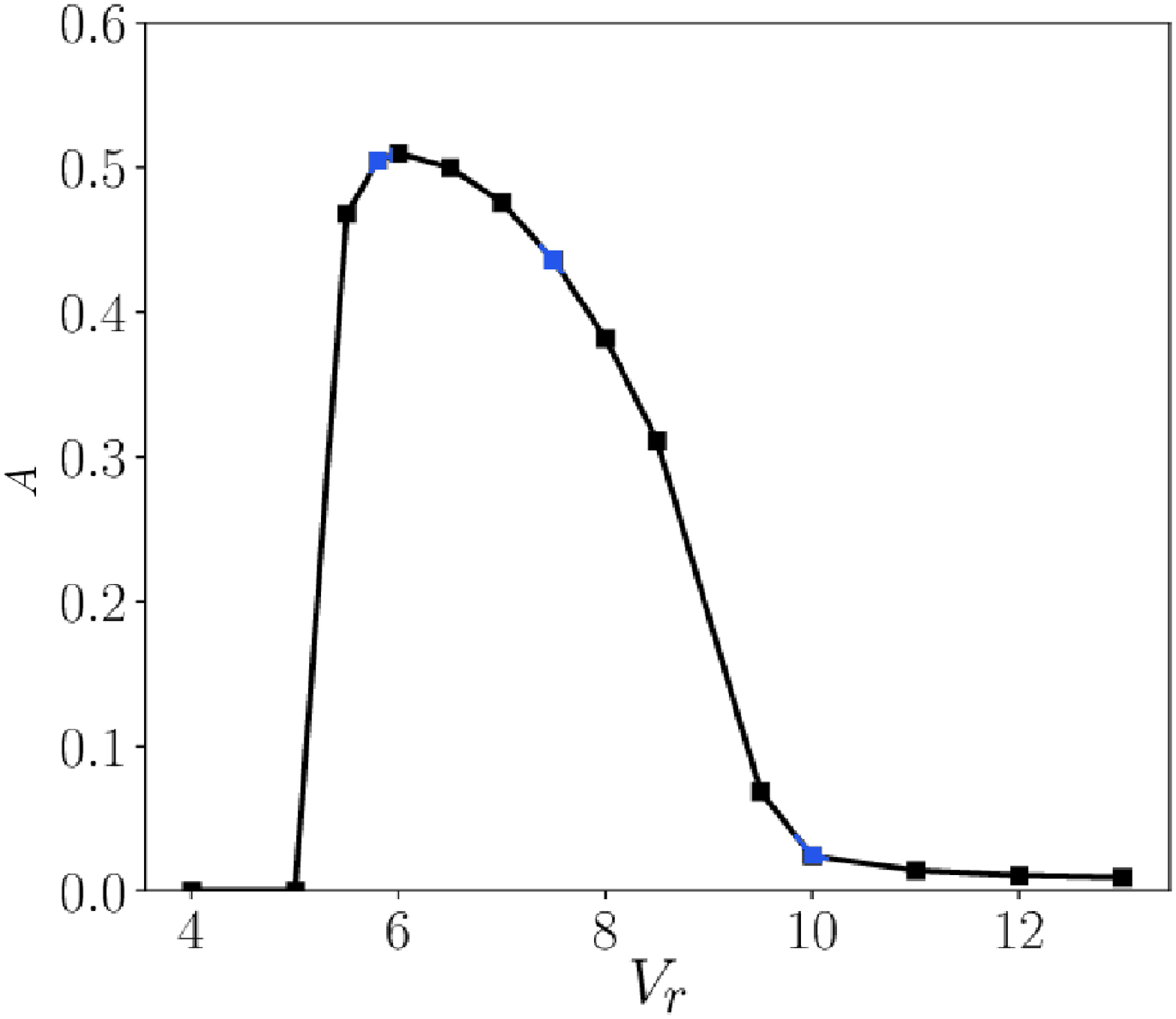}}
     \subfigure[\label{eigen_Re46_m7}]{\includegraphics[scale=0.3]{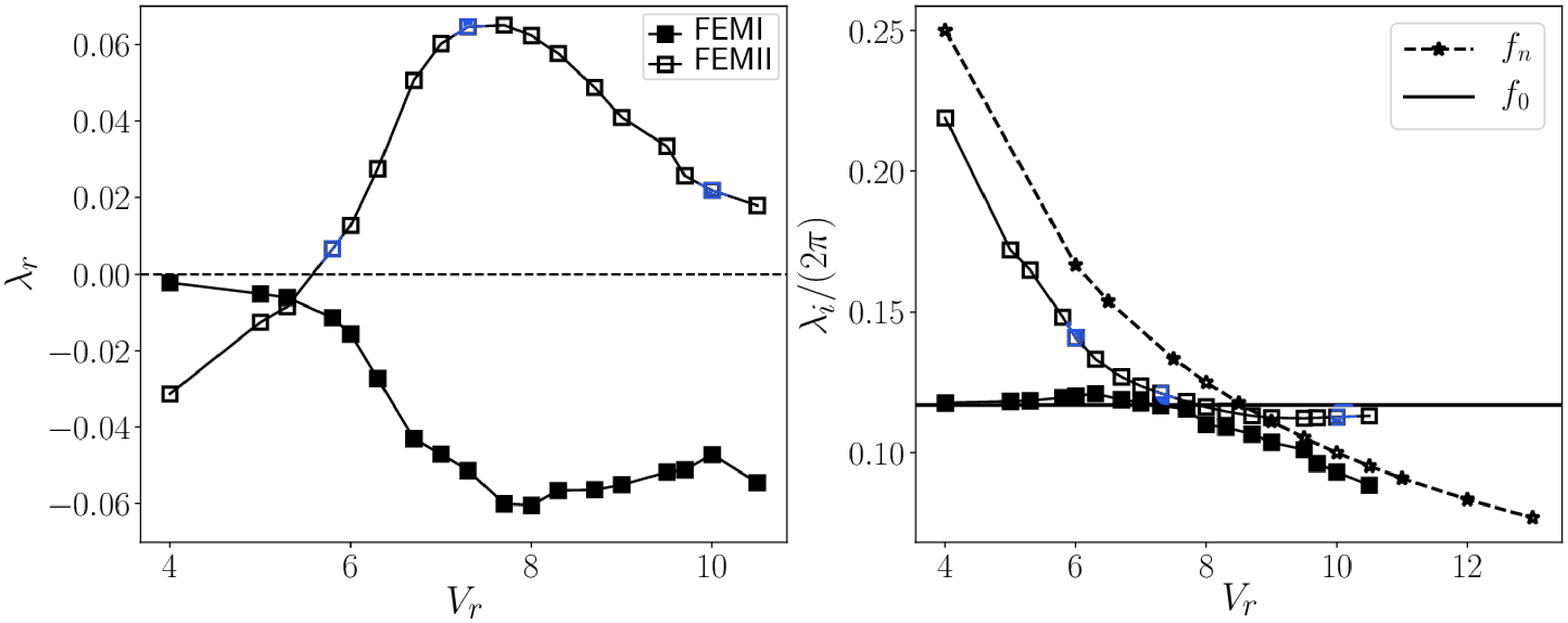}}
    \caption{Amplitude $A$ of cylinder oscillation for $(\Rey, m^{\ast}) = (46.6, 20)$ (a), and $(\Rey, m^{\ast}) = (46.6, 7)$ (c); Growth rate and eigen-frequency of the two least stable eigenvalues at $(\Rey, m^{\ast}) = (46.6, 20)$ (b), $(\Rey, m^{\ast}) = (46.6, 7)$ (d); $f_n$ is the natural frequency of the structure and $f_0$ is the eigen-frequency of the least stable mode of the flow around a fixed cylinder at $\Rey=46.6$. The blue points show the values of $V_r$ used in the sensitivity analyses.}
    \label{AmplXModes}
\end{figure}

In this work, comparisons of the sensitivities for the flow around fixed and elastically-mounted cylinders are performed, so the sensitivity fields of the fixed cylinder are used as benchmarks for comparisons. The critical Reynolds number $\Rey_c$ of the flow around a fixed cylinder is chosen. The value obtained in the current numerical simulations was $\Rey_c=46.6$, in agreement with previous works \citep{Jackson1987}. In the flow around an elastically-mounted cylinder, two values of mass ratio were chosen: relatively high $m^{\ast}=20$; and a lower $m^{\ast} = 7$. Next, the reduced velocity $V_r$ was chosen taking into account the results of the oscillation amplitude plotted in Figures~\ref{Amp_Re46_m20} and \ref{Amp_Re46_m7}. To do so, the two least stable modes were also assessed, and three values of $V_r$ inside of lock-in range were chosen. As introduced by \citet{Zhang2015, Mittal2016}, these two least stable modes are essential for linear stability analysis in the current FSI problem. 

By correlating the two least stable modes, whose corresponding eigenvalues are shown in Figures~\ref{eigen_Re46_m20} and \ref{eigen_Re46_m7}, and the amplitude of oscillation, we see that the initial branch starts at the same reduced velocity as that the mode FEMII shifts from stable to unstable. This behaviour was observed for both values of mass ratio: for $m^{\ast}=20$ and $m^{\ast}=7$ the lock-in ranges started at $V_r\approx 6.1$ and $V_r\approx 5.7$, respectively. Figures~\ref{eigen_Re46_m20} and \ref{eigen_Re46_m7}  also show a range where the eigen-frequencies matched (resonance range). For $m^{\ast}=20$ and $m^{\ast} = 7$, the resonance ranges were identified for $7.7< V_r < 8.6$ and $6< V_r < 9.8$, respectively. For higher $V_r$, when the eigen-frequencies depart, the elastically-mounted cylinder did not vibrate and the eigen-frequency of the stable mode tended to the frequency $f_0$ of the flow around a fixed cylinder for $\Rey = 46.6$. Considering these nonlinear and linear results, our objective was to choose one value of $V_r$ in the initial branch, one value in the final branch, and another $V_r$ in the range where the eigen-frequencies of the two least stable modes FEMI and FEMII matched. So, for $m^{\ast}=20$, reduced velocities $V_r=6.3$, $V_r=8$ and $V_r=9.5$ were used. For $m^{\ast}=7$, $V_r = 5.8$, $V_r = 7.5$ and $V_r = 10$ were chosen. These points are highlighted in Figures~\ref{eigen_Re46_m20} and \ref{eigen_Re46_m7} in blue. 

After choosing the cases to be evaluated in the sensitivity analyses, the least stable adjoint mode was verified by comparing its eigenvalue with the largest real part with the respective eigenvalue of the direct mode (they are expected to be equal). The comparisons are presented in Appendix~\ref{adj_mode_ver}, in which it can be seen that a good quantitative agreement was obtained.

\subsection{Sensitivity analyses}
Computations of sensitivity for a flow around an elastically-mounted cylinder were performed by considering the insertion of an external forcing (\ref{forcing}) in the momentum equation. According the framework presented in section~\ref{Sens_SteadyForceFSI}, this forcing can represent a small control device located at a point $\mathbf{x}_p = [x_p, y_p]^{T}$. The sensitivity analyses were carried out by evaluating separately the base flow sensitivity ($S1$) and the structural sensitivity ($S2$). Next, the total sensitivity ($S1 + S2$) was assessed. The coefficient $G$ in (\ref{forcing}) can be written using different formulations \citep{Pralitis:2010, Fani:2012} that represents the insertion of a small device in the domain. For simplicity we set $G=1$, always keeping in mind that the external forcing $\mathbf{G}$ has small amplitude. So, the sensitivities $S1$ and $S2$ analysed by the respective terms $\mathbf{U}\cdot \mathbf{U}^{\dagger}$ and $\widehat{\mathbf{u}}\cdot \mathbf{u}^{\dagger}$ can draw how the current FSI system responds to the forcing (\ref{forcing}). 

\begin{figure}
    \centering
    \includegraphics[width=0.5\textwidth]{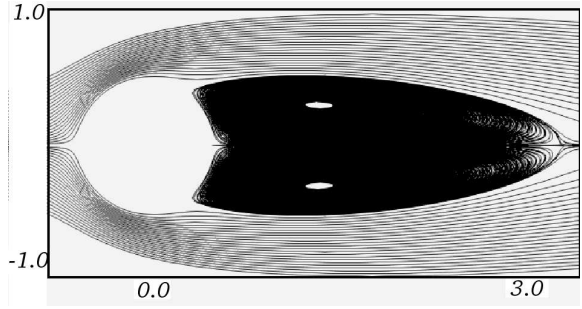}
    \caption{Streamlines of the steady base flow, at $\Rey = 46.6$. The separation bubble is the darker region immediately downstream of the cylinder and the separation points are the most upstream points of this bubble at the cylinder wall. }
    \label{Re46_BF1}
\end{figure}

The fields of sensitivity correspond to the least stable mode and its respective eigenvalue $\lambda_1 = \lambda_{1,r} + i \lambda_{1,i}$, where the real ($\lambda_{1,r}$) and imaginary ($\lambda_{1,i}$)  parts are referred to as growth rate and eigen-frequency, respectively. The results are evaluated always referencing the separation point and recirculation bubble of the base flow, shown in Figure~\ref{Re46_BF1}.
\subsubsection{Growth rate sensitivity}

\begin{figure}
    \centering
    \subfigure[ $S1_r = (\mathbf{U}\cdot \mathbf{U}^{\dagger})_{r}$\label{bf_fixed}] {\includegraphics[width=0.3\textwidth]{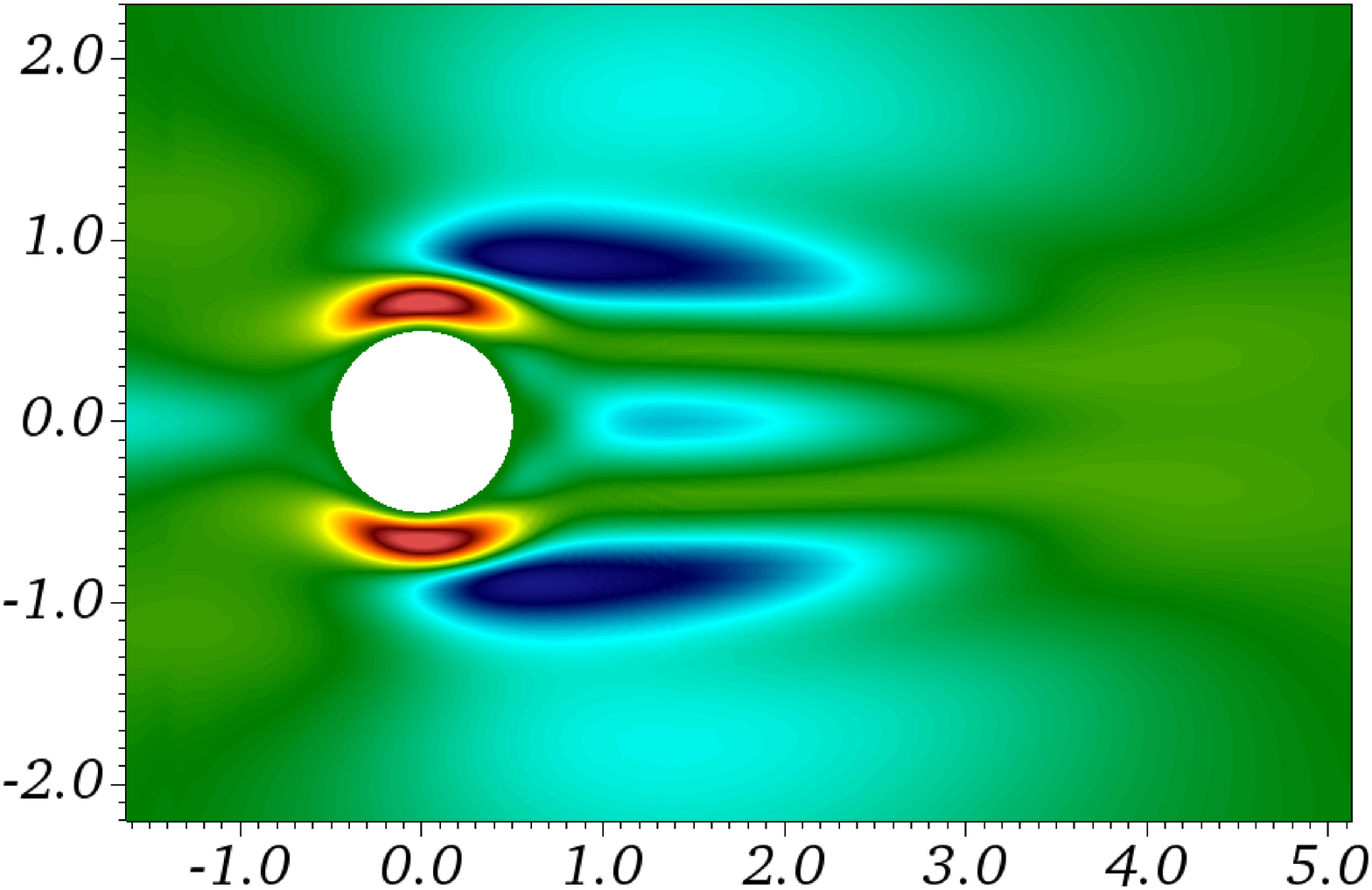}}
    \subfigure[ $S2_r = (\widehat{\mathbf{u}}\cdot \mathbf{u}^{\dagger})_{r}$\label{struc_fixed}] {\includegraphics[width=0.3\textwidth]{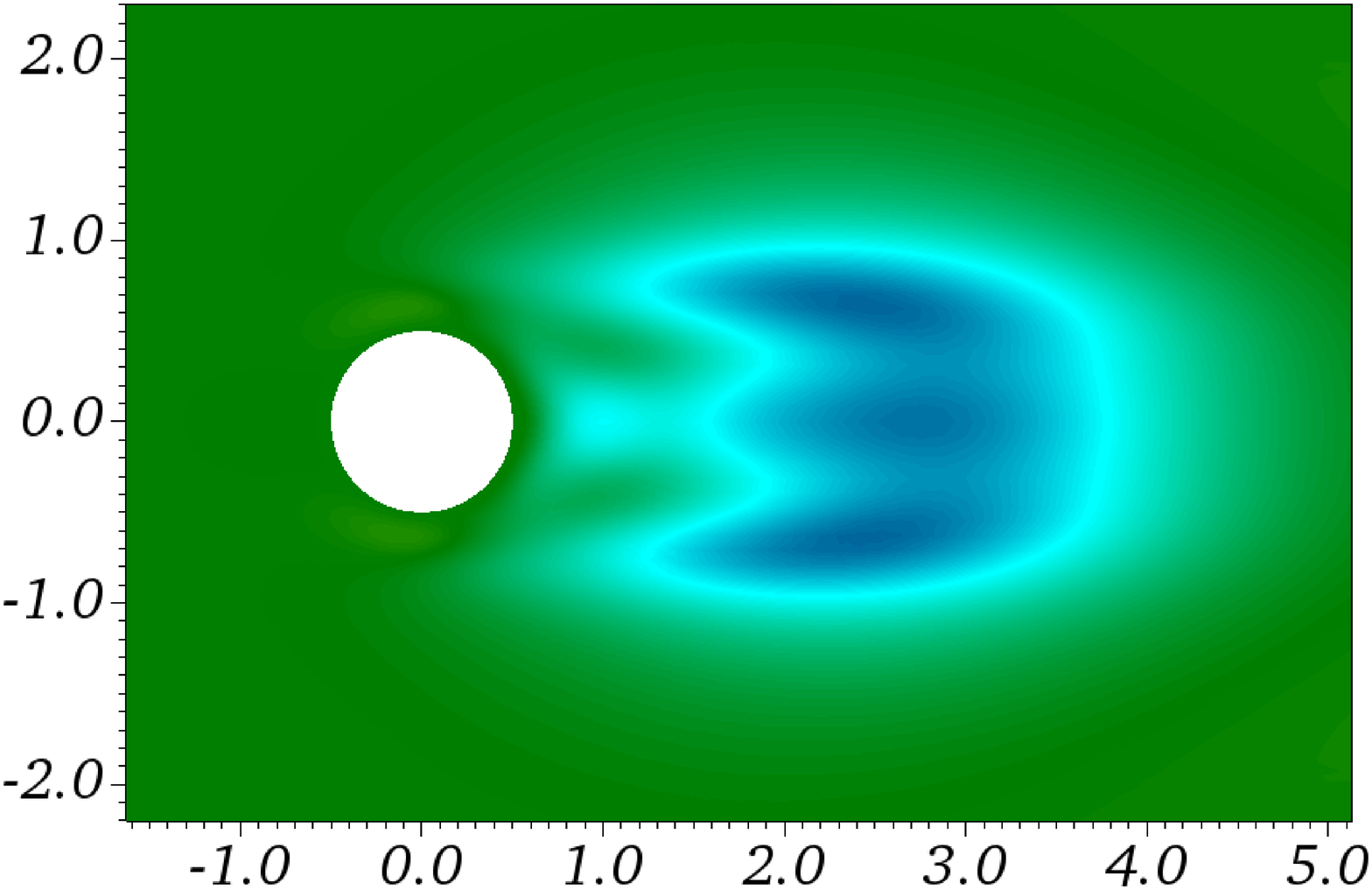}}
    \subfigure[ $(\mathbf{U}\cdot \mathbf{U}^{\dagger})_{r} + (\widehat{\mathbf{u}}\cdot \mathbf{u}^{\dagger})_{r}$\label{tot_fixed}] {\includegraphics[width=0.3\textwidth]{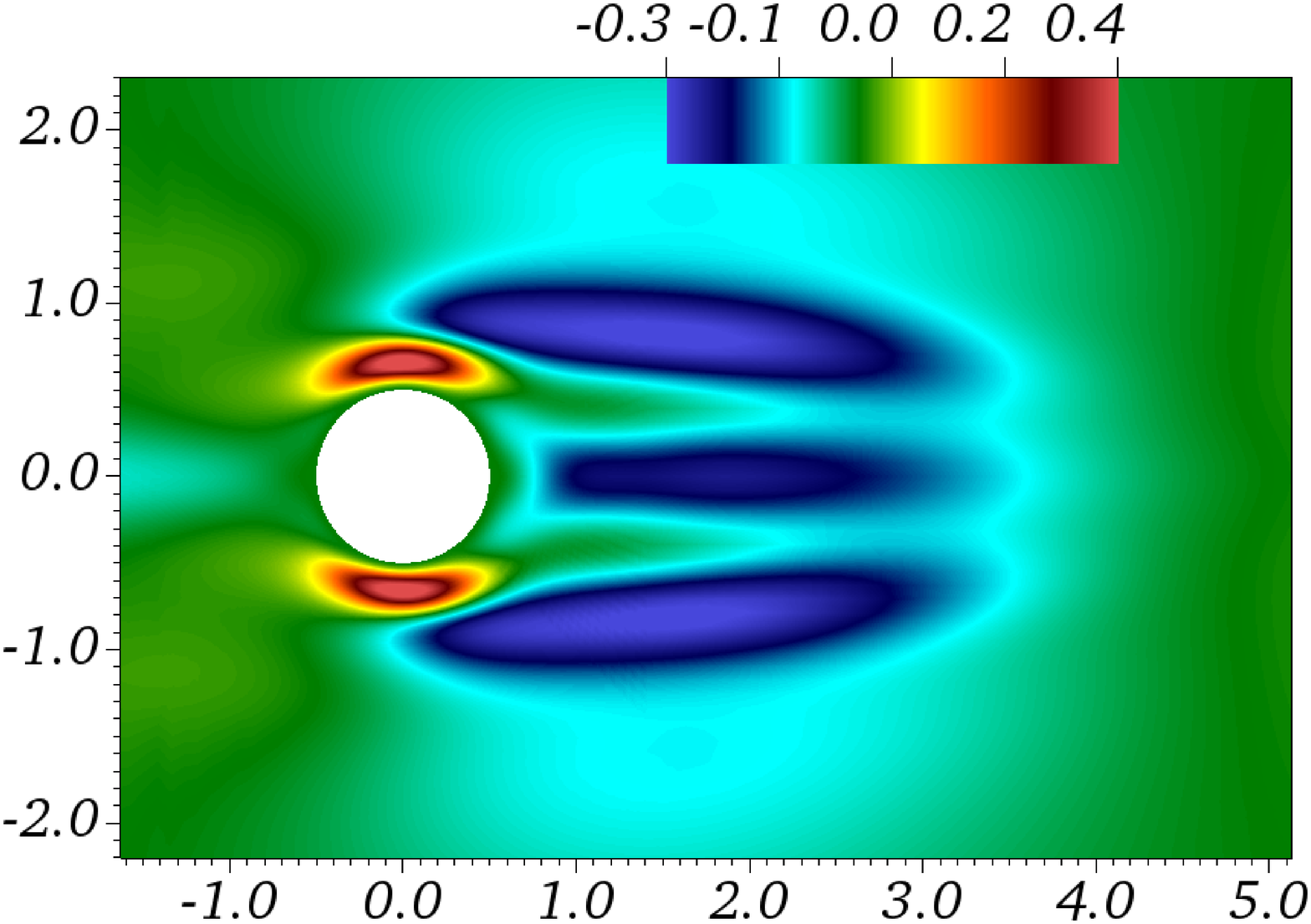}}
    \caption{Growth rate sensitivity fields for the flow around a fixed cylinder at $\Rey=46.6$.}
    \label{flowsens}
\end{figure}

Beginning with the growth rate sensitivity analysis, Figure~\ref{bf_fixed} shows the fields for the flow around a fixed cylinder, where the index $r$ indicates the real part of the sensitivity computation. These results can provide the regions where the forcing (\ref{forcing}) makes $\lambda_{1,r}$ either decrease (negative values) or increase (positive values). In this case, the growth rate increases if a local force is applied closer to the wall, at the upside and downside of the cylinder. On the other hand, the growth rate decreases when the local forcing is inserted at the edge and just out of the recirculation bubble (plotted in Figure~\ref{Re46_BF1}). It is worth highlighting that this field of $(\mathbf{U}^{\dagger} \cdot \mathbf{U})_r$ is in agreement with the results from \cite{Marquet2008}. Figure~\ref{struc_fixed} shows the field of growth rate structural sensitivity $S2_r$. Such field displays the maximal intensity symmetrically located downstream of the cylinder, across the separation bubble. In these regions, the inner product $(\mathbf{u}^{\dagger}\cdot \widehat{\mathbf{u}})_r$ shows that the forcing (\ref{forcing}) makes $\lambda_{1,r}$ to decrease. Figure~\ref{tot_fixed} shows the total growth rate sensitivity which is evaluated by $ S1_r + S2_r = (\mathbf{U}^{\dagger} \cdot \mathbf{U})_r + (\mathbf{u}^{\dagger}\cdot \widehat{\mathbf{u}})_r$. Looking at this field and comparing to the fields of $S1_r$ and $S2_r$, we notice that the base flow sensitivity $S1_r$ has an important role in the final results of $\delta \lambda_{1,r}$, mainly at the upside and downside of the cylinder, also at the limit and just out of the recirculation bubble. To conclude, the sensitivity of the flow around a fixed cylinder at $\Rey=46.6$ shows that the stabilisation of the flow system can be reached  by applying a local steady forcing (\ref{forcing}) at the upside and downside of the limits of the recirculation bubble, and across of the recirculation bubble. The regions far from the cylinder have small or null intensities, indicating that these regions do not have an important role on the dynamics of the instability.

\begin{figure}
    \centering
    \subfigure[$S1_r = (\mathbf{U}\cdot \mathbf{U}^{\dagger})_r$ \label{bf_sensm7Vr58}] {\includegraphics[width=0.3\textwidth]{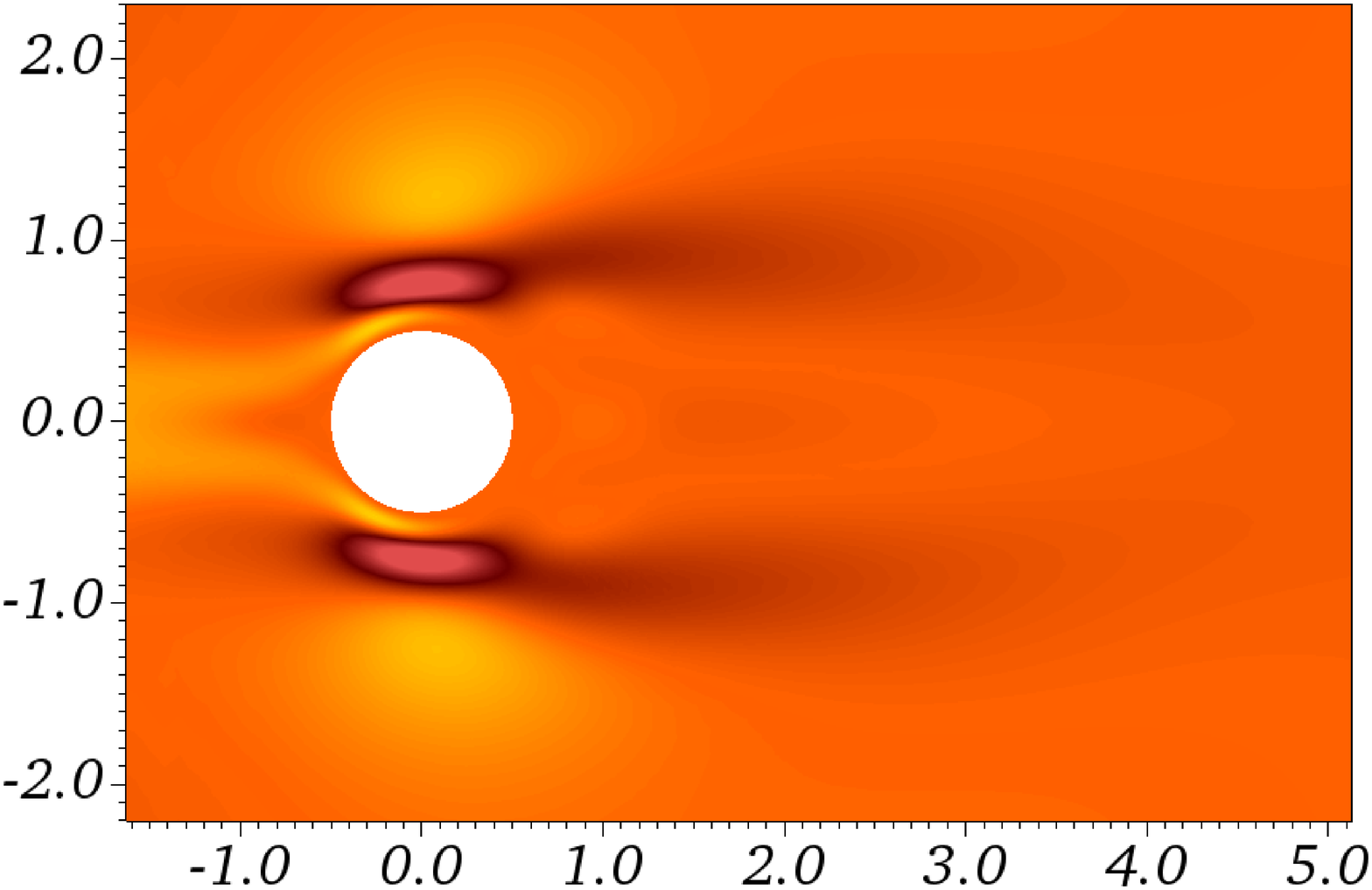}}
    \subfigure[$S2_r = (\widehat{\mathbf{u}}\cdot \mathbf{u}^{\dagger})_{r}$ \label{struc_sensm7Vr58}] {\includegraphics[width=0.3\textwidth]{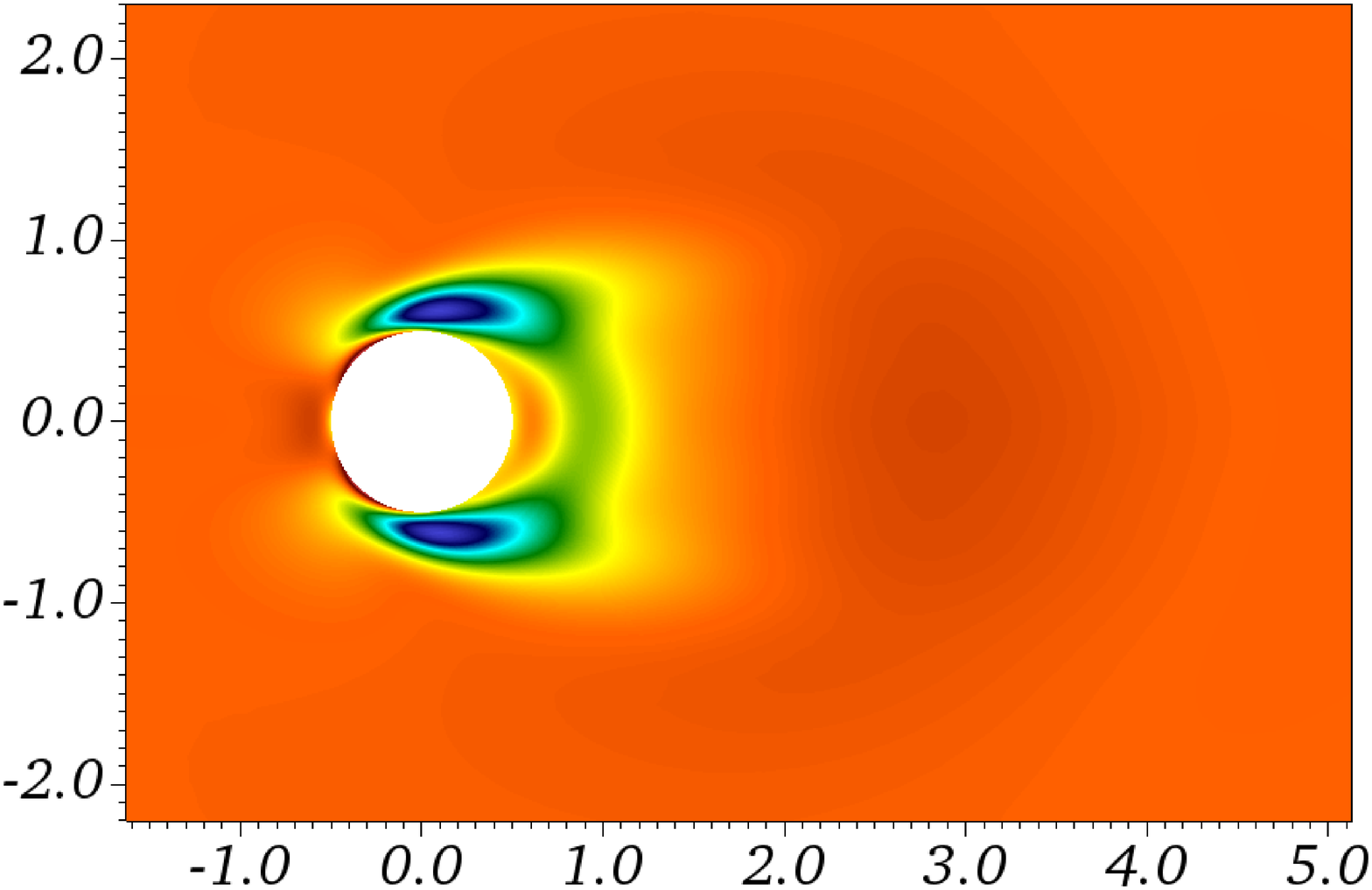}}
    \subfigure[$(\mathbf{U}\cdot \mathbf{U}^{\dagger})_{r} + (\widehat{\mathbf{u}}\cdot \mathbf{u}^{\dagger})_{r}$ \label{tot_sensm7Vr58}] {\includegraphics[width=0.3\textwidth]{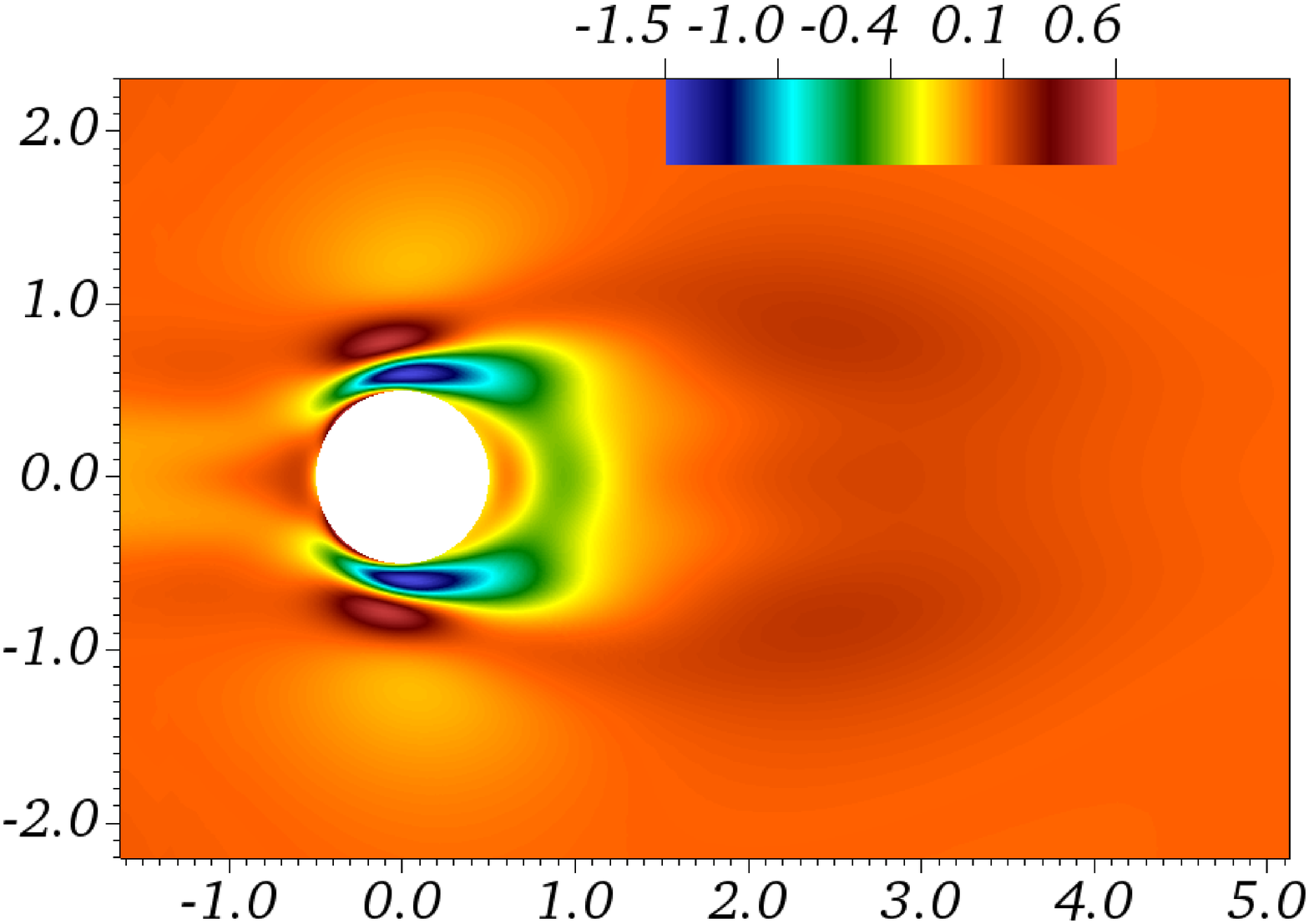}}
    \subfigure[$S1_r = (\mathbf{U}\cdot \mathbf{U}^{\dagger})_{r}$ \label{bf_sensm20Vr63}] {\includegraphics[width=0.3\textwidth]{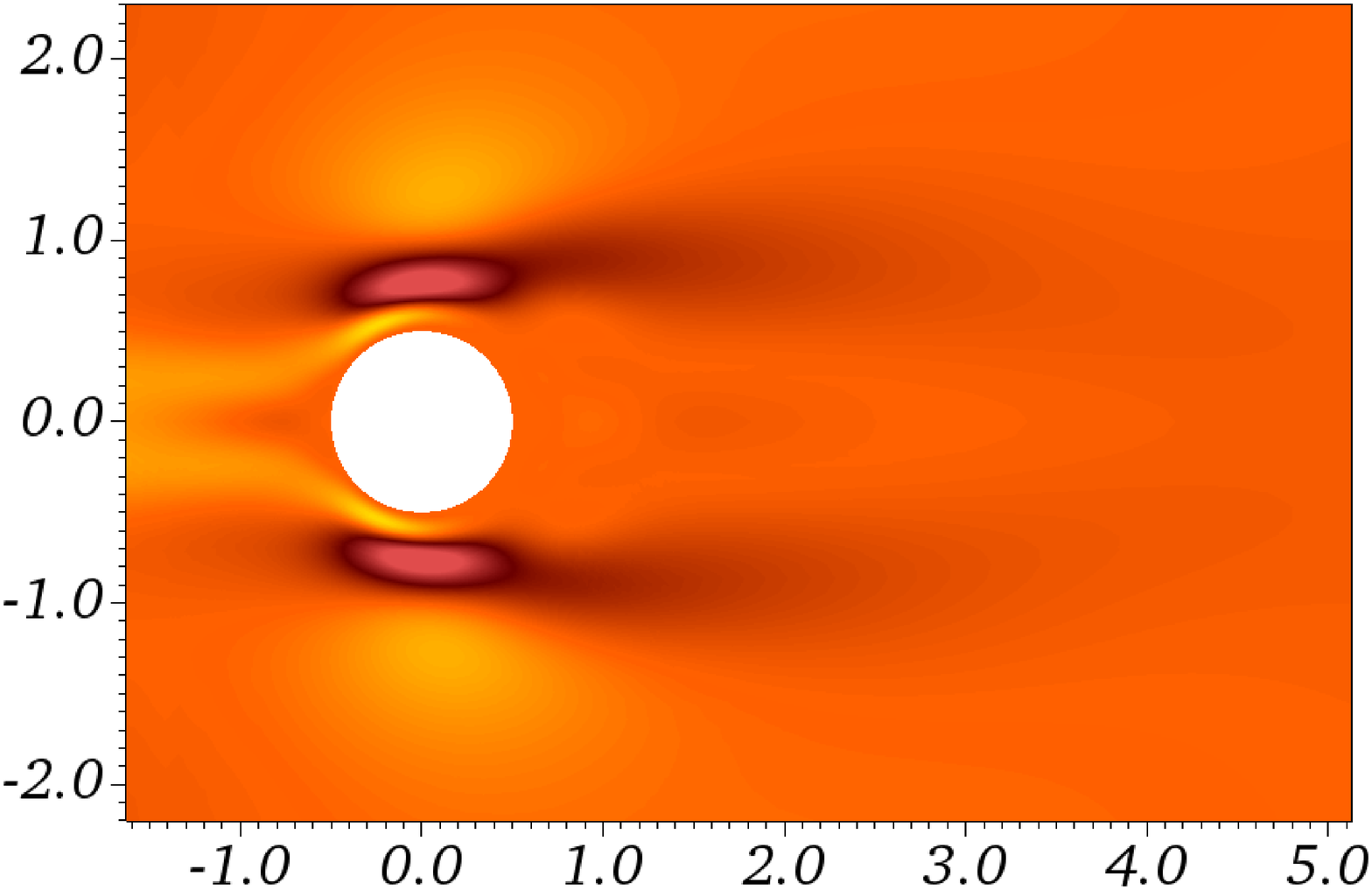}}
    \subfigure[$S2_r = (\widehat{\mathbf{u}}\cdot \mathbf{u}^{\dagger})_{r}$ \label{struc_sensm20Vr63}] {\includegraphics[width=0.3\textwidth]{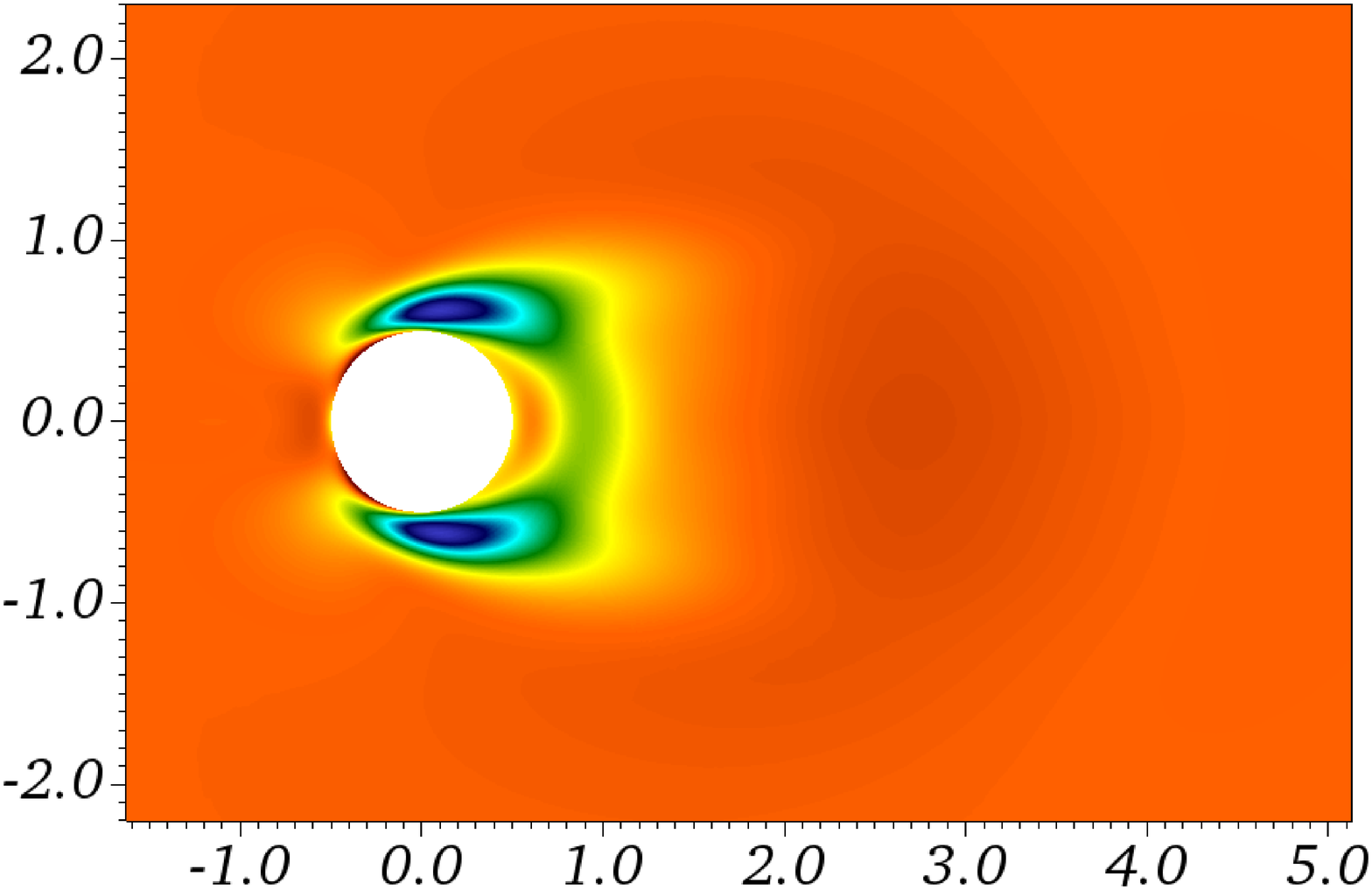}}
    \subfigure[$(\mathbf{U}\cdot \mathbf{U}^{\dagger})_{r} + (\widehat{\mathbf{u}}\cdot \mathbf{u}^{\dagger})_{r}$ \label{tot_sensm20Vr63}] {\includegraphics[width=0.3\textwidth]{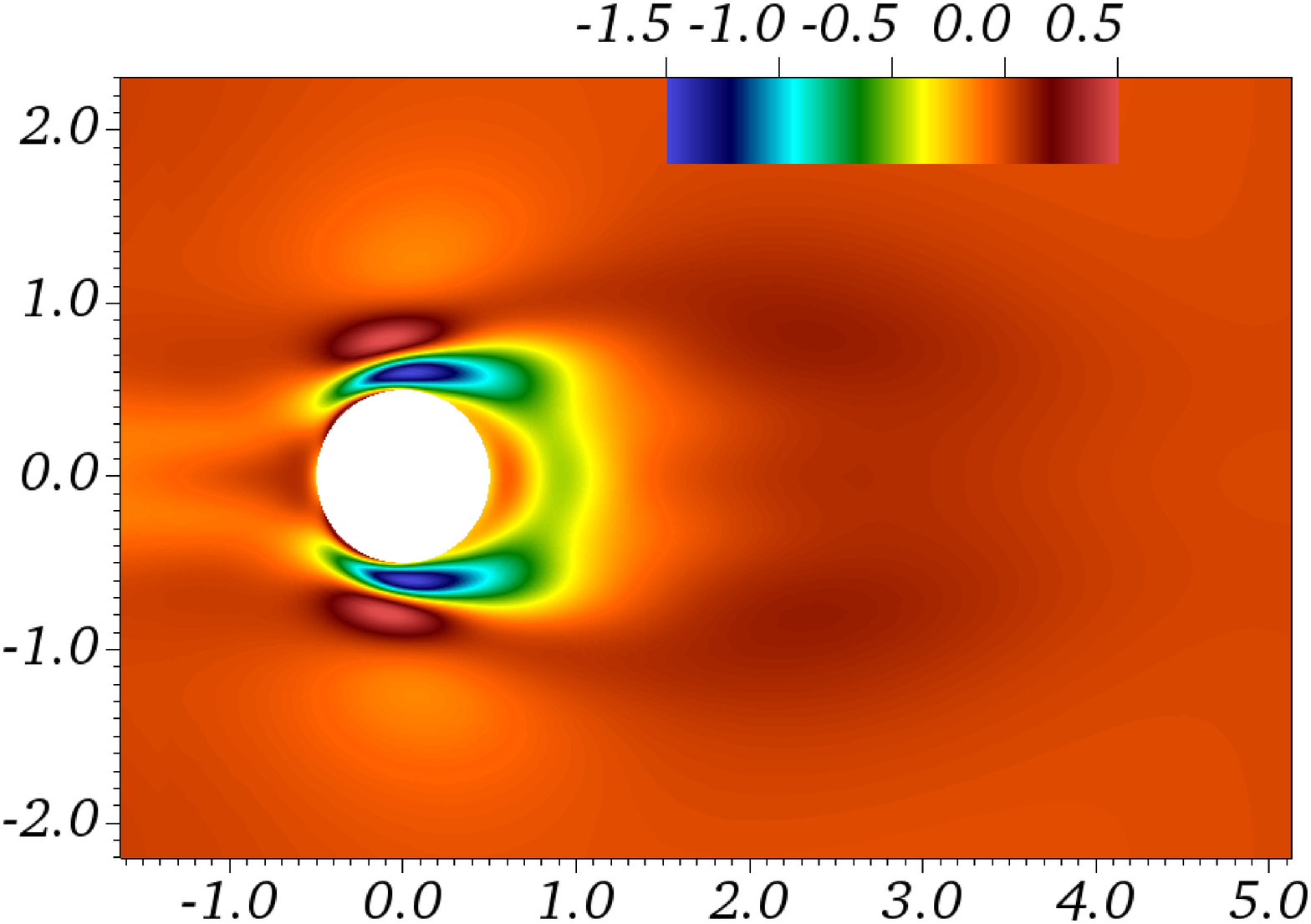}}
    \caption{Growth rate sensitivity fields for reduced velocity in the initial branch: $(m^{\ast},V_r) = (7, 5.8)$ (a), (b), and (c); $(m^{\ast},V_r) = (20, 6.3)$ (d), (e), and (f).}
    \label{sens_initbranch}
\end{figure}

In the flow around an elastically-mounted cylinder, the analyses started by evaluating the growth rate sensitivity for reduced velocities $V_r$ in the initial branch [$(m^{\ast},V_r) = (7, 5.8)$ and $(m^{\ast},V_r) = (20, 6.3)$]. Figure~\ref{sens_initbranch} shows the sensitivity fields for $m^{\ast} = 20$ and $m^{\ast} = 7$. In theses cases, marked differences are noticed when comparisons with the sensitivities of the fixed cylinder are made. Opposed to what is seen for the fixed cylinder, Figures~\ref{tot_sensm7Vr58} and \ref{tot_sensm20Vr63} show variation $\delta \lambda_{1,r}$ positive if the local forcing is added at the limit of the recirculation bubble and negative if inserted closer to the cylinder wall. As observed in Figures~\ref{bf_sensm7Vr58} and \ref{bf_sensm20Vr63}, at the limit of the recirculation bubble, the fields of $(\mathbf{U}^{\dagger} \cdot \mathbf{U})_r$ show that $S1_r$ has an important contribution in the total sensitivity. On the other hand, the structural sensitivity $S2_r$ has significant influence in the regions closer to the cylinder wall (see Figures~\ref{bf_sensm7Vr58} and \ref{bf_sensm20Vr63}), where the fields of $(\mathbf{u}^{\dagger}\cdot \widehat{\mathbf{u}})_r$ show the steady forcing (\ref{forcing}) making the growth rate $\lambda_{1,r}$ decrease. Different from the fixed cylinder, for $V_r$ in the initial branch, the maximal intensity is located at the upside and downside of the cylinder, and lower intensity are observed in the regions across of the recirculation bubble. Another result to highlight is about the magnitude of the sensitivity. In Figures~\ref{tot_sensm7Vr58} and \ref{tot_sensm20Vr63} we see that the sensitivity magnitudes are much higher than those for the fixed cylinder (shown in Figure~\ref{tot_fixed}).  Therefore, for $(m^{\ast},V_r) = (7, 5.8)$ and $(m^{\ast},V_r) = (20, 6.3)$, the results show that the FSI case is more sensitive to the external forcing (\ref{forcing}), also the responses to this kind of control can be very different from those of the flow around a fixed cylinder at $\Rey=46.6$.

\begin{figure}
    \centering
    \subfigure[$S1_r = (\mathbf{U}\cdot \mathbf{U}^{\dagger})_{r}$\label{bf_sensm7Vr75}] {\includegraphics[width=0.3\textwidth]{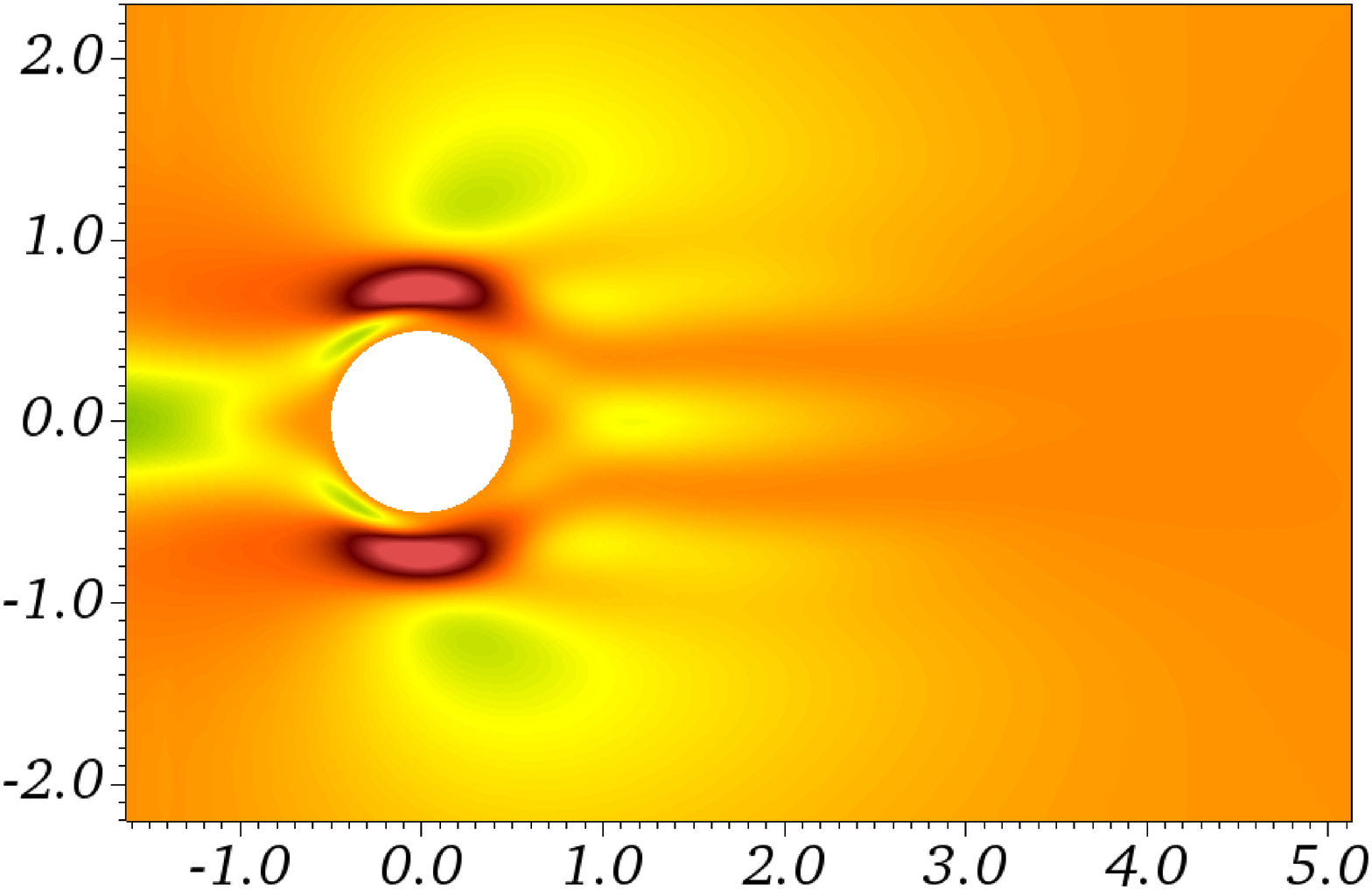}}
    \subfigure[$S2_r = (\widehat{\mathbf{u}}\cdot \mathbf{u}^{\dagger})_{r}$\label{struc_sensm7Vr75}] {\includegraphics[width=0.3\textwidth]{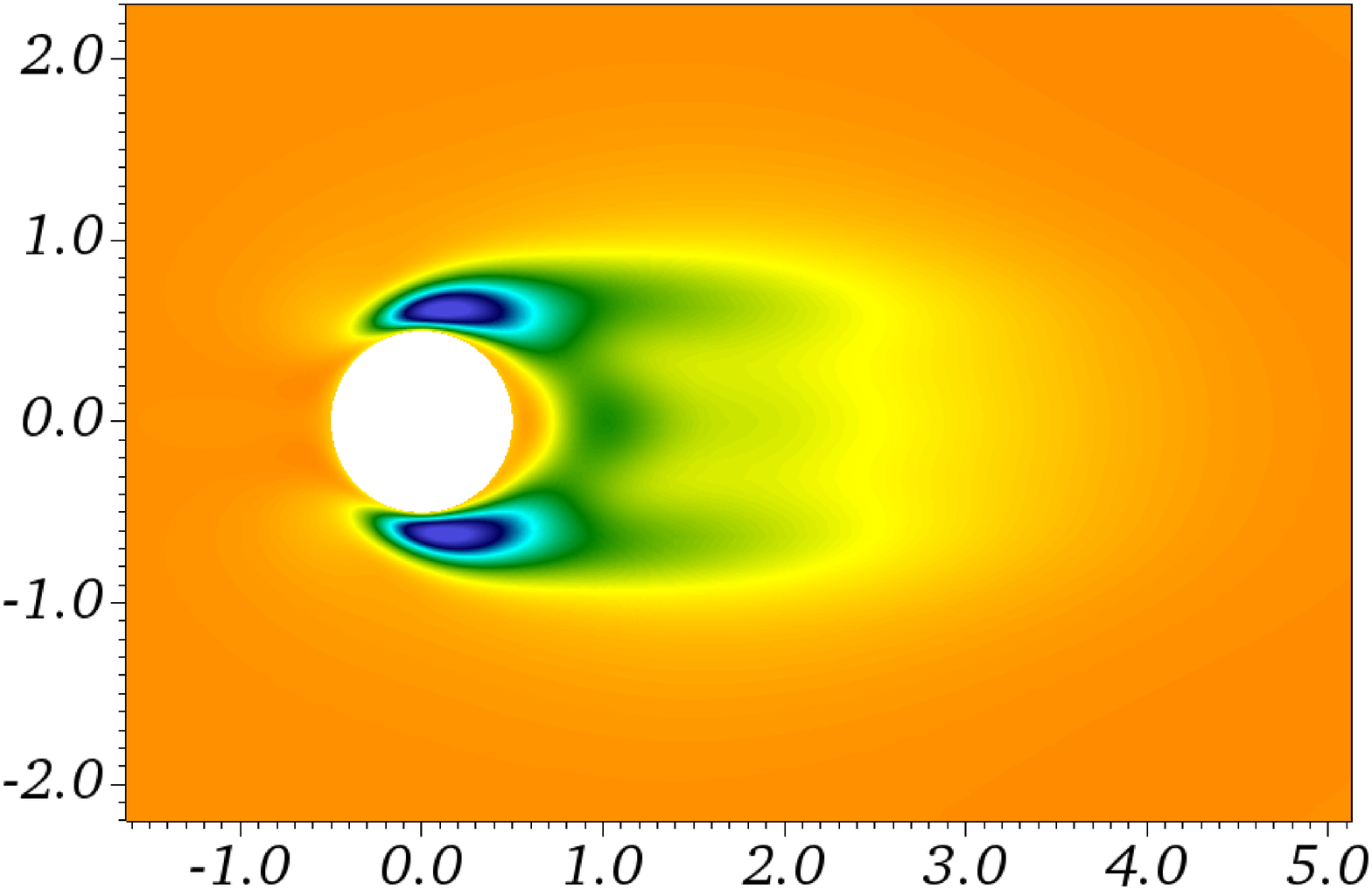}}
    \subfigure[$(\mathbf{U}\cdot \mathbf{U}^{\dagger})_{r} + (\widehat{\mathbf{u}}\cdot \mathbf{u}^{\dagger})_{r}$ \label{tot_sensm7Vr75}] {\includegraphics[width=0.3\textwidth]{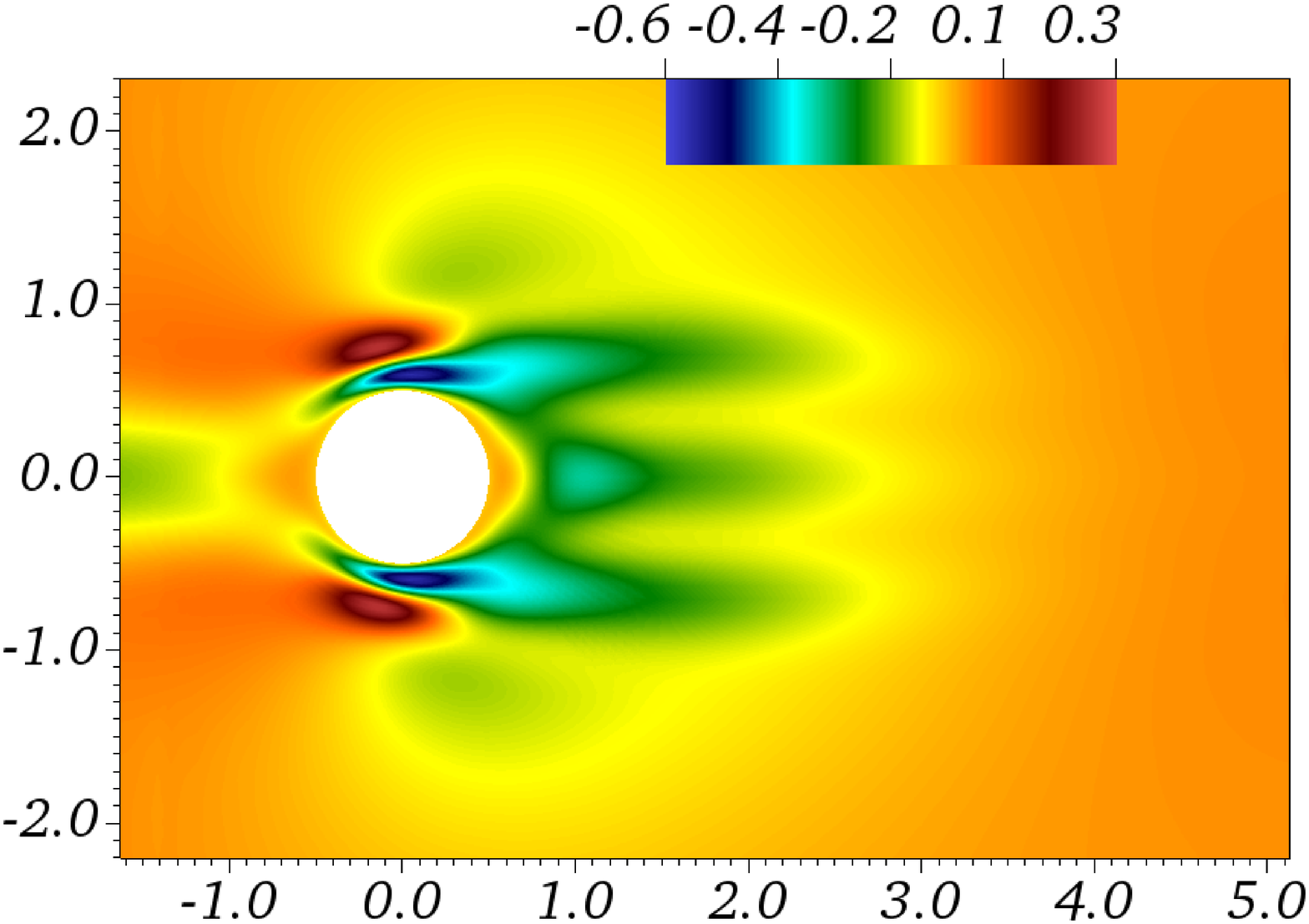}}
    \subfigure[$S1_r = (\mathbf{U}\cdot \mathbf{U}^{\dagger})_{r}$ \label{bf_sensm20Vr8}] {\includegraphics[width=0.3\textwidth]{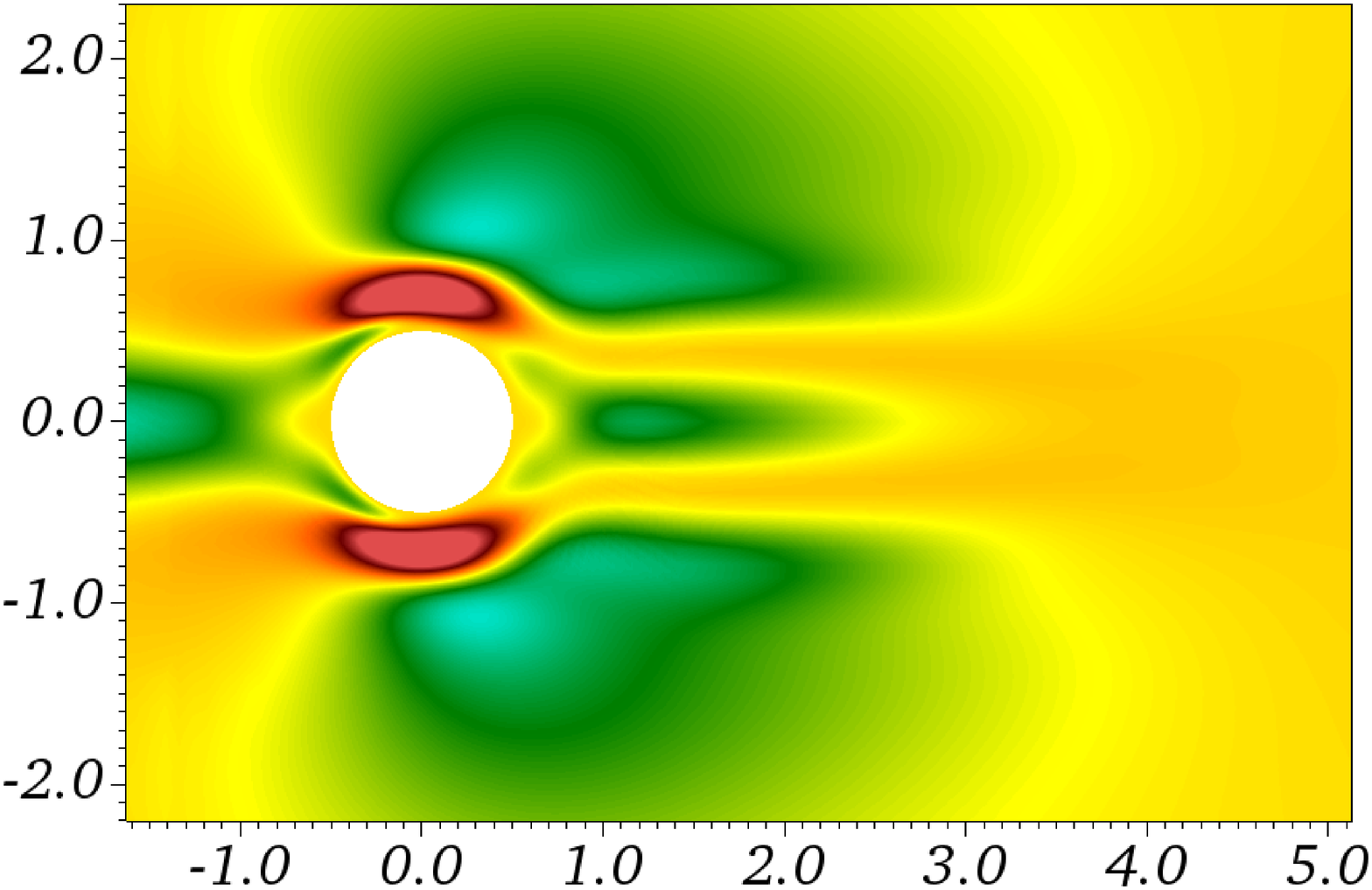}}
    \subfigure[$S2_r = (\widehat{\mathbf{u}}\cdot \mathbf{u}^{\dagger})_{r}$ \label{struc_sensm20Vr8}] {\includegraphics[width=0.3\textwidth]{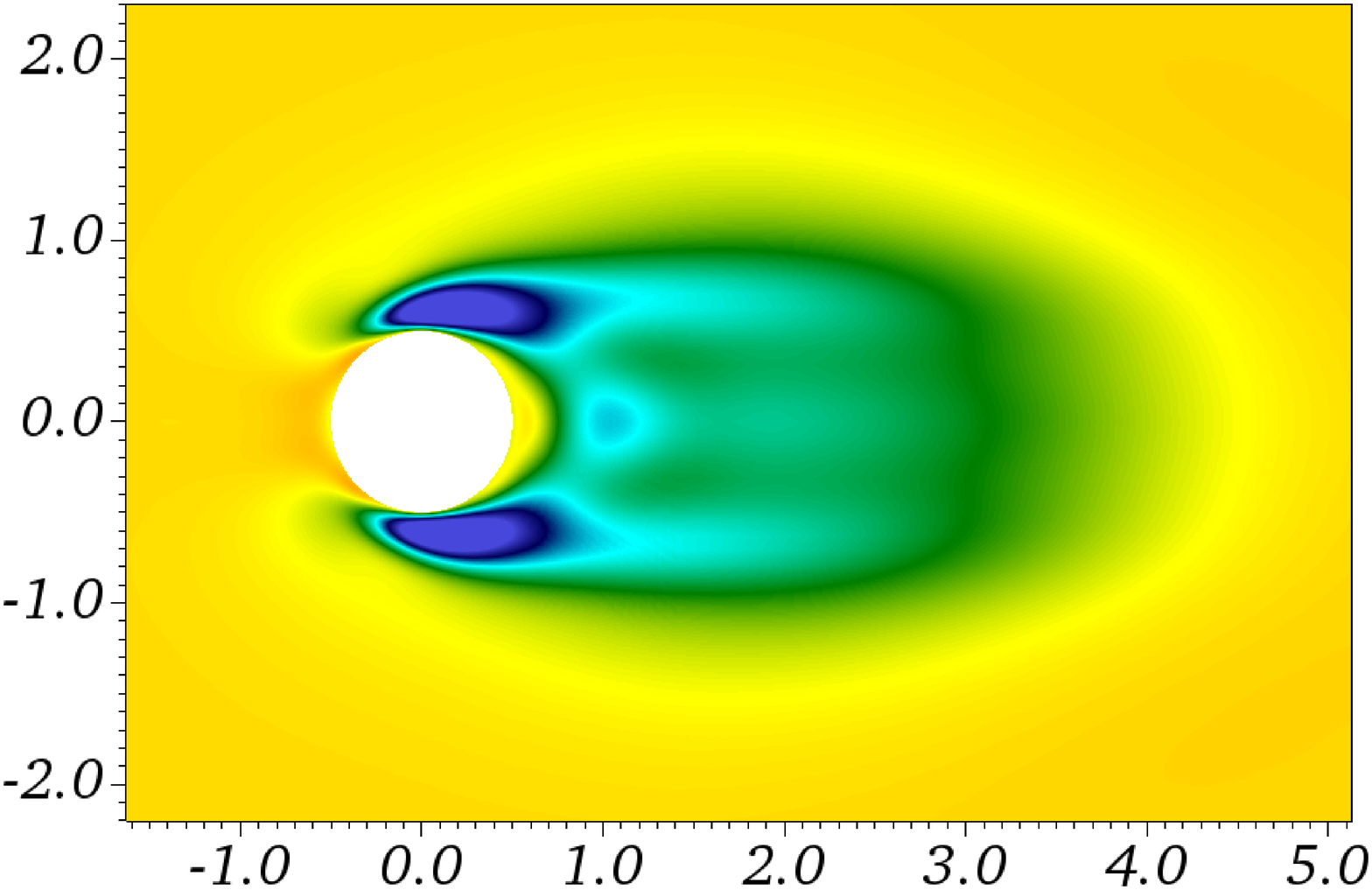}}
    \subfigure[$(\mathbf{U}\cdot \mathbf{U}^{\dagger})_{r} + (\widehat{\mathbf{u}}\cdot \mathbf{u}^{\dagger})_{r}$ \label{tot_sensm20Vr8}] {\includegraphics[width=0.3\textwidth]{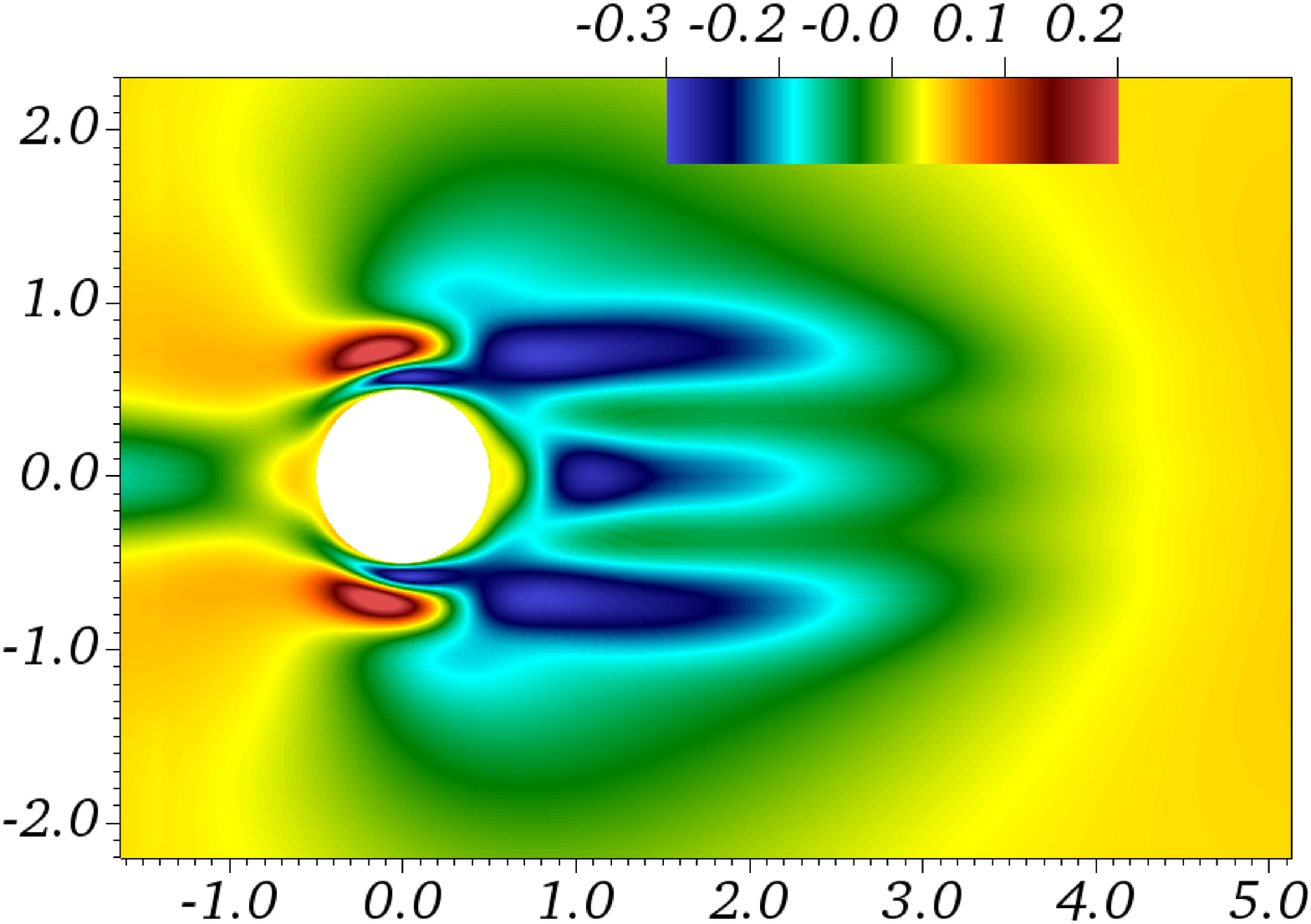}}
    \caption{Growth rate sensitivity fields for reduced velocity in the resonance range: $(m^{\ast},V_r) = (7, 7.5)$ (a), (b), and (c); $(m^{\ast},V_r) = (20,8)$ (d), (e), and (f).}
    \label{sens_resonance}
\end{figure}

\begin{figure}[!h]
    \centering
    \subfigure[$S1_r = (\mathbf{U}\cdot \mathbf{U}^{\dagger})_{r}$ \label{bf_sensm7Vr19}] {\includegraphics[width=0.3\textwidth]{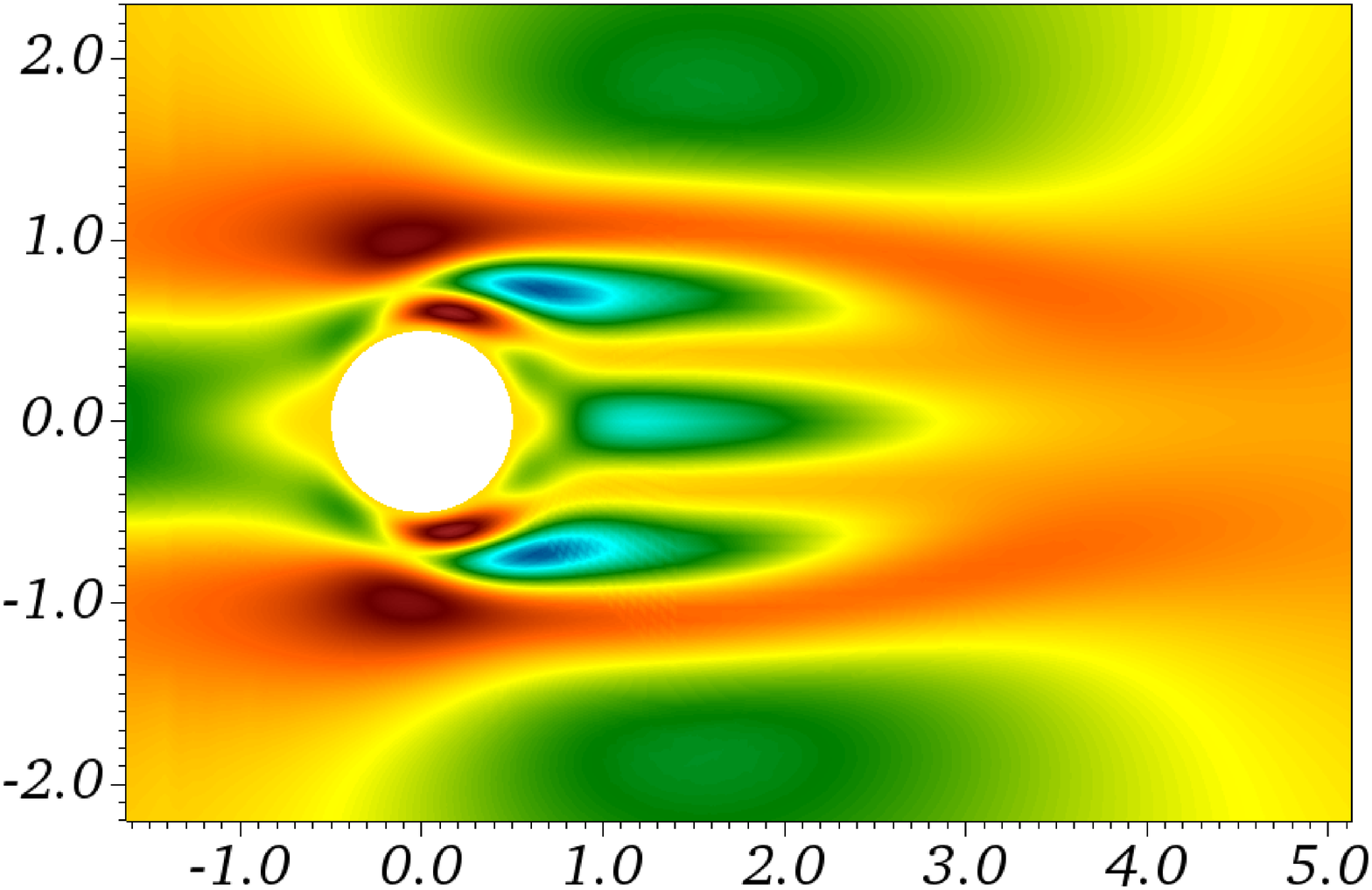}}
    \subfigure[$S2_r = (\widehat{\mathbf{u}}\cdot \mathbf{u}^{\dagger})_{r}$ \label{struc_sensm7Vr10}] {\includegraphics[width=0.3\textwidth]{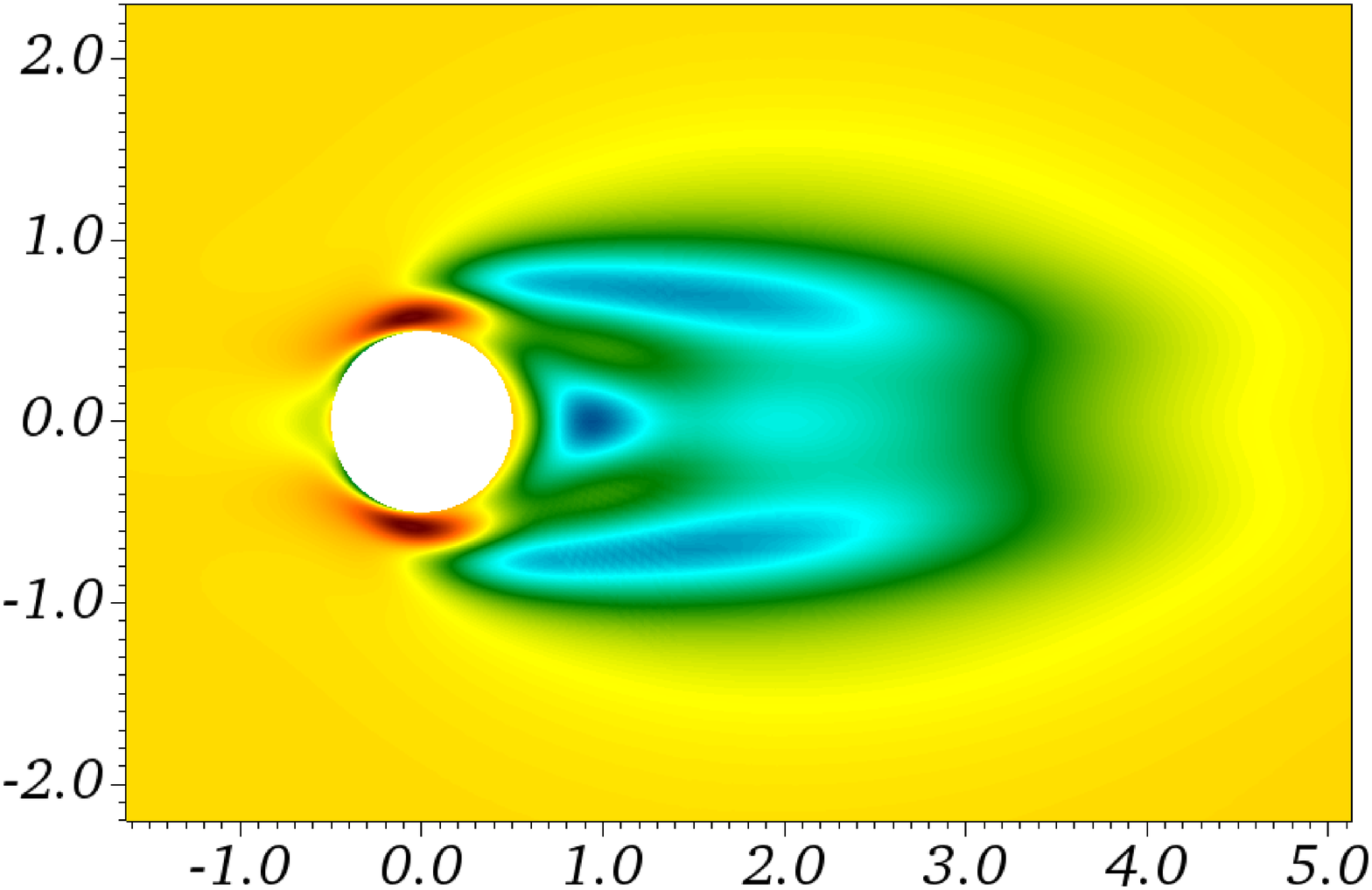}}
    \subfigure[$(\mathbf{U}\cdot \mathbf{U}^{\dagger})_{r} + (\widehat{\mathbf{u}}\cdot \mathbf{u}^{\dagger})_{r}$ \label{tot_sensm7Vr10}] {\includegraphics[width=0.3\textwidth]{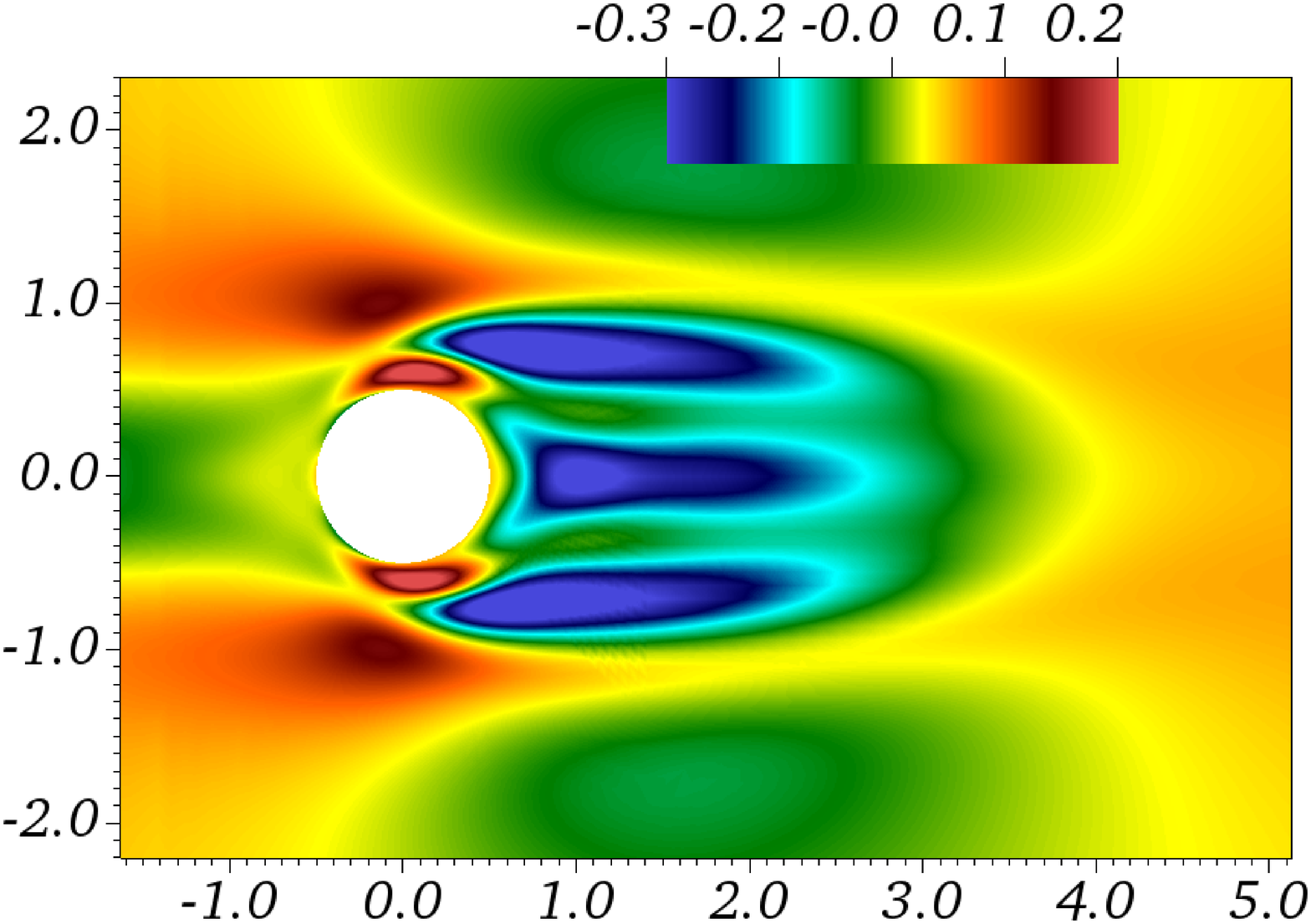}}
    \subfigure[$S1_r = (\mathbf{U}\cdot \mathbf{U}^{\dagger})_{r}$\label{bf_sensm20Vr95}] {\includegraphics[width=0.3\textwidth]{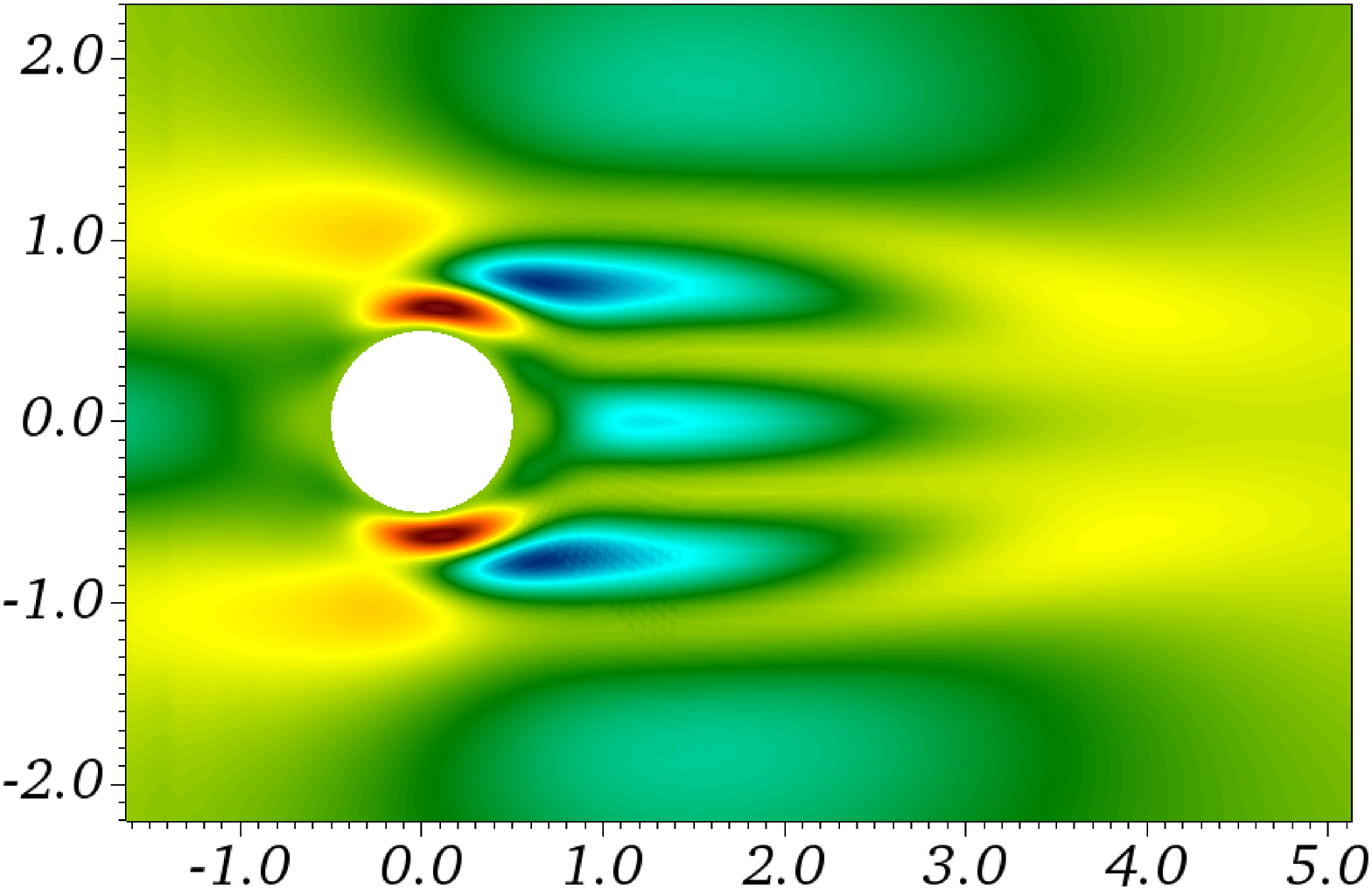}}
    \subfigure[$S2_r = (\widehat{\mathbf{u}}\cdot \mathbf{u}^{\dagger})_{r}$ \label{struc_sensm20Vr95}] {\includegraphics[width=0.3\textwidth]{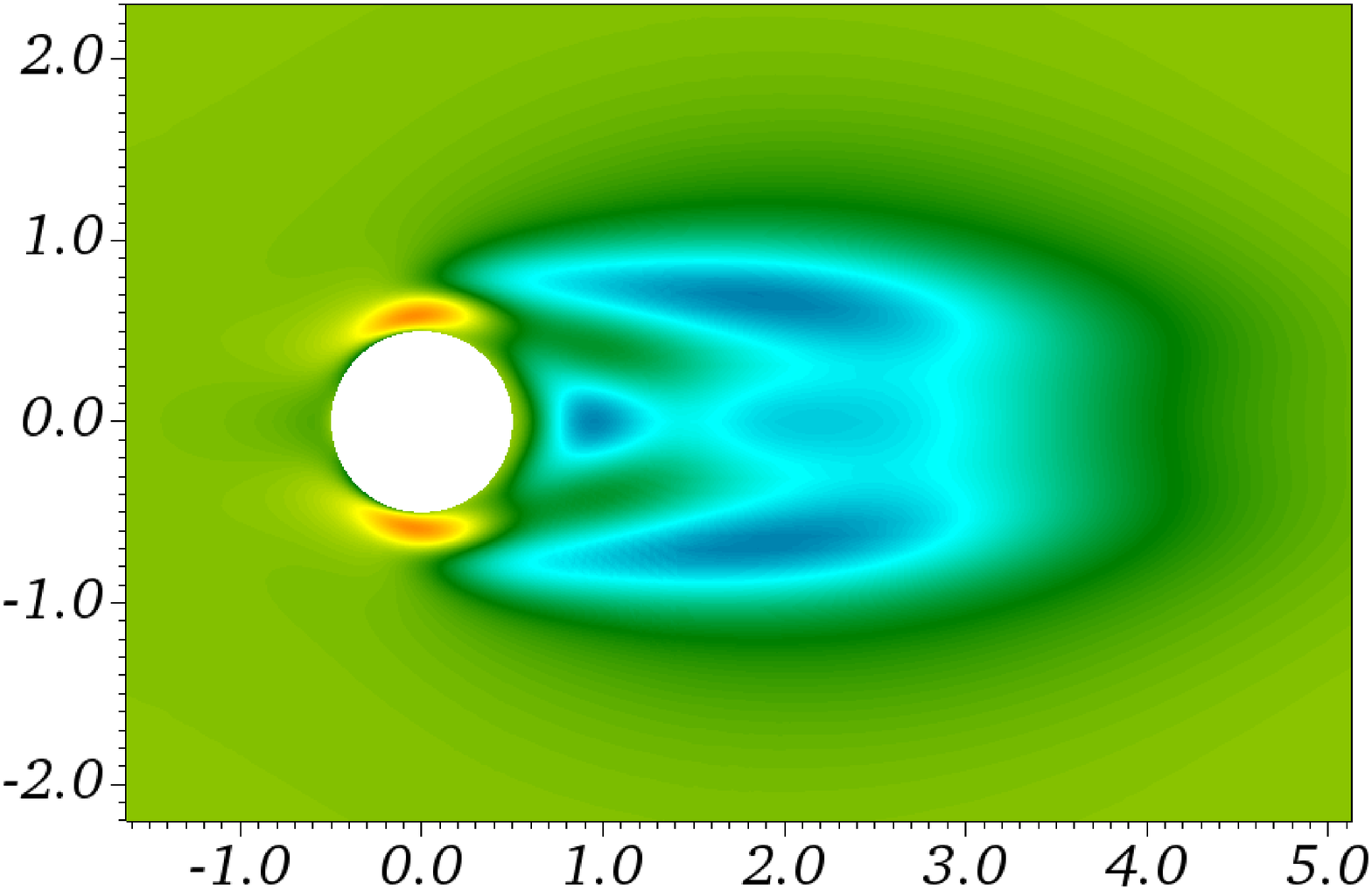}}
    \subfigure[$(\mathbf{U}\cdot \mathbf{U}^{\dagger})_{r} + (\widehat{\mathbf{u}}\cdot \mathbf{u}^{\dagger})_{r}$ \label{tot_sensm20Vr95}] {\includegraphics[width=0.3\textwidth]{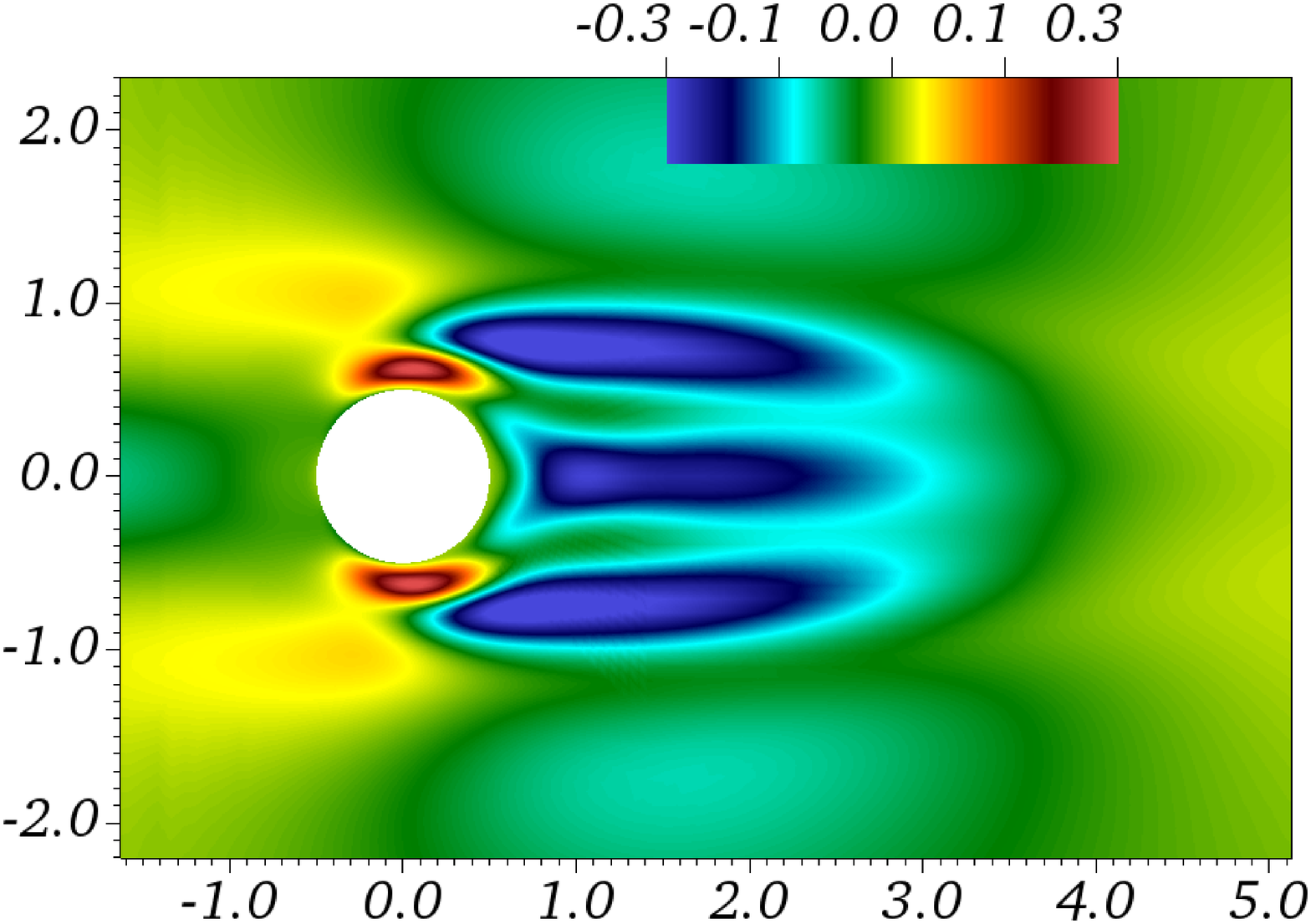}}
    \caption{Growth rate sensitivity fields for reduced velocity in the final branch. $(m^{\ast},V_r) = (7, 10)$ (a), (b), and (c); $(m^{\ast},V_r) = (20,9.5)$ (d), (e), and (f).}
    \label{sens_finalbranch}
\end{figure}

In the resonance range [$(m^{\ast},V_r) = (7, 7.5)$ and $(m^{\ast},V_r) = (20, 8)$], Figures~\ref{tot_sensm7Vr75} and \ref{tot_sensm20Vr8} show that the steady forcing (\ref{forcing}) has the role to decrease the growth rate of the least stable mode if applied downstream of the cylinder, across the separation bubble, also at the limit and just out of the recirculation bubble. Once again, the sensitivity $S2_r$ has an important role in the regions closer to the cylinder wall. In these regions, the field provided by $(\widehat{\mathbf{u}}\cdot \mathbf{u}^{\dagger})_{r}$ displays that the growth rate of the least stable mode has negative variation if the steady forcing (\ref{forcing}) is inserted. Opposed to it, in the final branch [$(m^{\ast},V_r) = (7, 10)$ and $(m^{\ast},V_r) = (20, 9.5)$], Figures~\ref{tot_sensm7Vr10} and \ref{tot_sensm20Vr95} show this same forcing makes the growth rate of the least stable mode to increase. Notice in Figure~\ref{sens_finalbranch} the fields of sensitivity $\delta \lambda_{1,r}$ showing more agreement with the growth rate sensitivity of the fixed cylinder. As observed in Figures~\ref{Amp_Re46_m20} and \ref{Amp_Re46_m7}, for $(m^{\ast},V_r) = (7, 10)$ and $(m^{\ast},V_r) = (20, 9.5)$ the amplitude of oscillations are smaller and the eigen-frequencies $\lambda_{1,i}$ are closer to the $\lambda_{1,i}$ of the fixed cylinder.

\subsubsection{Frequency sensitivity}
\begin{figure}
    \centering
    \subfigure[ $S1_i = (\mathbf{U}\cdot \mathbf{U}^{\dagger})_{i}$\label{bf_fixedf}] {\includegraphics[width=0.3\textwidth]{figures/sens_bf_fixed.eps}}
    \subfigure[ $S2_i = (\widehat{\mathbf{u}}\cdot \mathbf{u}^{\dagger})_{i}$\label{struc_fixedf}] {\includegraphics[width=0.3\textwidth]{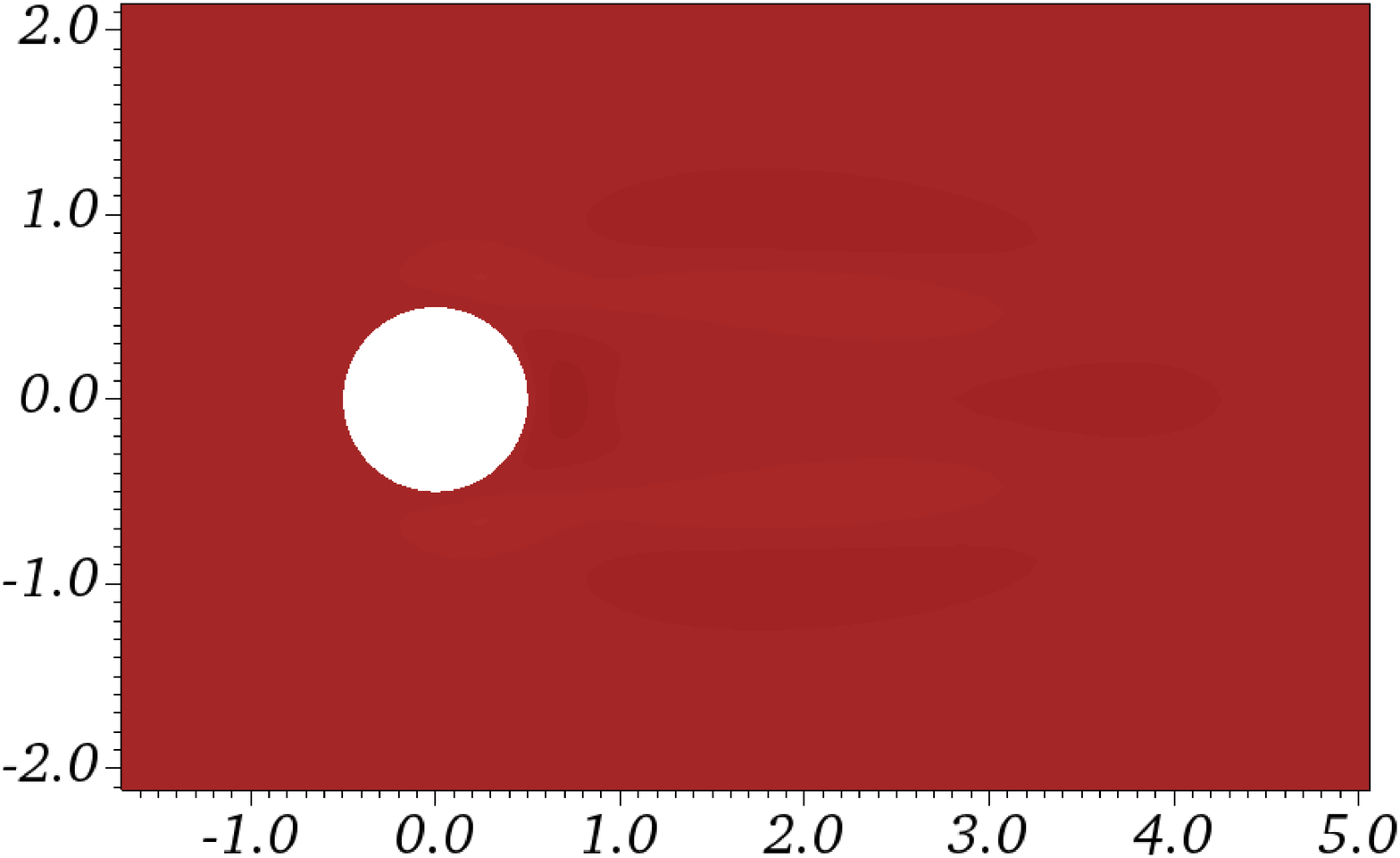}}
    \subfigure[ $(\mathbf{U}\cdot \mathbf{U}^{\dagger})_{i} + (\widehat{\mathbf{u}}\cdot \mathbf{u}^{\dagger})_{i}$\label{tot_fixedf}] {\includegraphics[width=0.3\textwidth]{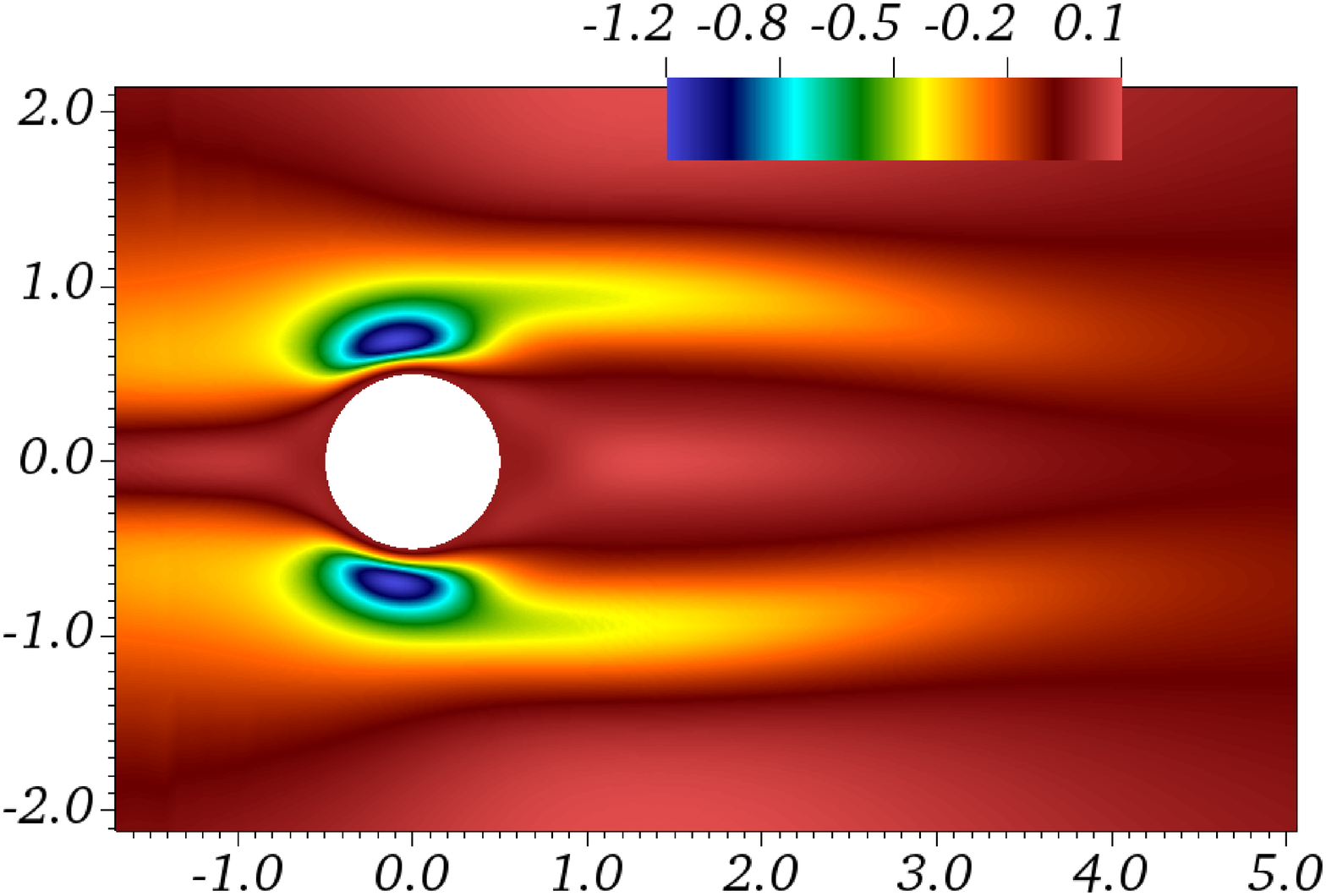}}
    \caption{Frequency sensitivity fields for the flow around a fixed cylinder at $\Rey=46.6$.}
    \label{flowsensf}
\end{figure}

\begin{figure}
    \centering
    \subfigure[$(m^{\ast},V_r) = (7, 7.5)$ \label{tot_sensm7Vr75f}] {\includegraphics[width=0.3\textwidth]{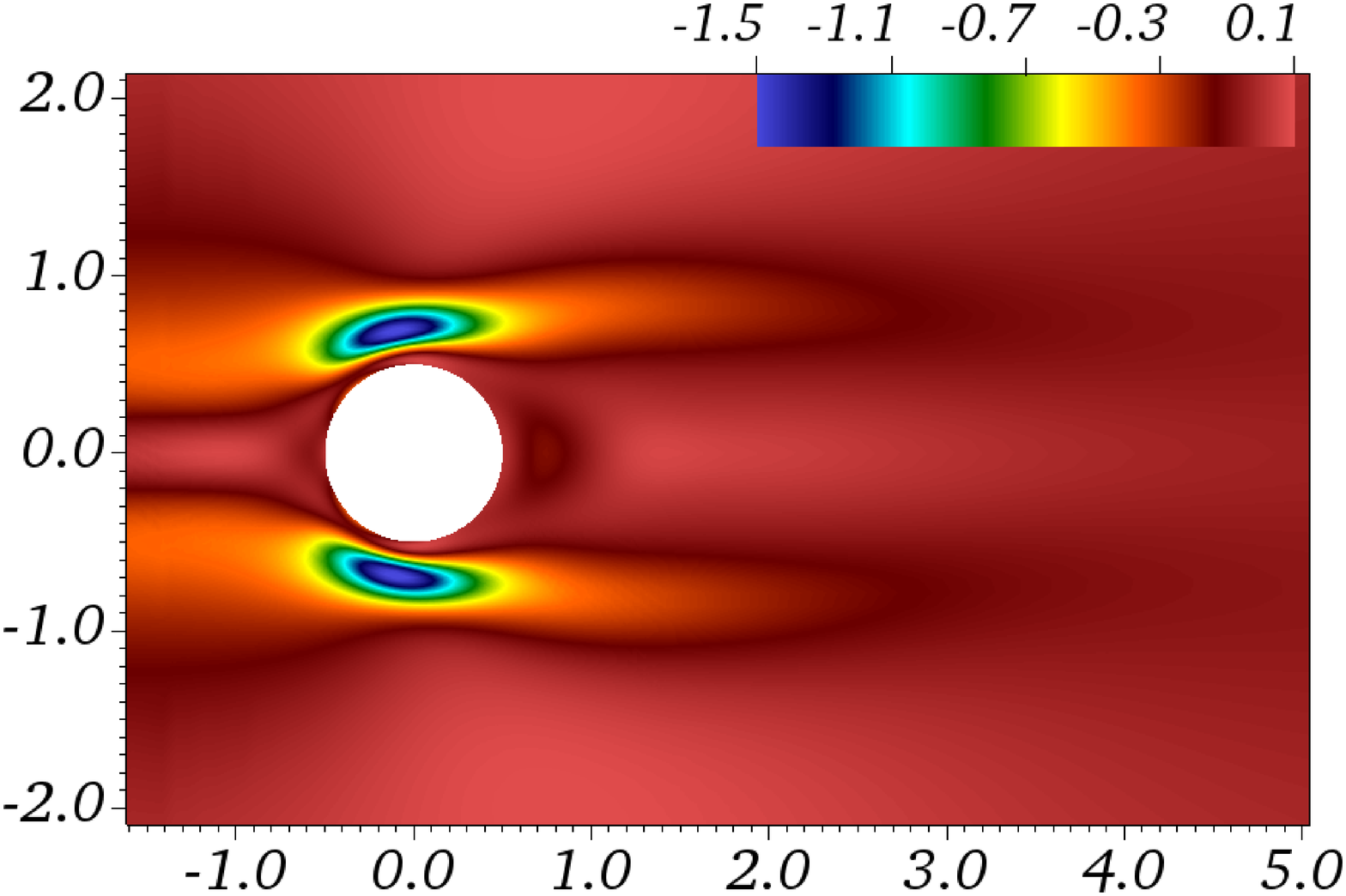}}
    \subfigure[$(m^{\ast},V_r) = (20,8)$ \label{tot_sensm20Vr8f}]  {\includegraphics[width=0.3\textwidth]{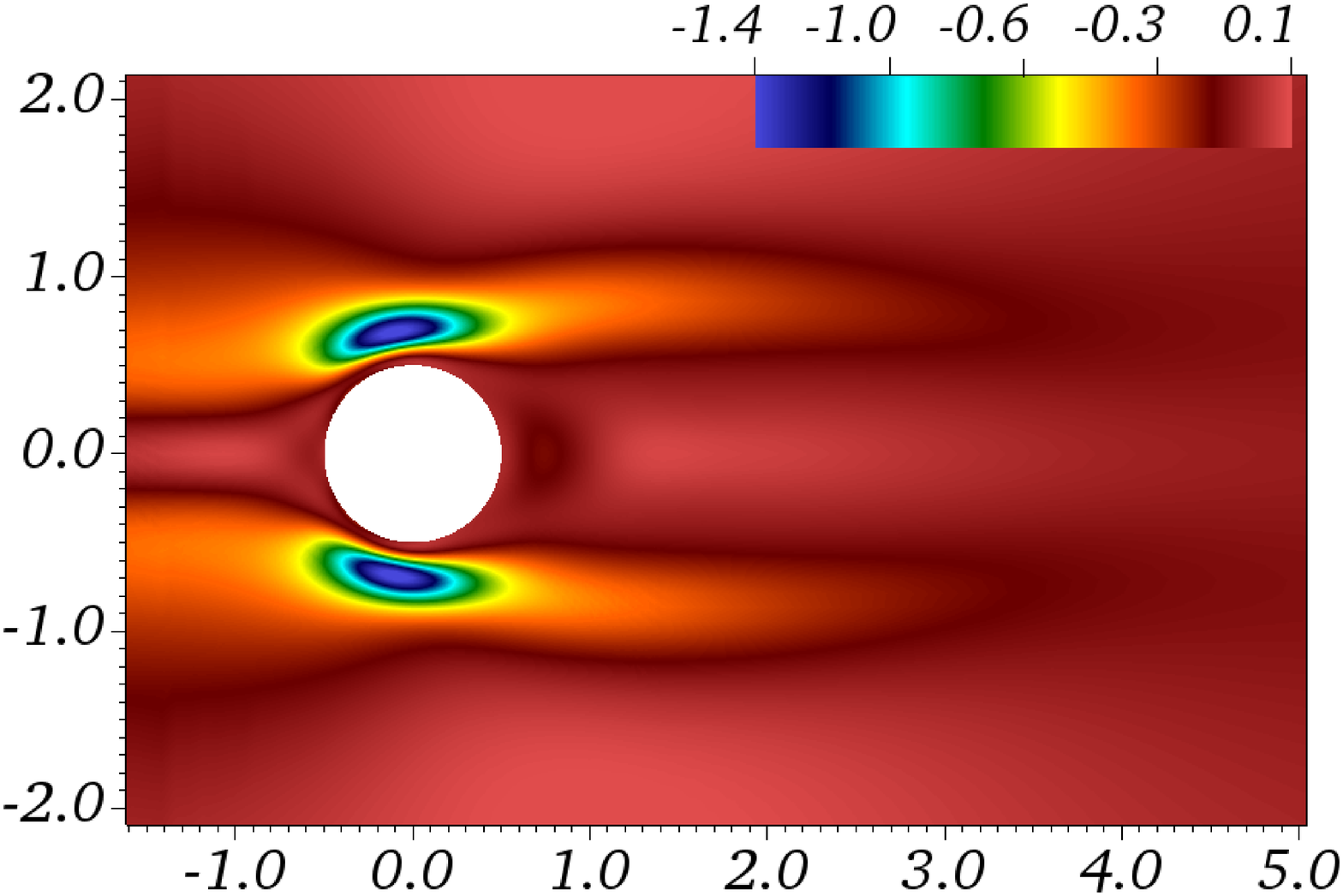}}\\
    \subfigure[$(m^{\ast},V_r) = (7, 10)$ \label{tot_sensm7Vr10f}] {\includegraphics[width=0.3\textwidth]{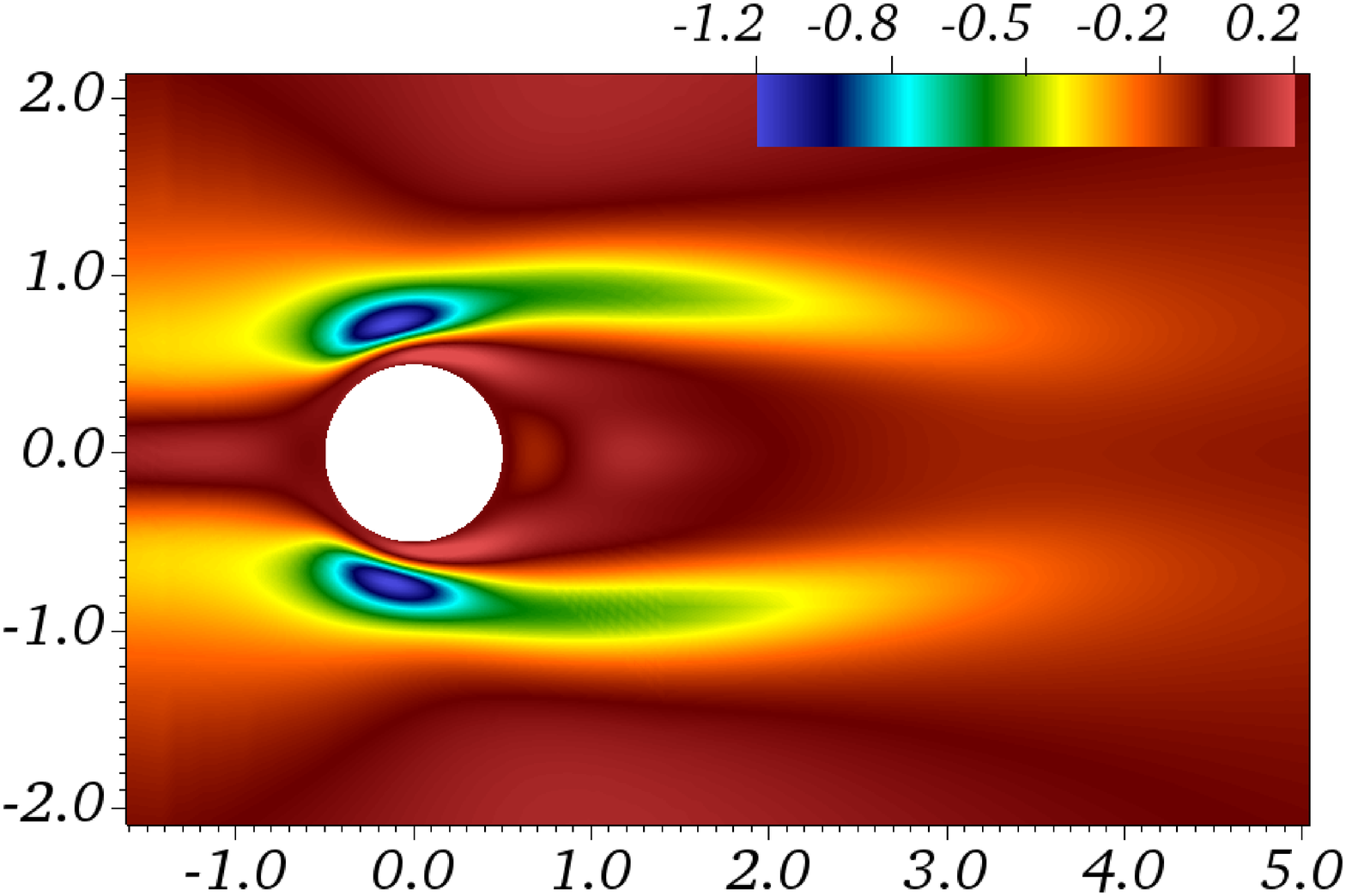}}
    \subfigure[$(m^{\ast},V_r) = (20,9.5)$ \label{tot_sensm20Vr95f}]  {\includegraphics[width=0.3\textwidth]{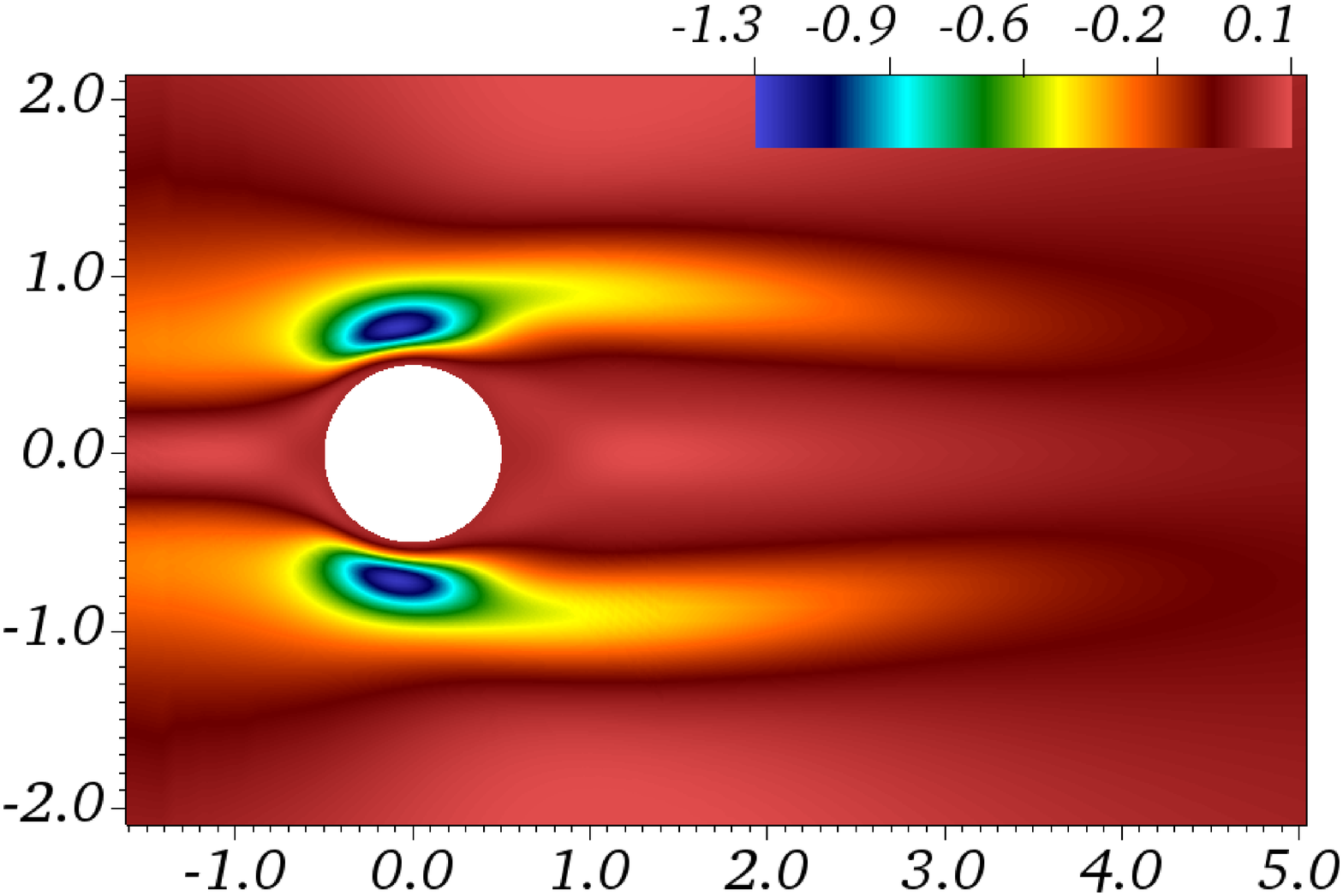}}
    \caption{Frequency sensitivity fields evaluated by $(\mathbf{U}\cdot \mathbf{U}^{\dagger})_{i} + (\widehat{\mathbf{u}}\cdot \mathbf{u}^{\dagger})_{i}$ for reduced velocity in the resonance range: $(m^{\ast},V_r) = (7, 7.5)$ (a) and $(m^{\ast},V_r) = (20,8)$ (b); and in the final branch $(m^{\ast},V_r) = (7, 10)$ (c) and $(m^{\ast},V_r) = (20,9.5)$ (d).}
    \label{sens_resonance_finalf}
\end{figure}

Figures~\ref{flowsensf}--\ref{sens_resonance_finalf} show the frequency sensitivity fields of the least stable mode in the flow around fixed and elastically-mounted cylinders for $Re=46.6$. For the fixed cylinder, Figure~\ref{struc_fixedf} shows that the frequency structural sensitivity $S2_i$ does not contribute to the total sensitivity, which is plotted in Figure~\ref{tot_fixedf}. This is also the case for the flow around an elastically-mounted cylinder when $V_r$ is in the resonance range or in the final branch. Therefore, only the total frequency sensitivity is plotted in Figure~\ref{sens_resonance_finalf}. 

\begin{figure}
    \centering
    \subfigure[$S1_i = (\mathbf{U}\cdot \mathbf{U}^{\dagger})_i$ \label{bf_sensm7Vr58f}] {\includegraphics[width=0.3\textwidth]{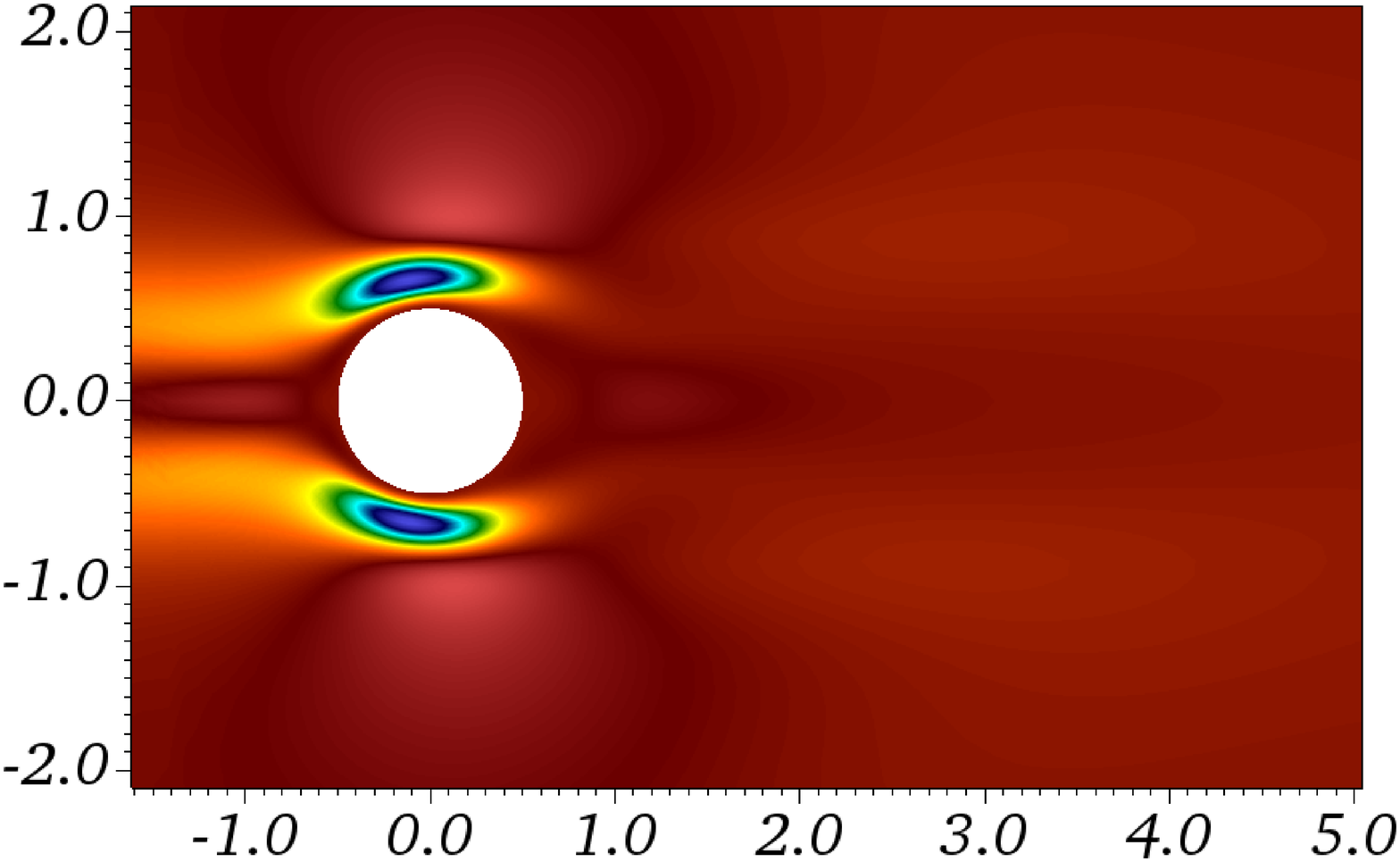}}
    \subfigure[$S2_i = (\widehat{\mathbf{u}}\cdot \mathbf{u}^{\dagger})_{i}$ \label{struc_sensm7Vr58f}] {\includegraphics[width=0.3\textwidth]{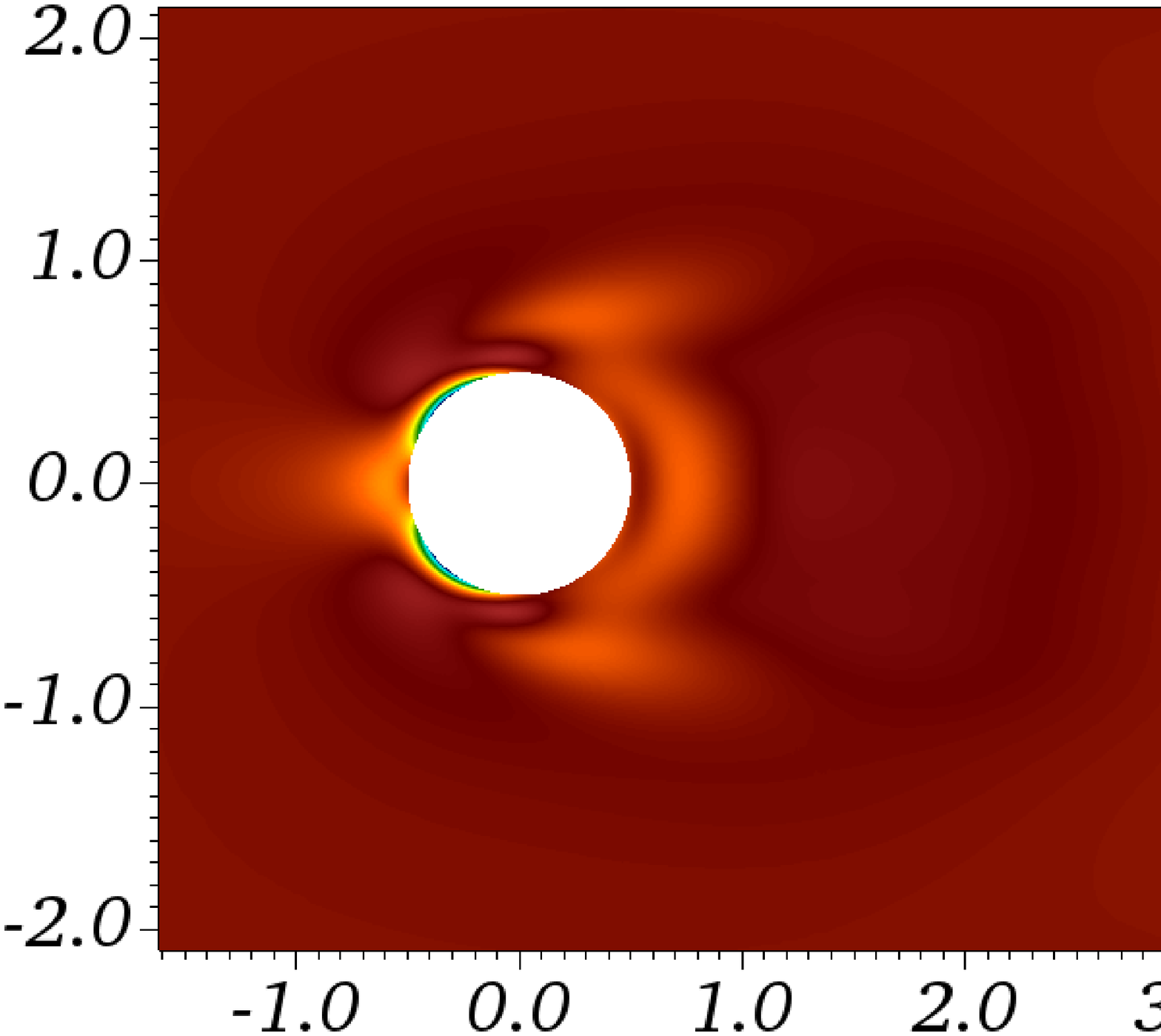}}
    \subfigure[$(\mathbf{U}\cdot \mathbf{U}^{\dagger})_{i} + (\widehat{\mathbf{u}}\cdot \mathbf{u}^{\dagger})_{i}$ \label{tot_sensm7Vr58f}] {\includegraphics[width=0.3\textwidth]{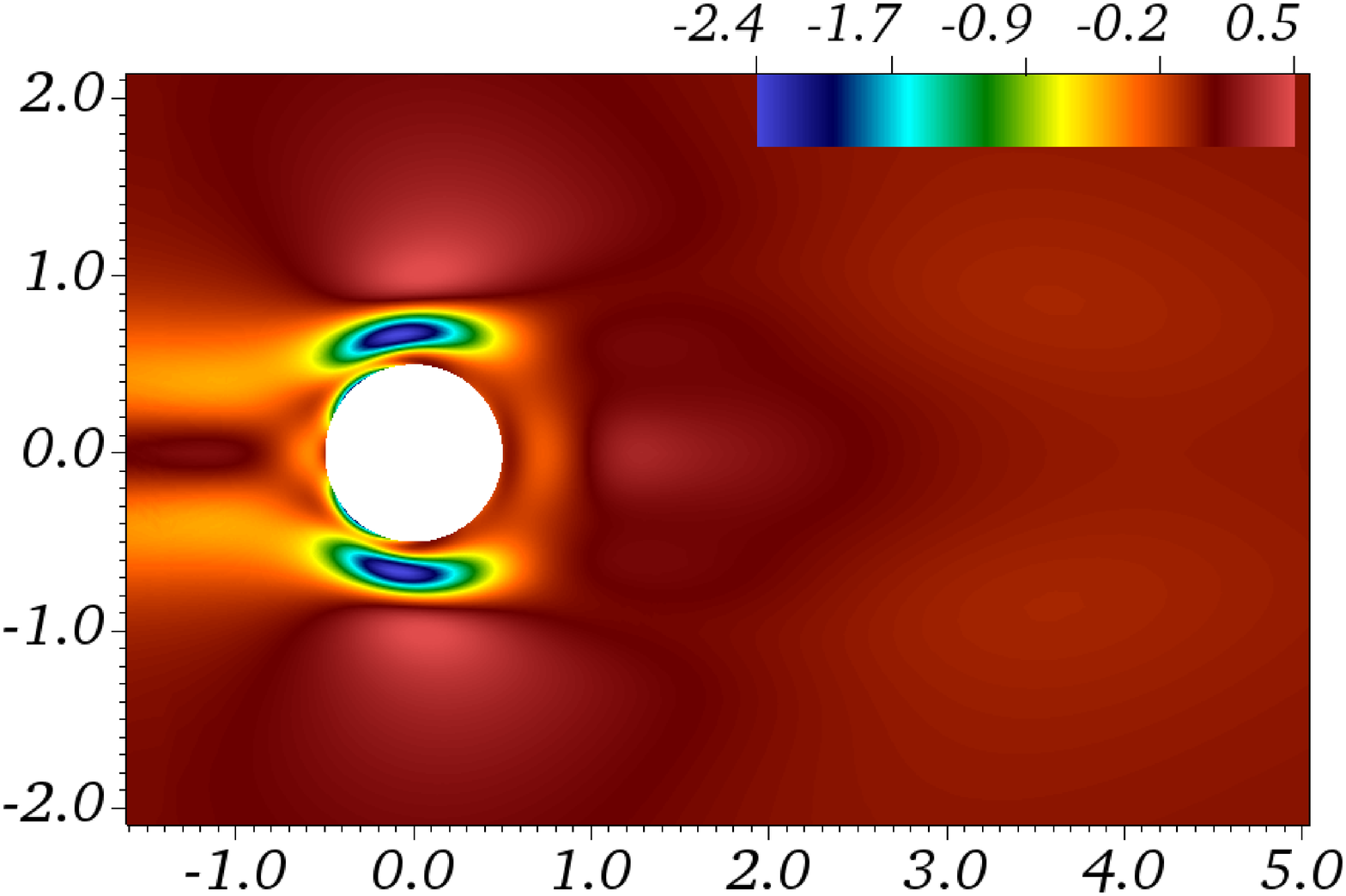}}
    \subfigure[$S1_i = (\mathbf{U}\cdot \mathbf{U}^{\dagger})_{i}$ \label{bf_sensm20Vr63f}] {\includegraphics[width=0.3\textwidth]{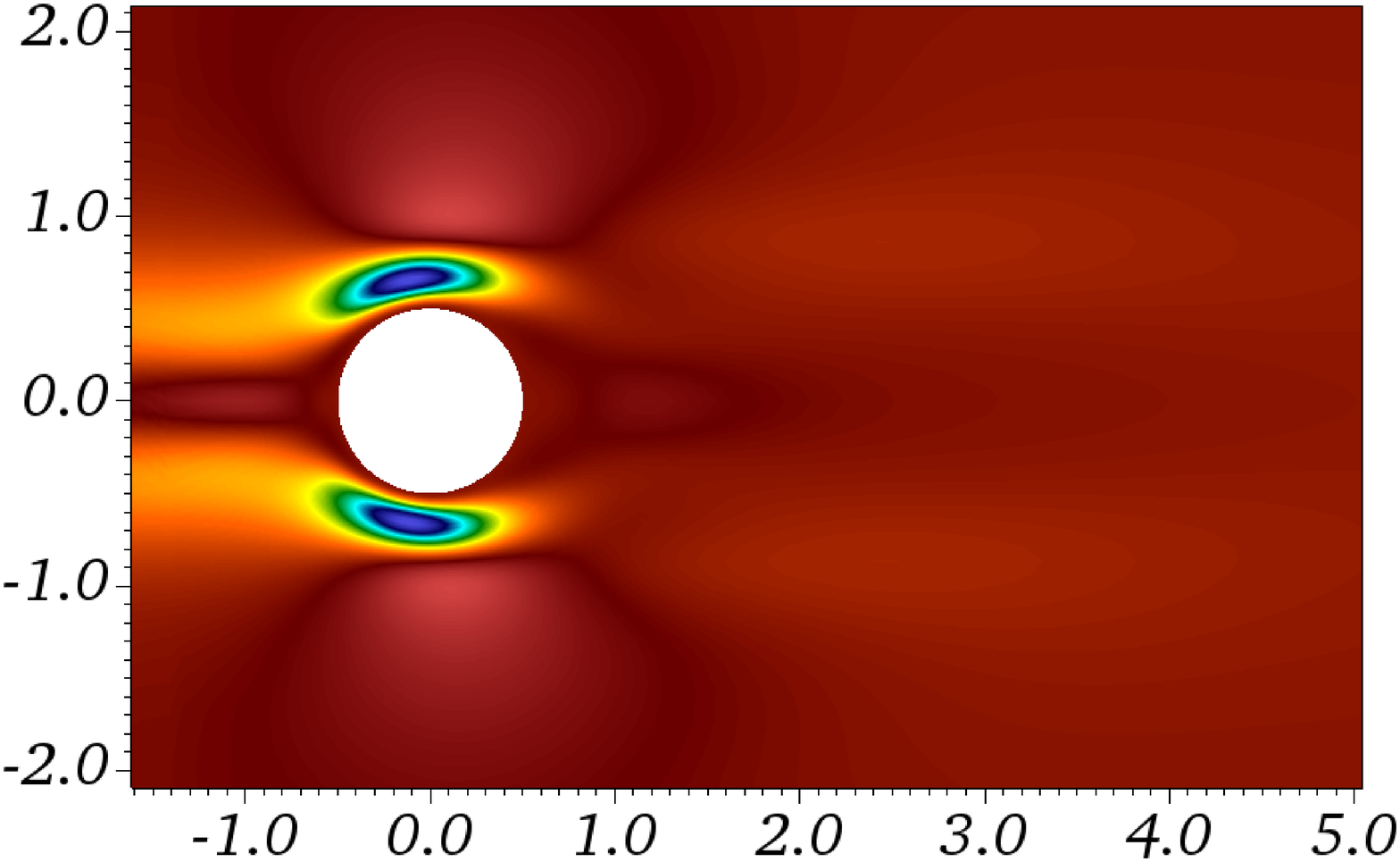}}
    \subfigure[$S2_i = (\widehat{\mathbf{u}}\cdot \mathbf{u}^{\dagger})_{i}$ \label{struc_sensm20Vr63f}] {\includegraphics[width=0.3\textwidth]{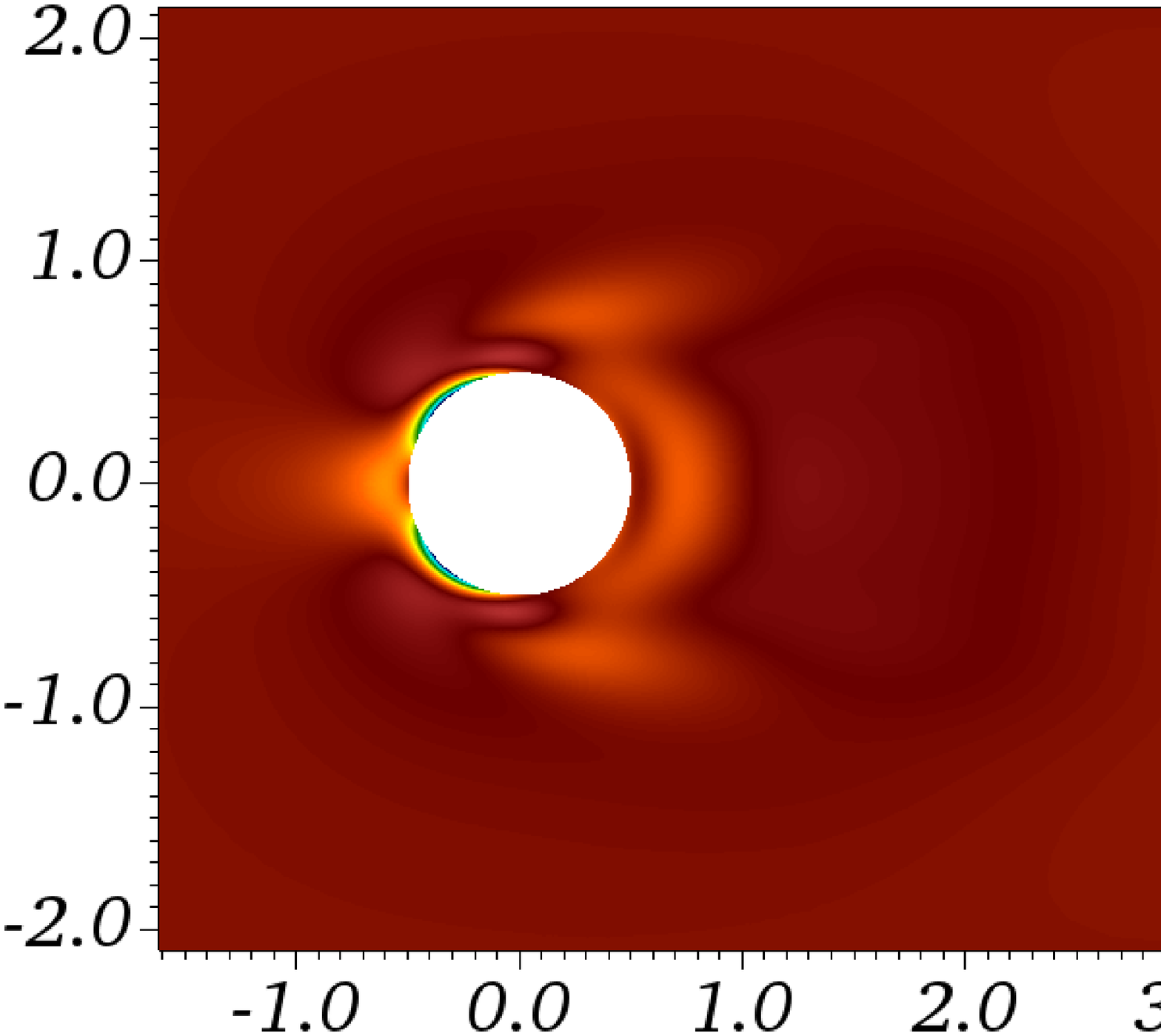}}
    \subfigure[$(\mathbf{U}\cdot \mathbf{U}^{\dagger})_{i} + (\widehat{\mathbf{u}}\cdot \mathbf{u}^{\dagger})_{i}$ \label{tot_sensm20Vr63f}] {\includegraphics[scale=0.12]{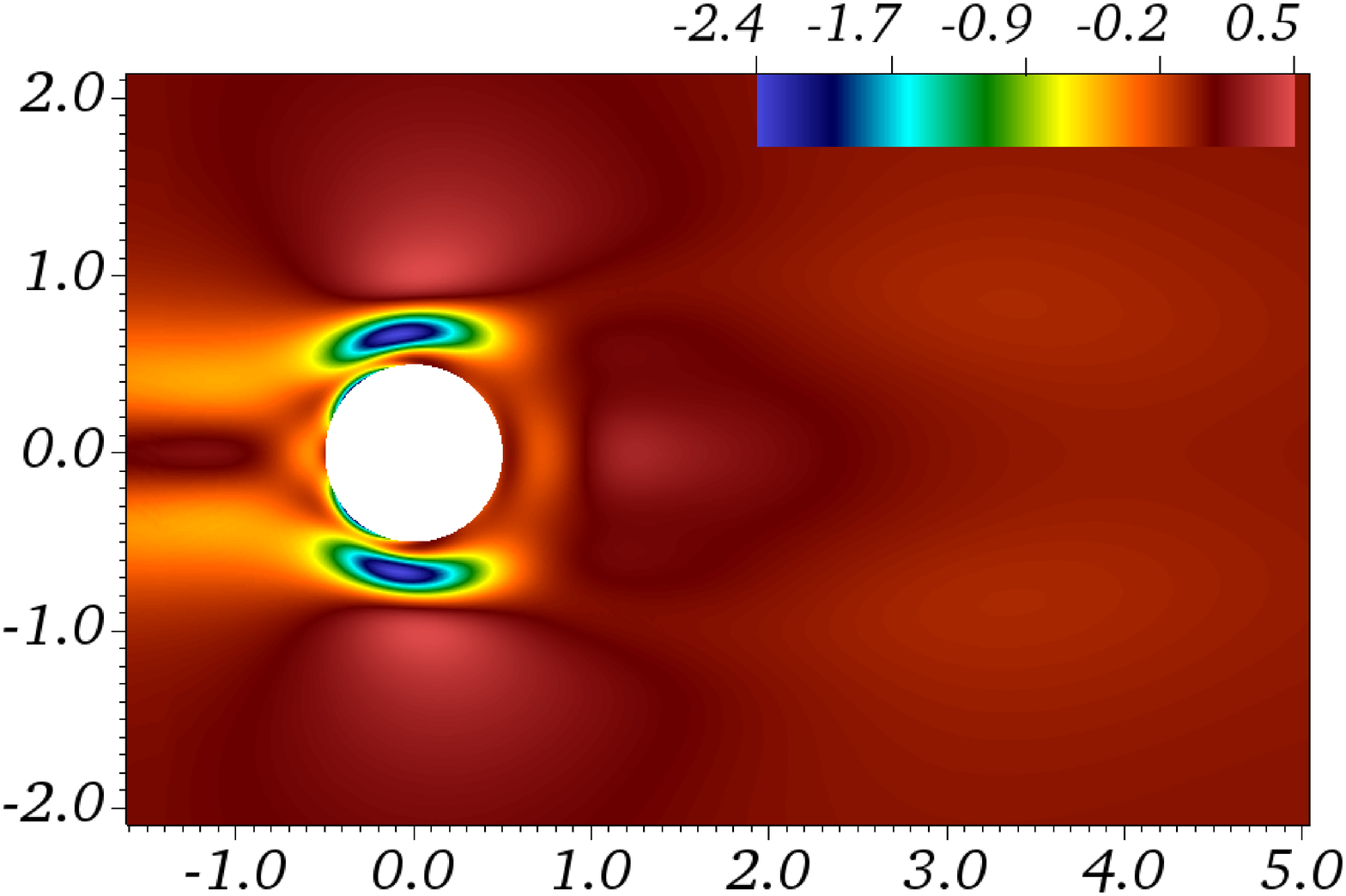}}
    \caption{Frequency sensitivity fields for reduced velocity in the initial branch: $(m^{\ast},V_r) = (7, 5.8)$ (a), (b), and (c); $(m^{\ast},V_r) = (20, 6.3)$ (d), (e), and (f).}
    \label{sens_initbranchf}
\end{figure}

For reduced velocities in the initial branch [$(m^{\ast},V_r) = (7, 5.8)$ and $(m^{\ast},V_r) = (20, 6.3)$], Figures~\ref{struc_sensm7Vr58f} and \ref{struc_sensm20Vr63f} show the structural sensitivity with important contribution in the total sensitivity evaluated by the field $(\mathbf{U}\cdot \mathbf{U}^{\dagger})_{i} + (\widehat{\mathbf{u}}\cdot \mathbf{u}^{\dagger})_{i}$, where larger intensities are identified at the cylinder wall, upstream of the separation point. In this case, by comparing the fields of $(\mathbf{U}\cdot \mathbf{U}^{\dagger})_{i} + (\widehat{\mathbf{u}}\cdot \mathbf{u}^{\dagger})_{i}$ against the fixed cylinder case, differences are observed in the magnitudes and in the spatial distribution of stronger regions of sensitivity. For $(m^{\ast},V_r) = (7, 5.8)$ and $(m^{\ast},V_r) = (20, 6.3)$, the results show the flow around an elastically-mounted cylinder is more sensitive to an external forcing (\ref{forcing}), mainly upstream of the separation points at the cylinder wall, and at the upside and downside of the cylinder. In addition, the intensity is not high at the limit of the recirculation bubble as has been noticed in the fixed cylinder sensitivity (Figure~\ref{tot_fixedf}), and for the elastically-mounted cylinder with $V_r$ in the resonance range and in the final branch (Figure~\ref{sens_resonance_finalf}).

In all cases, the variation $\delta \lambda_{1,i}$ is negative if the forcing (\ref{forcing}) is applied closer to the cylinder, at the upside and downside. When $V_r$ is in the initial branch, Figures~\ref{tot_sensm7Vr58f} and \ref{tot_sensm20Vr63f} show that the variation of the least stable mode is negative if the forcing (\ref{forcing}) is inserted close to the cylinder wall, upstream from the separation point.

\subsection{Verification of the sensitivity analyses with a passive control}

According to the results presented in the previous sections, the growth rate sensitivities of the current FSI problem exhibited the greatest difference from that of the flow around a fixed cylinder when $V_r$ was in the initial branch ($(m^{\ast},V_r) = (7, 5.8)$ and $(m^{\ast},V_r) = (20, 6.3)$). As shown in Figures~\ref{eigen_Re46_m20} and ~\ref{eigen_Re46_m7}, in these cases the least stable mode has eigen-frequencies closer to the natural frequency of the structure. Another result to highlight is the change of the sensitivity field when the reduced velocity is varied. For instance, for $(m^{\ast},V_r) = (7, 5.8)$ and $(m^{\ast},V_r) = (20, 6.3)$ the steady forcing (\ref{forcing}) located symmetrically at $(x_p, y_p) = (1.2, \pm 1)$ makes the growth rate $\lambda_{1,r}$ to increase. On the other hand, this same forcing makes the $\lambda_{1,r}$ to decrease for $(m^{\ast},V_r) = (7, 7.5)$ and $(m^{\ast},V_r) = (20, 8)$. 

\begin{figure}
    \centering
    \subfigure[] {\includegraphics[scale=0.15]{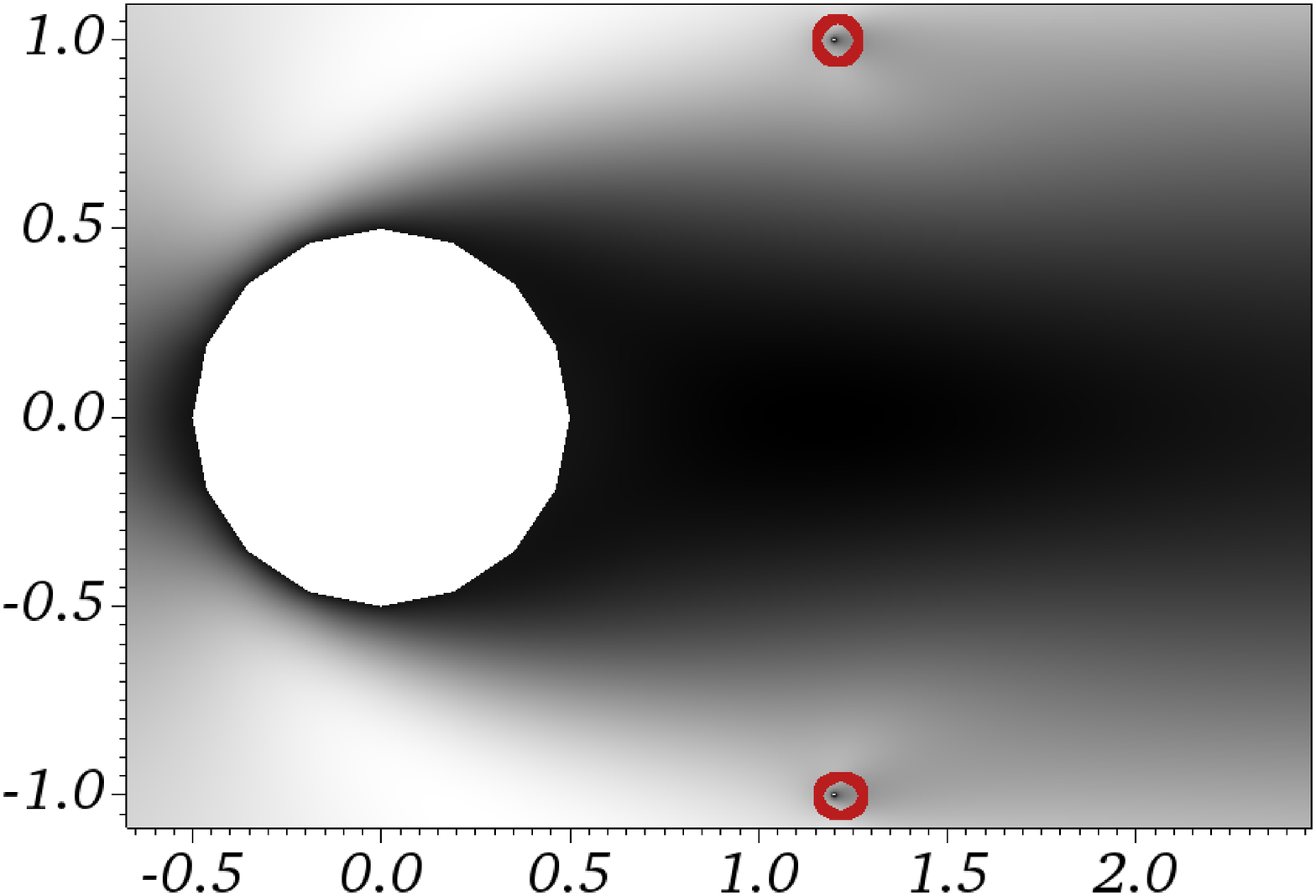}}
    \subfigure[] {\includegraphics[scale=0.15]{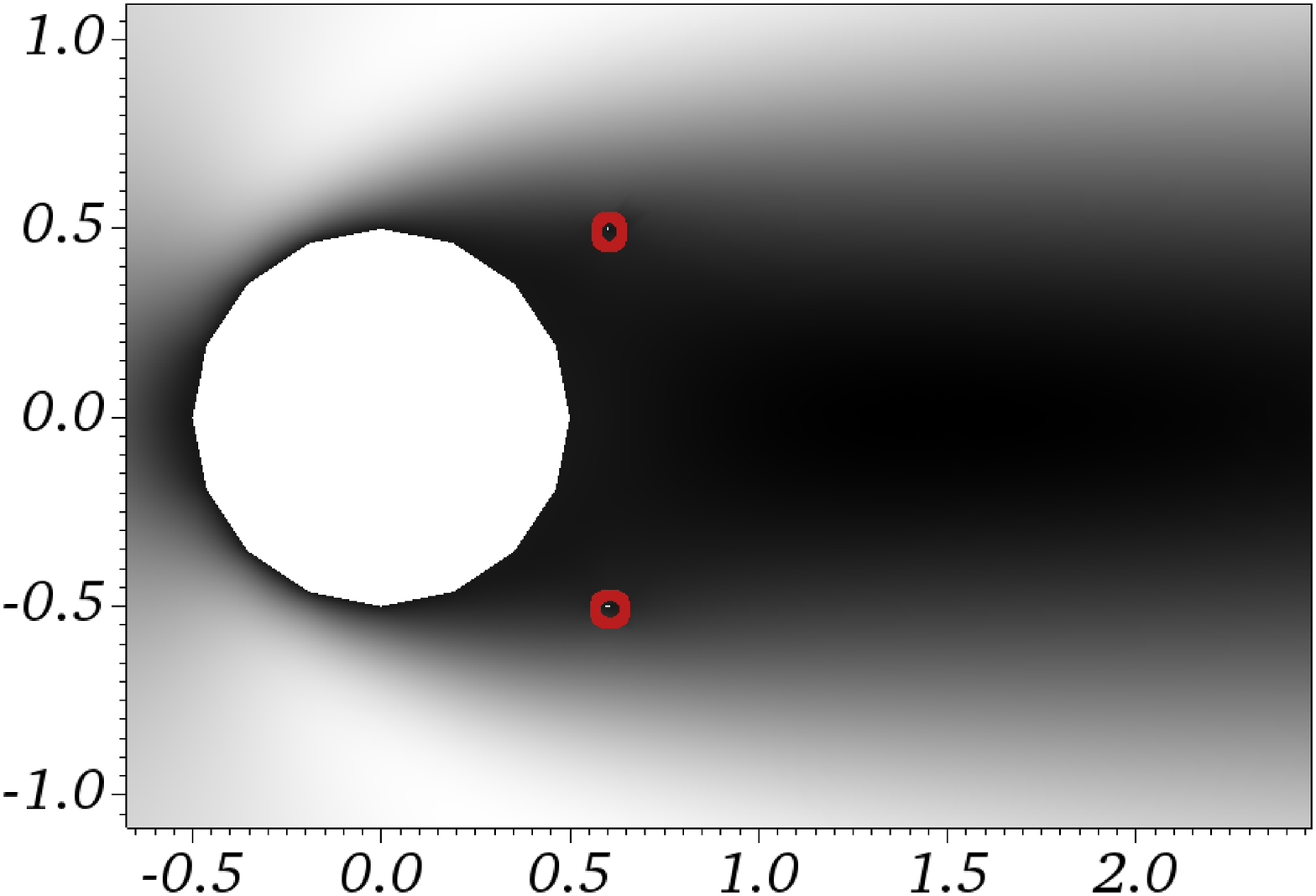}}
    \caption{Illustration of the circular cylinder with small control devices inserted in the positions $(x_p, y_p) = (1.2, \pm 1)$ (a), and $(x_p, y_p) = (0.6, \pm 0.5)$ (b).}
    \label{controls}
\end{figure}

\begin{figure}[!h]
    \centering
    \includegraphics[width = 0.8\textwidth]{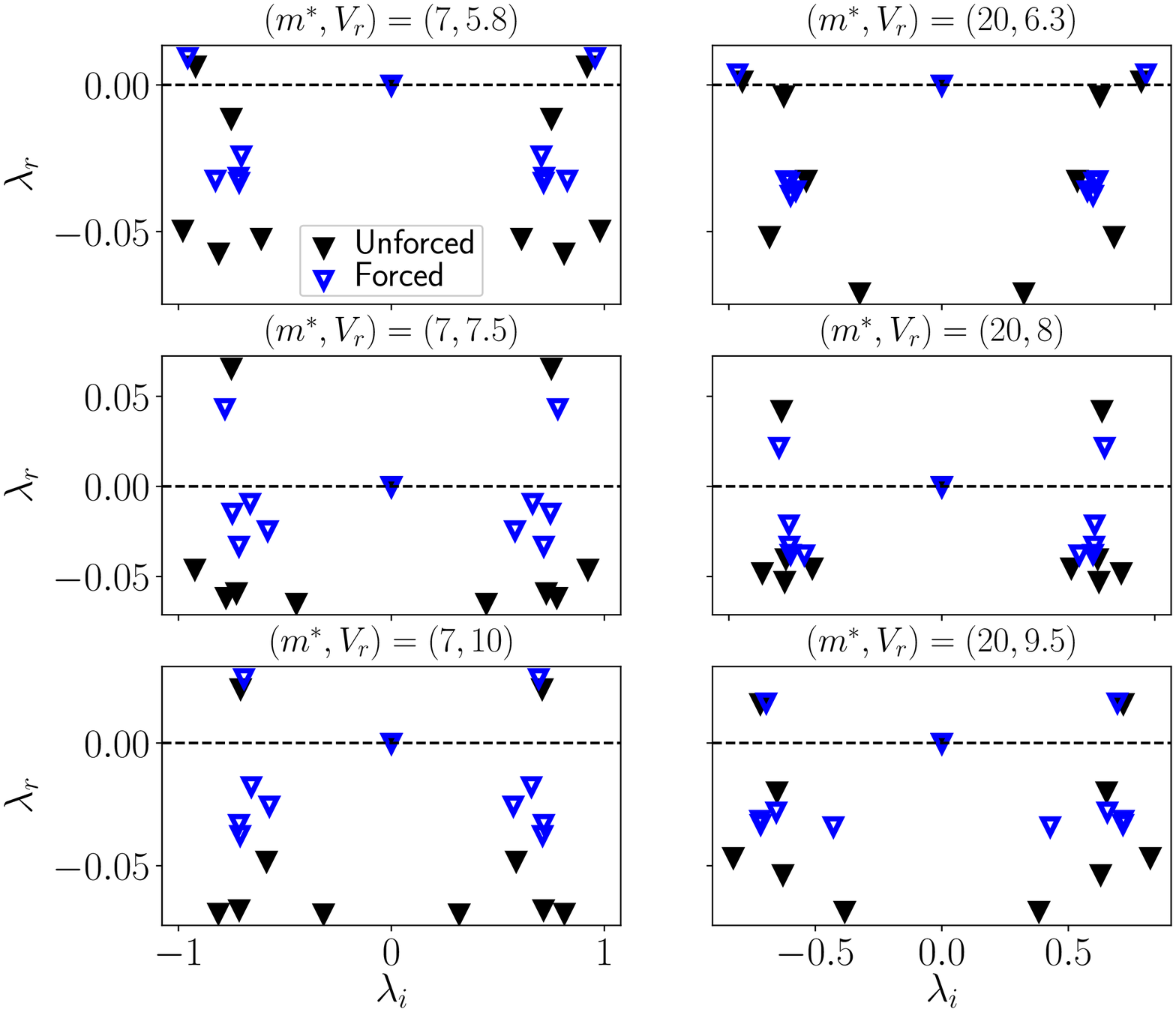}
    \caption{Five least stable eigenvalues pairs for the forced and unforced FSI systems. The insertion of small cylinders with the diameter $d=0.01$ located at $(x_p, y_p) = (1.2, \pm 1)$ is refereed as forced.}
    \label{eigen_control0}
\end{figure}

To verify the results provided by the sensitivity fields, small cylinders with diameter $d = 0.01$ were included in the field to serve as steady forcing in nonlinear simulations. Firstly, they were inserted symmetrically at the positions $(x_p, y_p) = (1.2, \pm 1)$ as illustrated in Figure~\ref{controls}(a). Next, the steady base flow was computed, and the modes and respective eigenvalues were obtained. Figure~\ref{eigen_control0} shows the five least stable eigenvalues for unforced and forced systems with reduced velocity in the initial branch [$(m^{\ast},V_r) = (7, 5.8)$ and $(m^{\ast},V_r) = (20, 6.3)$], in the resonance range [$(m^{\ast},V_r) = (7, 7.5)$ and $(m^{\ast},V_r) = (20, 8)$], and in the final branch [$(m^{\ast},V_r) = (7, 10)$ and $(m^{\ast},V_r) = (20, 9.5)$]. By forced cases we mean the cases with the insertion of the small cylinder.

Figure~\ref{eigen_control0} shows the insertion of small cylinders at $(x_p, y_p) = (1.2, \pm 1)$, which was expected to provide different responses. For reduced velocities $V_r$ in the initial branch, the real part of the least stable eigenvalue $\lambda_{1,r}$ showed positive variation. On the other hand, $\lambda_{1,r}$ exhibited negative variation for $V_r$ in the resonance range. In the final branch, there were different responses for $m^{\ast} = 20$ and $m^{\ast} = 7$. In the first case, the insertion of small cylinders did not modify $\lambda_{1,r}$, whereas $\lambda_{1,r}$ showed positive variation for $m^{\ast}=7$. These results are in agreement with the sensitivity fields, Figures~\ref{tot_sensm7Vr58} and \ref{tot_sensm20Vr63} show the sensitivity fields with positive values on the vicinity of $(x_p, y_p) = (1.2, \pm 1)$ for $V_r$ on the initial branch. On the other hand, for $V_r$ in the resonance range Figures~\ref{tot_sensm7Vr75} and \ref{tot_sensm20Vr8}) show that the sensitivity fields have negative values. In the final branch, it is necessary to make the comparison with more attention: Figure~\ref{tot_sensm7Vr10} shows for $(m^{\ast},V_r) = (7, 10)$ that the point $(x_p, y_p) = (1.2, \pm 1)$ is in the threshold region where the sensitivity field has positive values. 

Regarding the imaginary part $\lambda_{1,i}$, Figure~\ref{eigen_control0} shows that the insertion of the small cylinder centred at $(x_p, y_p) = (1.2, \pm 1)$ did not cause significant variations. Even though, by comparing the variation of $\lambda_{1,i}$ with the sensitivity field, we see a good agreement for $V_r$ in the initial branch and in the final branch. In these cases, the sensitivity fields present, respectively, positive and negative magnitudes in the vicinity of the point $(x_p, y_p) = (1.2, \pm 1)$ as shown in Figures~\ref{tot_sensm7Vr58f} and \ref{tot_sensm20Vr63f} for $V_r$ in the initial branch, and in Figures~\ref{tot_sensm7Vr10f} and \ref{tot_sensm20Vr95f} for $V_r$ in the final branch. For $(m^{\ast},V_r) = (20, 7.5)$, Figure~\ref{eigen_control0} shows $\lambda_{1,i}$ with positive variation when the cylinder was inserted. The sensitivity fields show a transition from negative to positive values near the point $(x_p, y_p) = (1.2, \pm 1)$.

\begin{figure}
    \centering
    \includegraphics[width=0.8\textwidth]{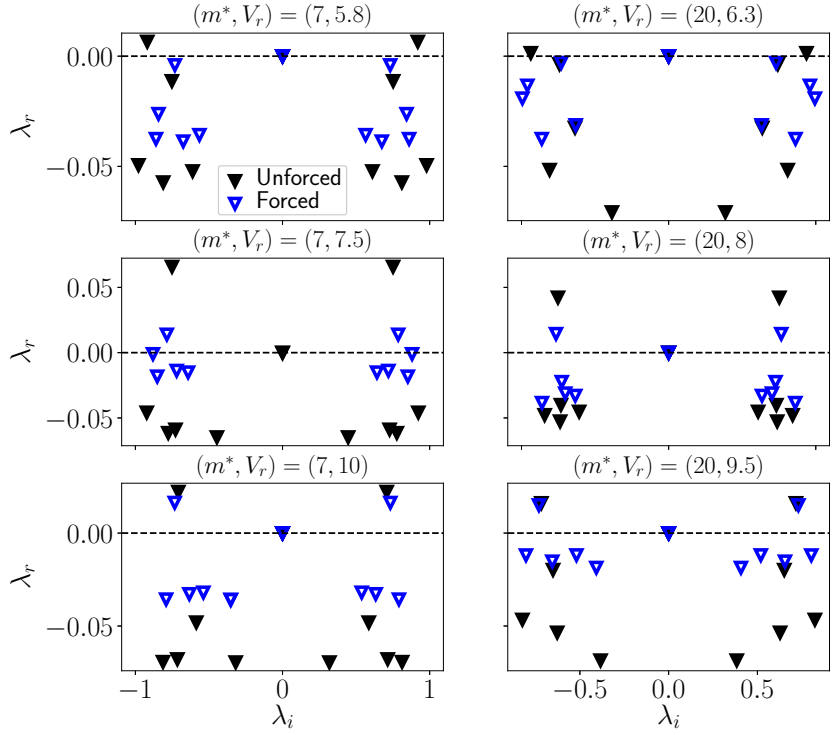}
    \caption{Five least stable eigenvalues pairs for the forced and unforced FSI systems. The insertion of small cylinders with the diameter $d=0.01$ located at $(x_p, y_p) = (0.6, \pm 0.5)$ is referred as forced.}
    \label{eigens}
\end{figure}

\begin{figure}
    \centering
    \subfigure[] {\includegraphics[scale=0.18
    ]{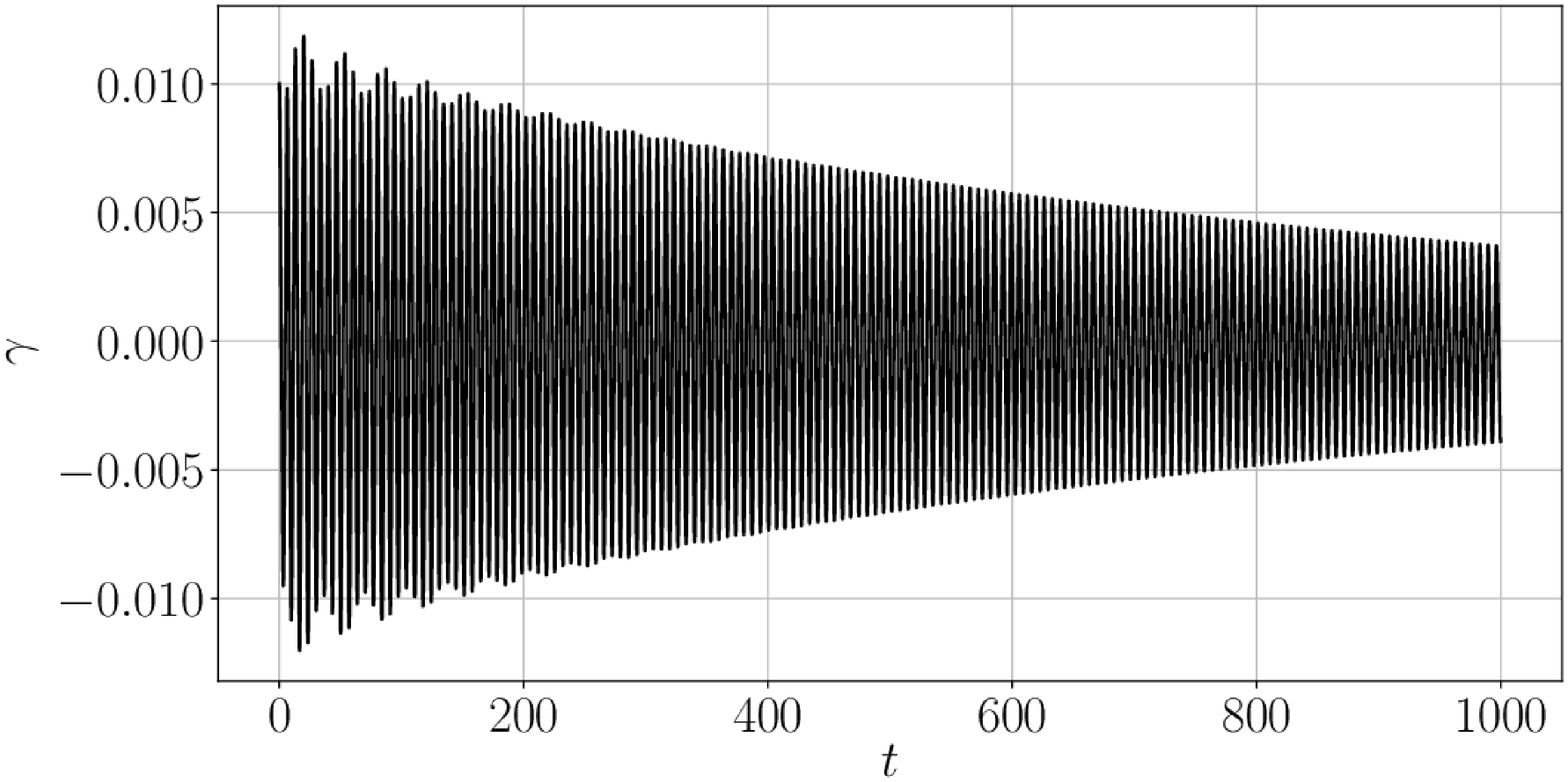}}
    \subfigure[] {\includegraphics[scale=0.15
    ]{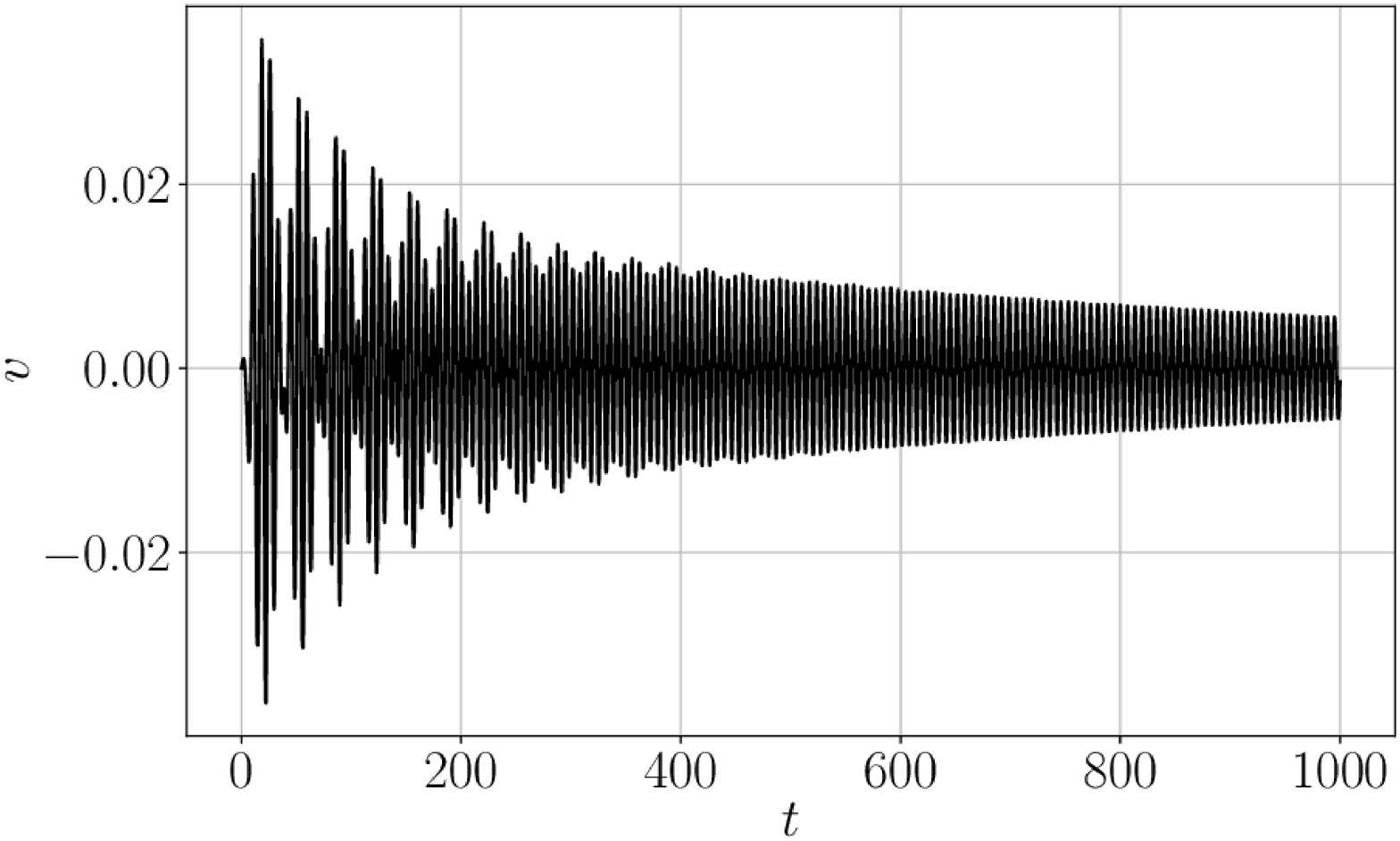}}
    \subfigure[] {\includegraphics[scale=0.12
    ]{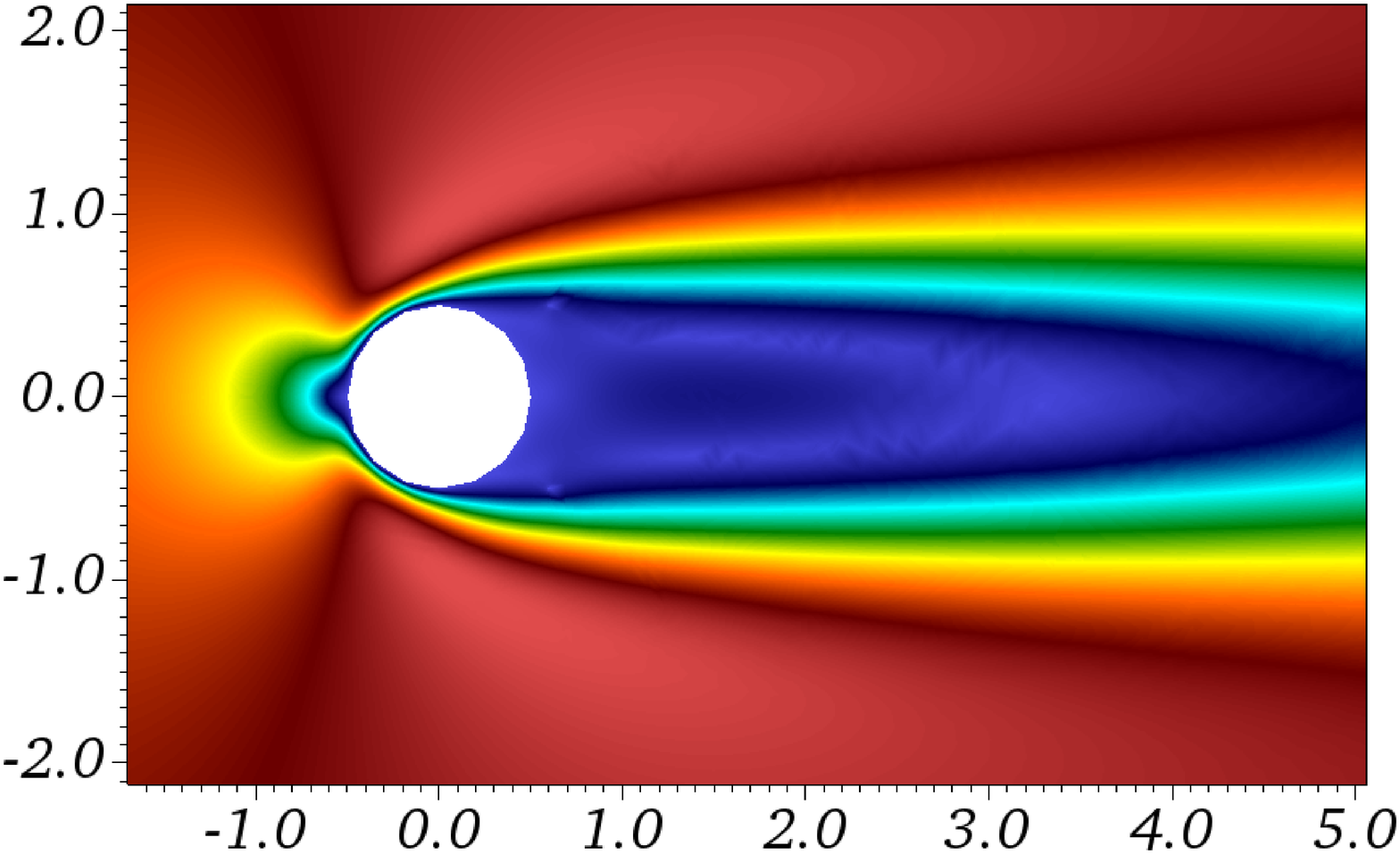}}
    \subfigure[] {\includegraphics[scale=0.18
    ]{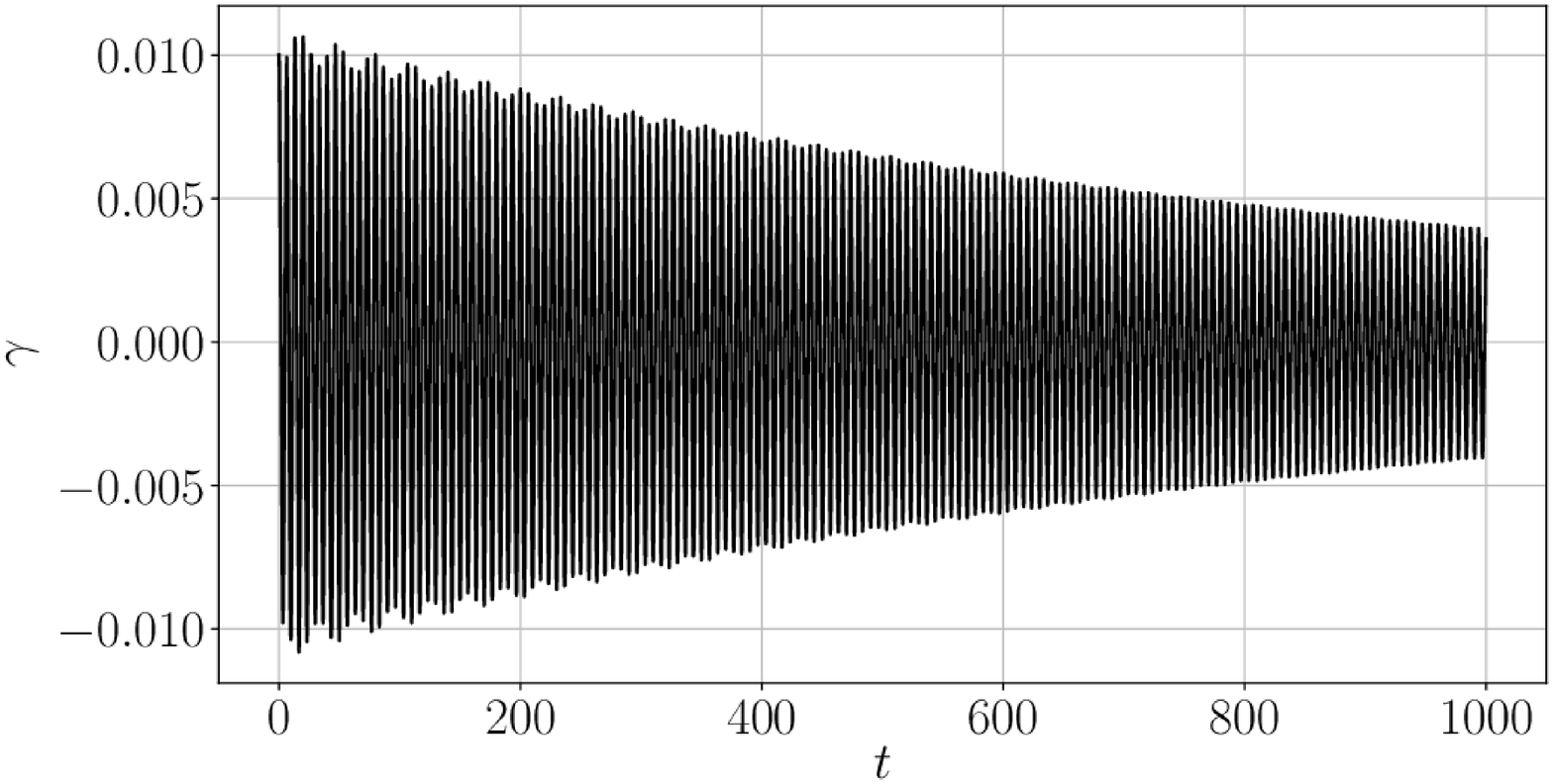}}
    \subfigure[] {\includegraphics[scale=0.15
    ]{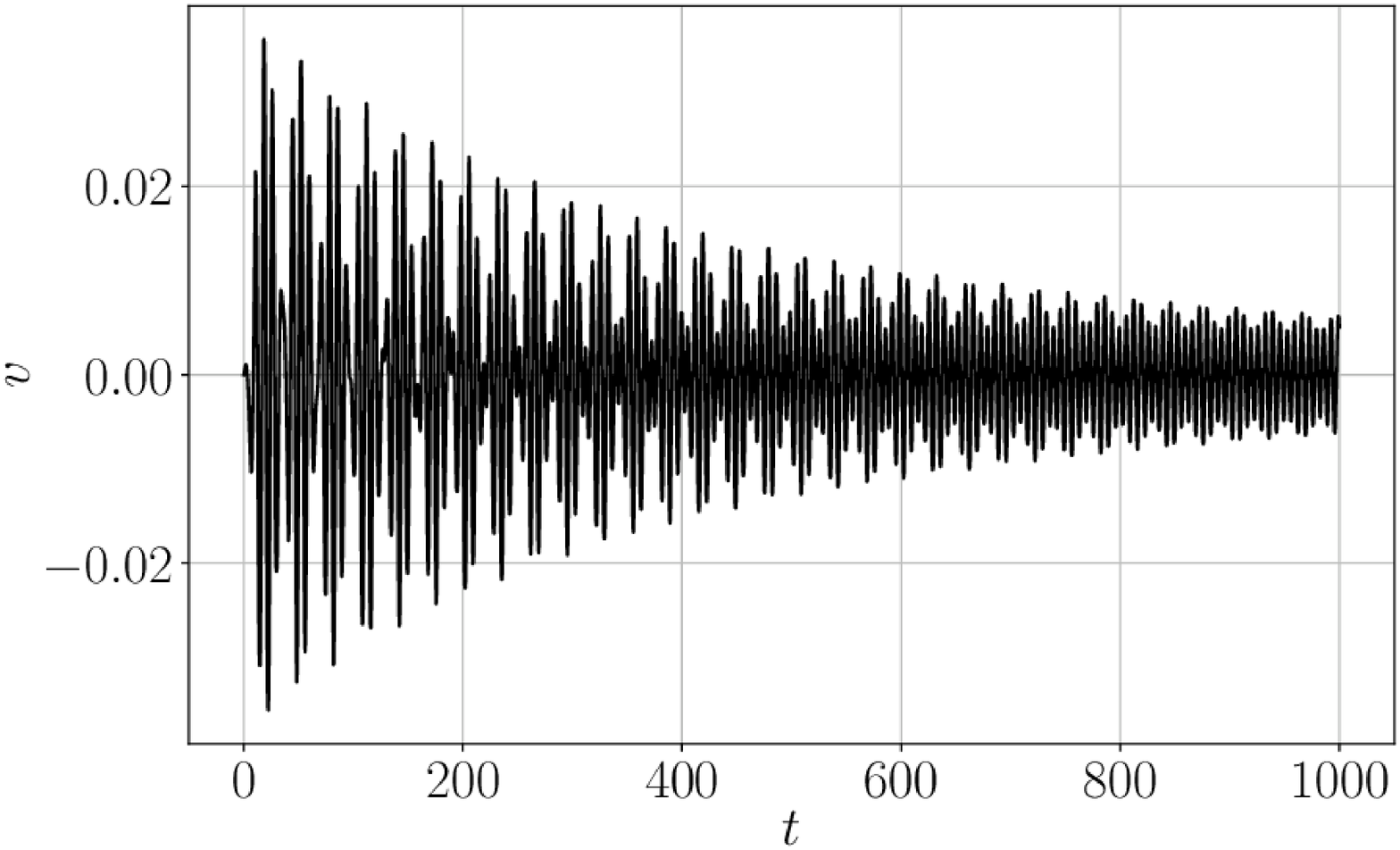}}
    \subfigure[] {\includegraphics[scale=0.12
    ]{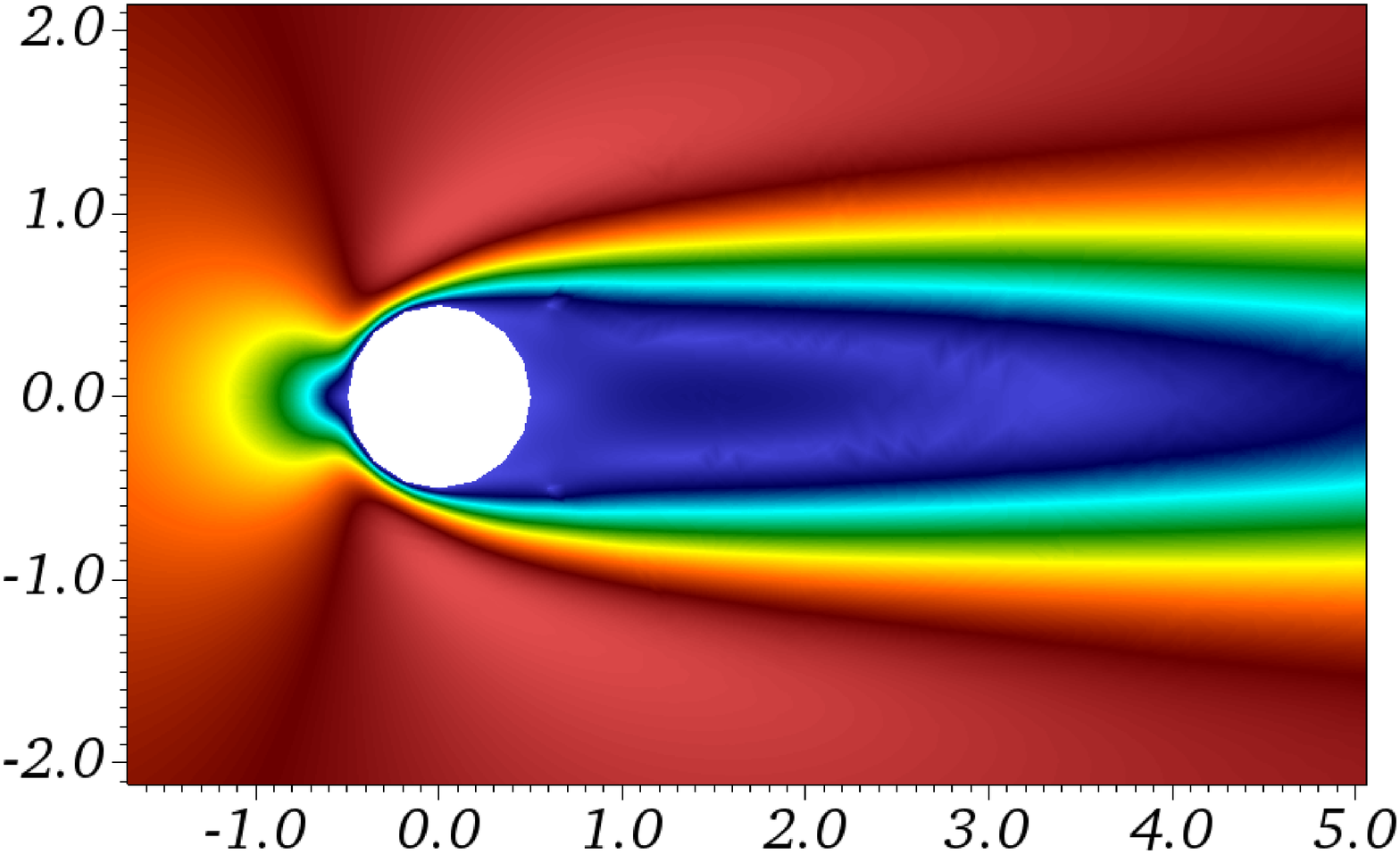}}
    \caption{Displacement of the cylinder $\gamma$, velocity component $v$ at the point (3, 0), and final velocity magnitude for $(m^{\ast},V_r) = (7, 5.8)$ (a), (b) and (c); and for $(m^{\ast},V_r) = (20, 6.3)$ (d), (e) and (f). Both cases considered a small control cylinder placed at $(x_p, y_p) = (0.6, \pm 0.5)$.}
    \label{disp_control}
\end{figure}

After evaluating the sensitivity fields, we chose to place the small cylinder at a point which would result in a decrease of the growth rate of the least stable mode. Therefore, small cylinders with a diameter of $d = 0.01$ were placed at $(x_p, y_p) = (0.6, \pm 0.5)$, as shown in Figure~\ref{controls}(b). As seen in Figure~\ref{eigens}, the growth rate $\lambda_{1,r}$ decreased in all cases. For $(m^{\ast},V_r) = (7, 5.8)$ and $(m^{\ast},V_r) = (20, 6.3)$, the FSI systems were stabilised. Figures~\ref{disp_control}(a) and  \ref{disp_control}(b) display the cylinder oscillation $\gamma$, the velocity component $v$ at the point $(x, y) = (3,0)$, and the final velocity magnitude field for $(m^{\ast},V_r) = (7, 5.8)$ and $(m^{\ast},V_r) = (20, 6.3)$. The displacement $\gamma$ and the component $v$ decreased with time. Therefore, the linear analyses provided the correct response, as it shows that the current FSI system is stable when the control cylinders are inserted at $(x_p, y_p) = (0.6, \pm 0.5)$. 

\begin{figure}
    \centering
    \subfigure[$(m^{\ast},V_r) = (7, 7.5)$] {\includegraphics[scale=0.24
    ]{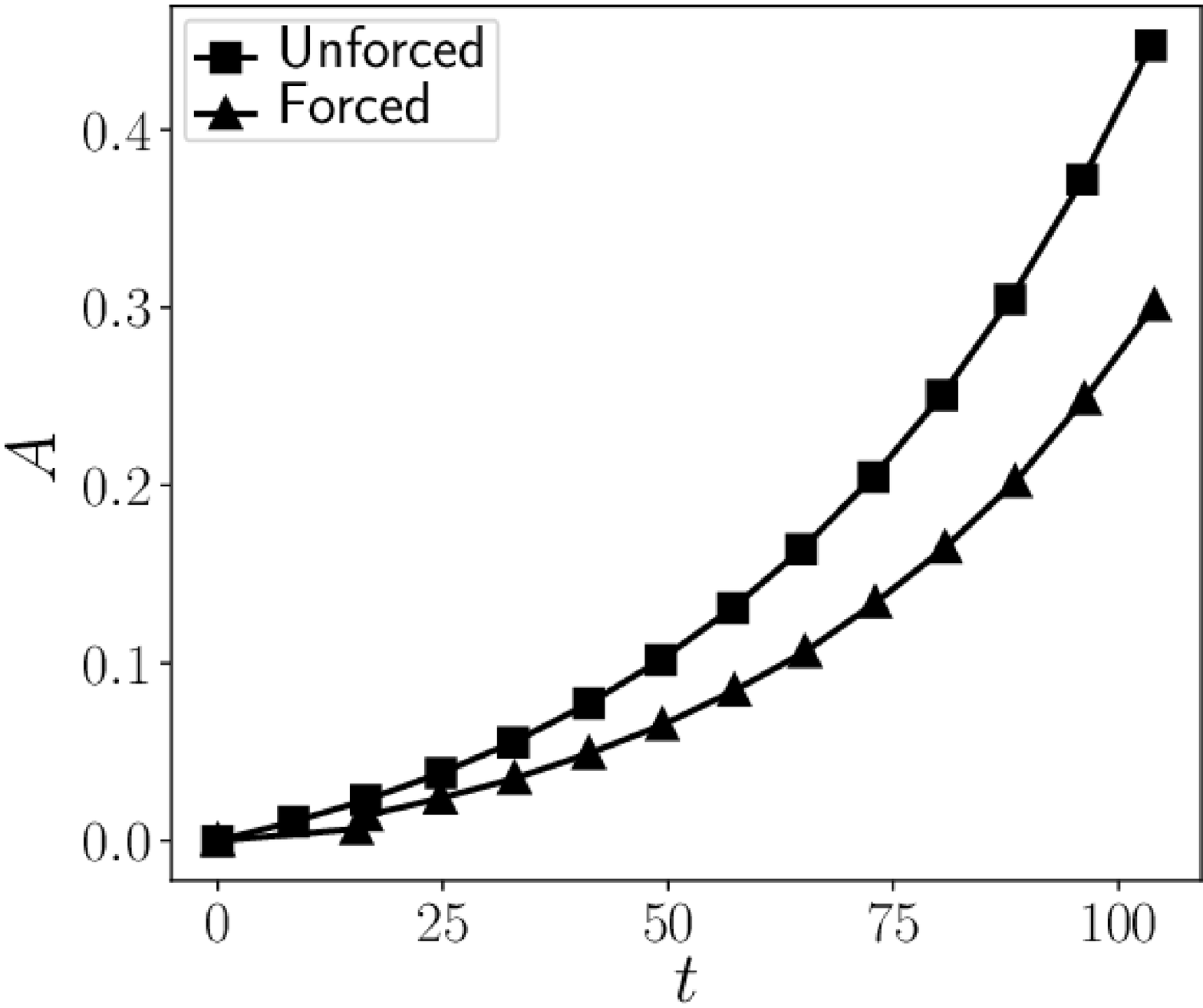}}
    \subfigure[$(m^{\ast},V_r) = (20, 8)$] {\includegraphics[scale=0.24]{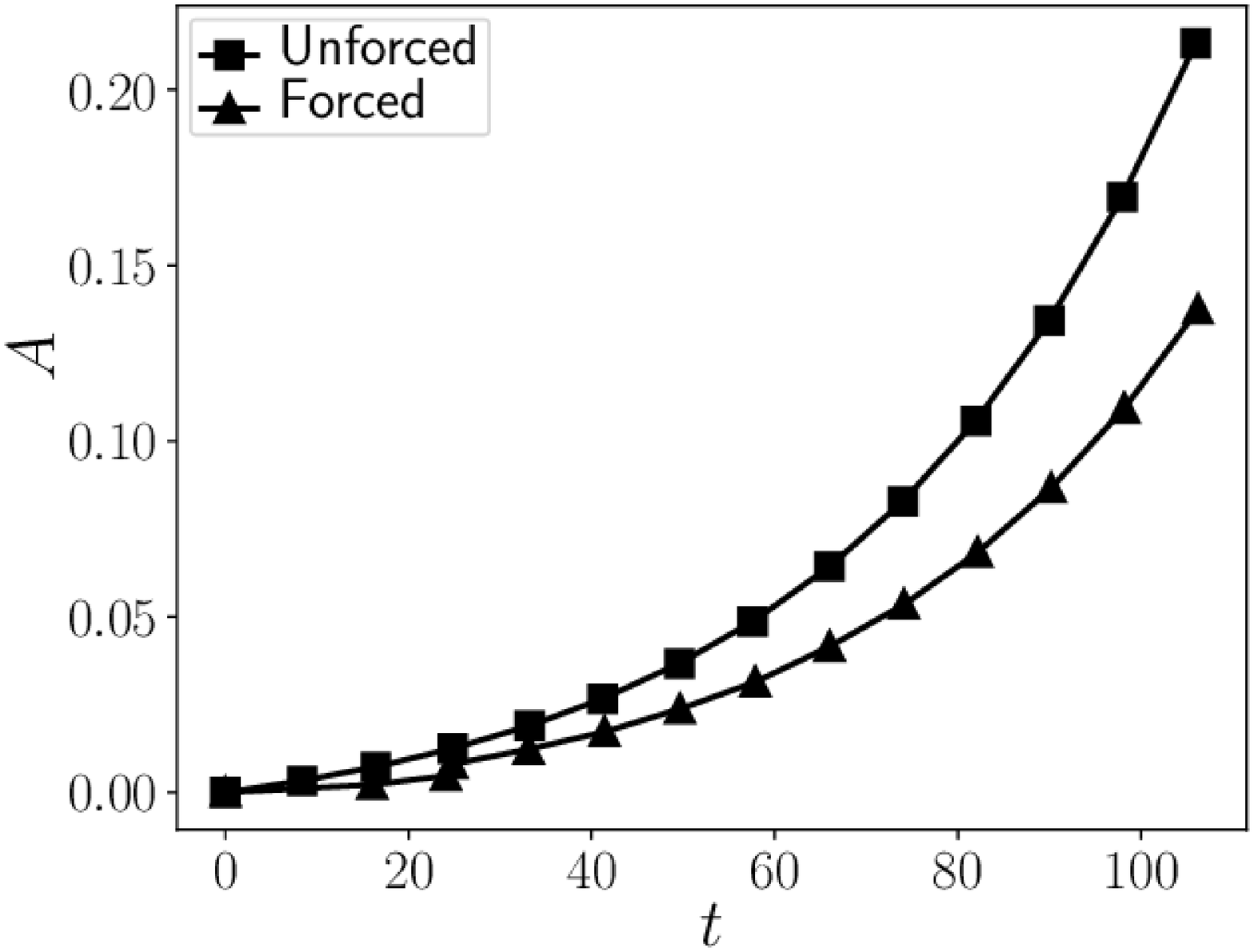}}
    \caption{Fitted curve of the maximal cylinder displacement in each period of oscillation.}
    \label{fit_exp}
\end{figure}

\begin{figure}
    \centering
    \subfigure[] {\includegraphics[scale=0.2
    ]{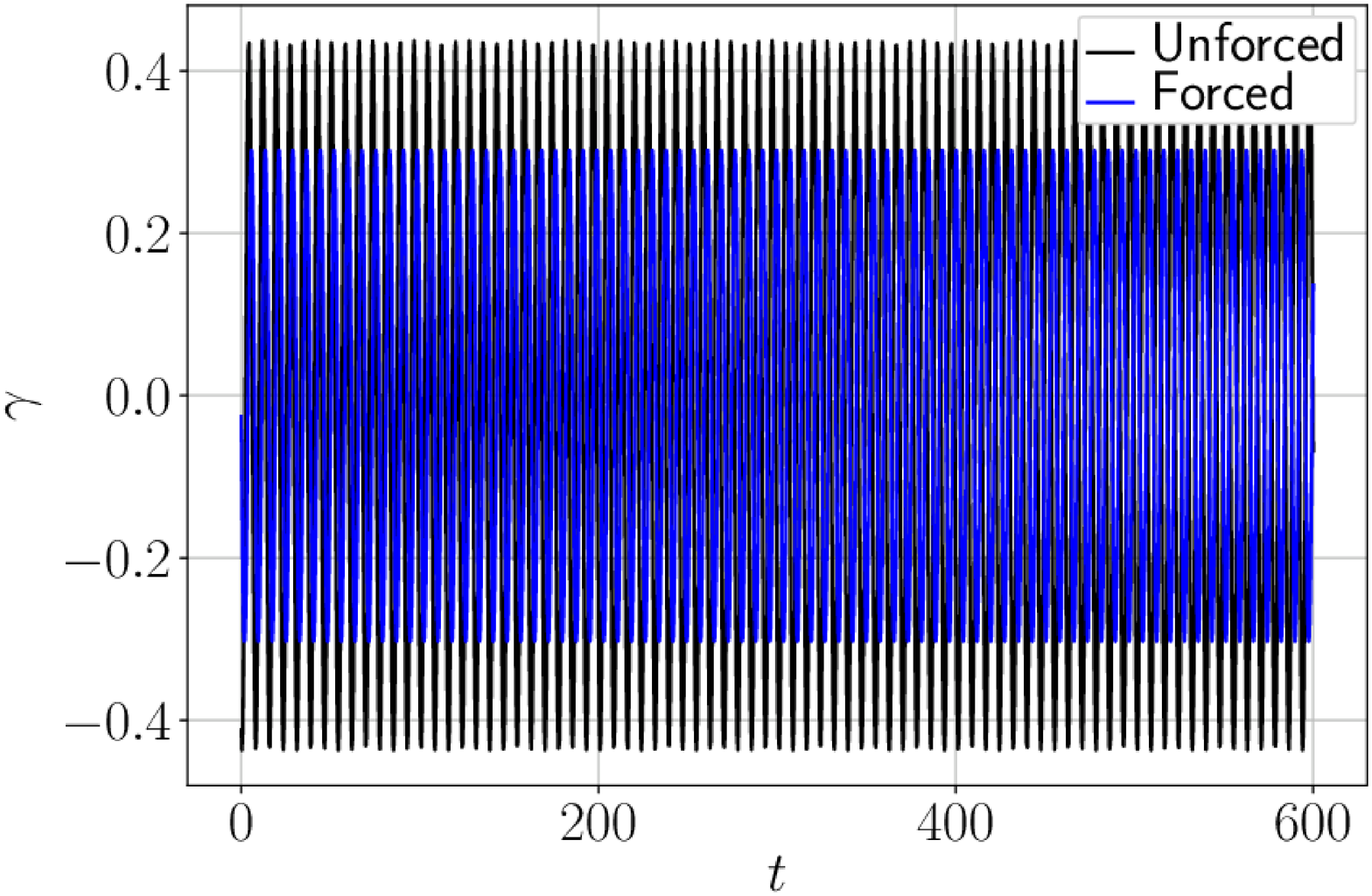}}
    \subfigure[] {\includegraphics[scale=0.19
    ]{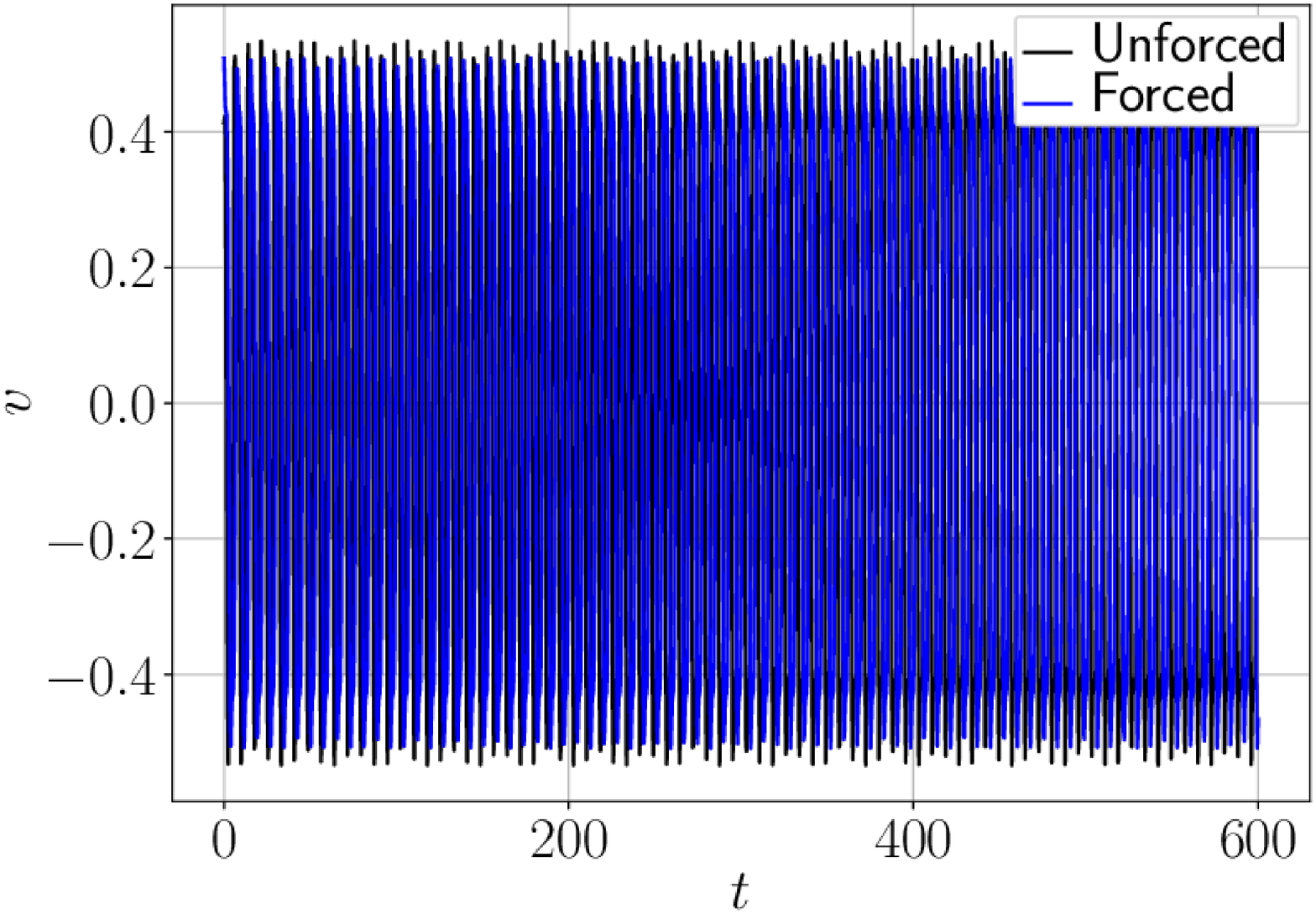}}
    \subfigure[] {\includegraphics[scale=0.19]{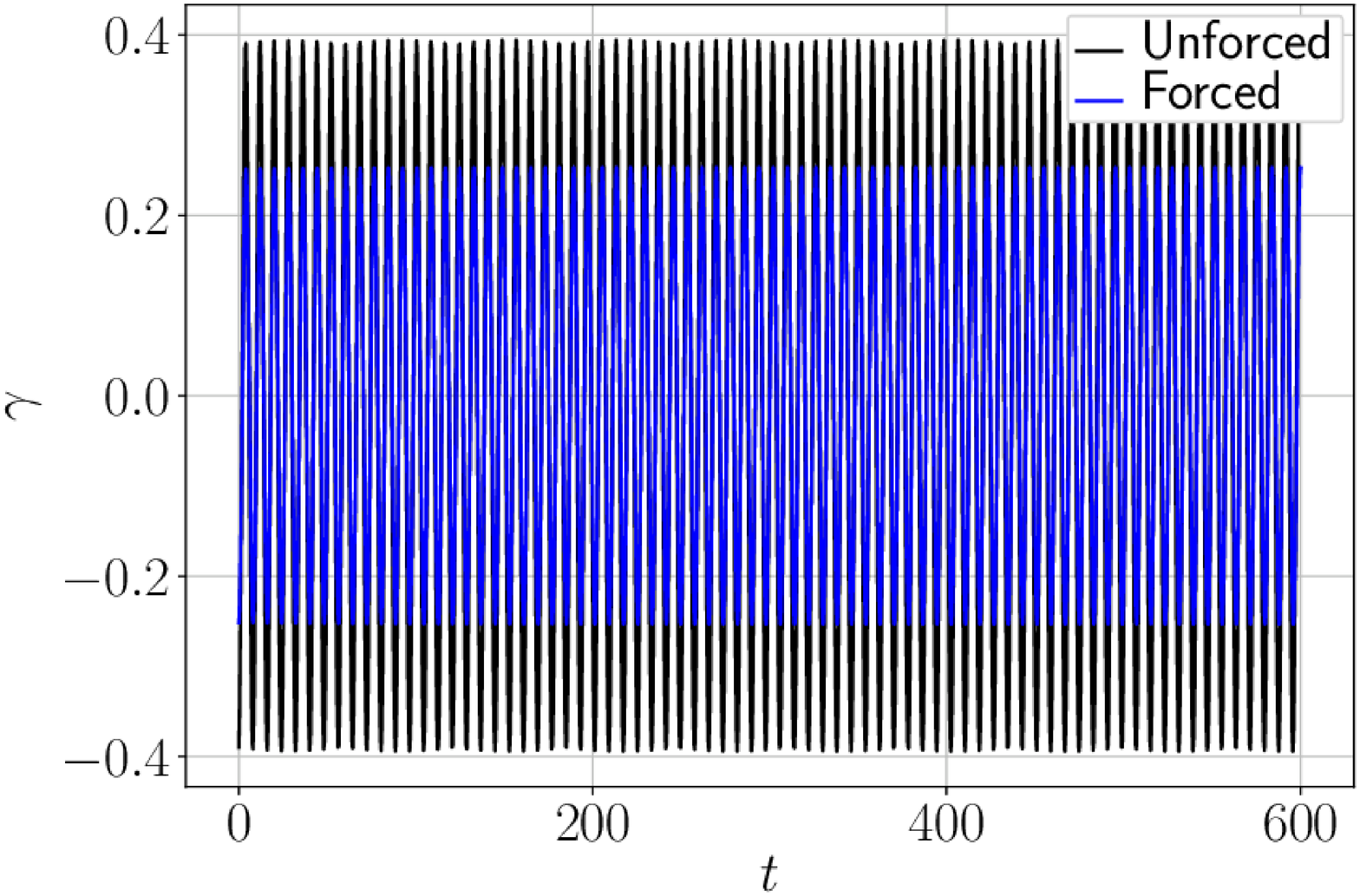}}
    \subfigure[] {\includegraphics[scale=0.2
    ]{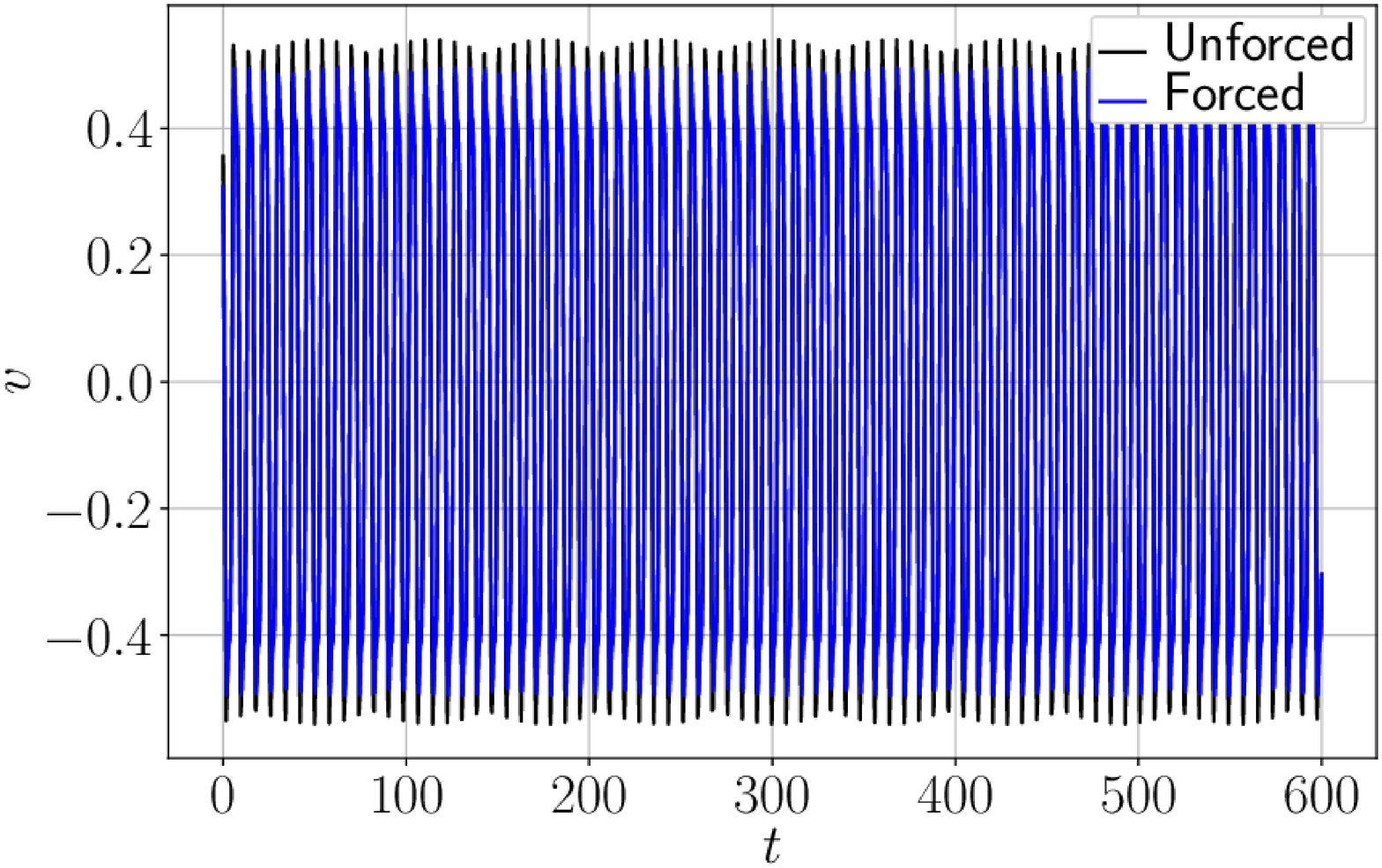}}
    \caption{Cylinder displacement $\gamma$ and the velocity component $v$ at the point (3, 0)  for $(m^{\ast},V_r) = (7, 7.5)$ (a) and (b); and for $(m^{\ast},V_r) = (20, 8)$ (c) and (d).  The insertion of small cylinders with the diameter $d=0.01$ located at $(x_p, y_p) = (0.6, \pm 0.5)$ is referred as forced.}
     \label{cyl_disp}
\end{figure}

Still for the cases in which small cylinders with a diameter of $d = 0.01$ placed at $(x_p, y_p) = (0.6, \pm 0.5)$, Figure~\ref{controls} shows negative variation of $\lambda_{1,r}$ for $V_r$ in the resonance range and in the final branch. However, the least stable mode remains with a positive growth rate ($\lambda_{1,r}>0$). To verify if the growth rate decreases, nonlinear numerical simulations were carried out by using the steady base flow with the control cylinders as the initial condition, and the cylinder displacement was perturbed with an amplitude of $0.01$ at the initial time. After that, in each period of oscillation, the maximal value of $\gamma$ was used to perform the fit of the exponential function $a \exp{(b t)}$. Figure~\ref{fit_exp} shows the fitted curve, for which we can observe the forced FSI system with lower growth rate for $(m^{\ast},V_r) = (7, 7.5)$ and $(m^{\ast},V_r) = (20, 8)$. So these results show that the nonlinear results are in agreement with the sensitivity analyses. Other interesting results are related to the nonlinear response of the system. Figures~\ref{cyl_disp}(a) and \ref{cyl_disp}(c) show that the amplitude of oscillation decreases for $(m^{\ast},V_r) = (7, 7.5)$ and $(m^{\ast},V_r) = (20, 8)$, respectively. Figures~\ref{cyl_disp}(b) and \ref{cyl_disp}(d) show the fluctuations of the velocity component $v$ at the point $(x_v, y_v) = (3,0)$ did not show significant amplitude change as observed in the amplitude of oscillation $\gamma$.

\section{Conclusions} \label{Conclusions}

This work presents a methodology to calculate the sensitivity of the least stable modes to local forces in incompressible laminar steady flows around elastically-mounted bodies. The method makes use of the adjoint equations of the flow-structure coupled system to obtain the gradients and was implemented using a high-order, spectral/\emph{hp} element spatial discretization framework. Results were obtained for the flow around a circular cylinder allowed to move in the cross-stream direction, which is a fluid-structure interaction setup widely investigated, at low Reynolds numbers. Moderate and low mass ratio values were employed in the calculations, and the reduced velocity values were selected based on preliminary nonlinear simulations, in order to include parameter sets inside the lock-in range, but which lead to different amplitudes of vibration. The gradients and sensitivity fields were compared to those obtained for a fixed cylinder, in order to understand the changes the flexibility of the structure imposes to the behaviour of the system. The two least stable modes, corresponding to the two pairs of eigenvalues with the largest real part, were considered in all calculations, as previous works have shown that the dynamics of such fluid-structure systems require these two modes to be taken into account for an adequate description. The sensitivity results were verified by placing actual small cylinders, which emulate the type of local force considered in the analyses, at selected points of the domain and checking if these forced flows would exhibit the behaviour expected. Linear and nonlinear calculations were performed, and in all cases the results confirmed the prediction, with the eigenvalues shifting as anticipated in the linear analyses and the amplitude of vibration increasing or decreasing accordingly in the nonlinear simulations. In some cases, it was possible to suppress the vibration completely.  

The results show that the sensitivity fields obtained for the FSI cases are remarkably different from those obtained for the fixed cylinder. This means that calculations carried out to obtain a passive control for vortex shedding around a fixed body will probably not work to mitigate flow-induced vibrations of the same body when mounted on an elastic base. Moreover, the sensitivity field of the FSI system can be very different depending on the value of reduced velocity and mass ratio, which leads to the conclusion that a general effective passive control of the type we studied in this paper will be very difficult, perhaps impossible, to devise. Other types of passive control might work, though, and the methodology presented here can be employed to investigate other ideas.

This paper is an important contribution to a line of recent research works that have shown that the flexibility of the structure, however simple (linear, one degree-of-freedom), leads to significant differences in the dynamic behaviour of the system. Previous works focused on the nonlinear response, linear stability, and nonlinear characterisation of the bifurcation, and this paper adds results and analyses of the sensitivity fields and passive control implementation. The methodology presented here is a solid starting point for analyses of more complex systems, such as multi degree-of-freedom structures, transient laminar flows, and turbulent flows. Notwithstanding the fact that there are major challenges to overcome in order to extend the formulation for those cases, this work is an important step towards a better understanding of the dynamic behaviour of fluid-structure interaction engineering and natural systems and their control.

\section*{Acknowledgments}

DID acknowledges financial support from the Brazilian National Council for Scientific and Technological Development (CNPq) in the form of PhD student (grant number 140519/2015-7). BSC acknowledges financial support from the Brazilian National Council for Scientific and Technological Development (CNPq) in the form of a productivity grant (grant number 312951/2018-3).

\section*{Declaration of Interests}
The authors report no conflict of interest.
\section{Appendix} \label{Appendix}

\subsection{Adjoint equations}\label{app:adjoint}
The gradients with respect to direct mode $\widehat{\mathbf{q}}$ and with respect to the base field $\mathbf{Q}$ provide the adjoint systems. The algebraic development is presented below.

\subsubsection{$\dfrac{\partial \mathcal{L}}{\partial \widehat{\mathbf{q}}} \delta \widehat{\mathbf{q}}$}

Starting to work out with the gradient $\dfrac{\partial \mathcal{L}}{\partial \widehat{\mathbf{q}}} \delta \widehat{\mathbf{q}}$, in which the derivative is defined by eq.~(\ref{Gateaux}), we have:
\begin{eqnarray*}
    \dfrac{\partial \mathcal{L}}{\partial \widehat{\mathbf{q}}} \delta \widehat{\mathbf{q}} = - \int_{\mat{\Omega}_0} \left[\lambda \delta \widehat{\mathbf{u}}  + \mathbf{U} \cdot \nabla_0 \delta \widehat{\mathbf{u}} + \nabla_0 \mathbf{U}\cdot\delta \widehat{\mathbf{u}} - Re^{-1}\nabla_0^{2} \delta \widehat{\mathbf{u}} - \nabla_0 \delta \hat{p} \right] \cdot \widehat{\mathbf{u}}^{\dagger} \text{d}\mathcal{V}_0 + \\
    - \int_{\mat{\Omega}_0} \left[ \nabla_0 \cdot \delta \widehat{\mathbf{u}} \right] \cdot \widehat{p}^{\dagger} \text{d}\mathcal{V}_0 +  \\ 
    -  \left[ \lambda \delta \widehat{\gamma} - \delta \widehat{\gamma}_1\right] \cdot \widehat{\gamma}^{\dagger} 
    -  \left[ M^{\ast} \lambda \delta \dot{\widehat{\gamma}}_1  + C^{\ast} \delta \widehat{\gamma}_1 + K^{\ast}_1 \delta \widehat{\gamma} - F(\delta \widehat{\mathbf{u}}, \delta \widehat{p})\right]  \cdot \widehat{\gamma}_1^{\dagger}
\end{eqnarray*}  
Next, applying integration by parts and the divergence theorem, we arrive at:
\begin{eqnarray*}
    \dfrac{\partial \mathcal{L}}{\partial \widehat{\mathbf{q}}} \delta \widehat{\mathbf{q}} = - \int_{\mat{\Omega}_0} \delta \widehat{\mathbf{u}} \cdot\left[  -\lambda \widehat{\mathbf{u}}^{\dagger}  - \mathbf{U}\cdot \nabla_0 \widehat{\mathbf{u}}^{\dagger} + \nabla_0 \mathbf{U}\cdot\widehat{\mathbf{u}}^{\dagger}  - Re^{-1}\nabla_0^{2} \widehat{\mathbf{u}}^{\dagger}  - \nabla_0 \hat{p}^{\dagger} \right] \text{d}\mathcal{V}_0 + \\
    -\int_{\mat{\Omega}_0} \delta \widehat{p}\cdot \left[ \nabla_0 \cdot \widehat{\mathbf{u}}^{\dagger} \right] \text{d}\mathcal{V}_0  + \\
    - \int_{\partial \Omega_0} \mathbf{n} \cdot
    \left\{  (\mathbf{U} \cdot \delta \widehat{\mathbf{u}}) \widehat{\mathbf{u}}^{\dagger}  - \sigma (\delta \widehat{\mathbf{u}}, \delta \widehat{p}) \cdot \widehat{\mathbf{u}}^{\dagger} + \delta \widehat{\mathbf{u}} \cdot \sigma (\widehat{\mathbf{u}}^{\dagger}, - \widehat{p}^{\dagger})  \right\} \text{d}\mathcal{S}_0 +\\
    - \delta \widehat{\gamma} \cdot \left[ \lambda \widehat{\gamma}^{\dagger} + K^{\ast}\widehat{\gamma}_1^{\dagger}\right] 
    -  \delta \widehat{\gamma}_1 \cdot  \left[ -M^{\ast} \dot{\widehat{\gamma}}_1^{\dagger}  + C^{\ast} \widehat{\gamma}_1^{\dagger} -  \widehat{\gamma}^{\dagger}\right]\\
    - \int_{\partial \mat{\Omega}_{w, 0}} -\left[\mathbf{n}_y\cdot \sigma (\widehat{\mathbf{u}}, \widehat{p})\right] \cdot \widehat{\gamma}^{\dagger}_1  \text{d}S_{w, 0} 
\end{eqnarray*}  

To reach $\dfrac{\partial \mathcal{L}}{\partial \widehat{\mathbf{q}}} \delta \widehat{\mathbf{q}} = 0$, first let us to the boundary conditions of the adjoint system. Using the boundary conditions (\ref{inlet})--(\ref{outlet}) of the direct FSI system, the integrals along the boundary surfaces are reduced to:
\begin{eqnarray*}
    -  \int_{\partial \Omega_{o,0}} \mathbf{n} \cdot
    \left\{ \delta \widehat{\mathbf{u}} \cdot \left[ (\mathbf{U} \cdot \mathbf{n})\widehat{\mathbf{u}}^{\dagger} + Re^{-1} \nabla_0 \widehat{\mathbf{u}}^{\dagger}\cdot \mathbf{n} - p^{\dagger} \right] \right\} \text{d}\mathcal{S}_{o,0} +\\
    -  \int_{\partial \Omega_{i,0}} \mathbf{n} \cdot
    \left\{ - \sigma (\delta \widehat{\mathbf{u}}, \delta \widehat{p}) \cdot \widehat{\mathbf{u}}^{\dagger}  \right\} \text{d}\mathcal{S}_{i,0} \\
    -  \int_{\partial \Omega_{w,0}} \mathbf{n} \cdot
    \left\{ - \sigma (\delta \widehat{\mathbf{u}}, \delta \widehat{p}) \cdot \widehat{\mathbf{u}}^{\dagger} + \delta \widehat{\mathbf{u}} \cdot \sigma (\widehat{\mathbf{u}}^{\dagger}, - \widehat{p}^{\dagger})  \right\} \text{d}\mathcal{S}_{w,0}+\\
    - \int_{\partial \mat{\Omega}_{w, 0}} -\left[\mathbf{n}_y\cdot \sigma (\widehat{\mathbf{u}}, \widehat{p})\right] \cdot \widehat{\gamma}^{\dagger}_1  \text{d}S_{w, 0}  
\end{eqnarray*}
Next, the following boundary conditions are set:
\begin{eqnarray}
    \widehat{\mathbf{u}}^{\dagger} &=& \mathbf{0} \quad \text{at} \quad \partial\mat{\Omega}_{i, 0},\\
    (\widehat{u}^{\dagger}, \widehat{v}^{\dagger}) &=& (0,  -\widehat{\gamma}_1^{\dagger}) \quad \text{at} \quad \partial\mat{\Omega}_{w, 0},\\
    (\mathbf{U} \cdot \mathbf{n})\widehat{\mathbf{u}}^{\dagger} + Re^{-1} \nabla_0 \widehat{\mathbf{u}}^{\dagger}\cdot \mathbf{n} &=& p^{\dagger} = \mathbf{0} \quad \text{at} \quad \partial\mat{\Omega}_{o, 0}.
\end{eqnarray}
Thus, the gradient $\dfrac{\partial \mathcal{L}}{\partial \widehat{\mathbf{q}}} \delta \widehat{\mathbf{q}}$ is reduced to:
\begin{eqnarray*}
    \dfrac{\partial \mathcal{L}}{\partial \widehat{\mathbf{q}}} \delta \widehat{\mathbf{q}} = - \int_{\mat{\Omega}_0} \delta \widehat{\mathbf{u}} \cdot\left[  -\lambda \widehat{\mathbf{u}}^{\dagger}  - \mathbf{U} \nabla_0 \widehat{\mathbf{u}}^{\dagger} + \nabla_0 \mathbf{U}\widehat{\mathbf{u}}^{\dagger}  - Re^{-1}\nabla_0^{2} \widehat{\mathbf{u}}^{\dagger}  - \nabla_0 \hat{p}^{\dagger} \right] \text{d}\mathcal{V}_0 + \\
    -\int_{\mat{\Omega}_0} \delta \widehat{p}\cdot \left[ \nabla_0 \cdot \widehat{\mathbf{u}}^{\dagger} \right] \text{d}\mathcal{V}_0  + \\
    -  \delta \widehat{\gamma} \cdot \left[ \lambda \widehat{\gamma}^{\dagger} + K^{\ast}\widehat{\gamma}_1^{\dagger} - \int_{\partial \mat{\Omega}_{w, 0}} \frac{\partial V}{\partial y_0} \sigma(\widehat{\mathbf{u}}^{\dagger}, - p^{\dagger}) \cdot \mathbf{n}_y\, \text{d}S_{w, 0}\right]\\
    -  \delta \widehat{\gamma}_1 \cdot  \left[ -M^{\ast} \dot{\widehat{\gamma}}_1^{\dagger}  + C^{\ast} \widehat{\gamma}_1^{\dagger} -  \widehat{\gamma}^{\dagger} + \int_{\partial \Omega_{w,0}}  \sigma (\widehat{\mathbf{u}}^{\dagger}, - \widehat{p}^{\dagger})\cdot \mathbf{n}_y \, \text{d}\mathcal{S}_{w,0} \right]
\end{eqnarray*}  

The adjoint GEP is then given by: 
\begin{eqnarray}
    -\lambda \widehat{\mathbf{u}}^{\dagger}  - \mathbf{U} \cdot \nabla_0 \widehat{\mathbf{u}}^{\dagger} + \nabla_0 \mathbf{U}\cdot \widehat{\mathbf{u}}^{\dagger}  - Re^{-1}\nabla^{2} \widehat{\mathbf{u}}^{\dagger}  - \nabla_0 p^{\dagger} &= \mathbf{0}, \\
    \nabla_0 \cdot \widehat{\mathbf{u}}^{\dagger} &= 0,\\
    \lambda \widehat{\gamma}^{\dagger} + K^{\ast}\widehat{\gamma}_1^{\dagger} - \int_{\partial \mat{\Omega}_{w, 0}} \frac{\partial V}{\partial y_0} \sigma(\widehat{\mathbf{u}}^{\dagger}, - p^{\dagger}) \cdot \mathbf{n}_y\, \text{d}S_{w, 0}&= \mathbf{0},\\
  - M^{\ast} \lambda\dot{\widehat{\gamma}}_{1}^{\dagger} + C^{\ast} \widehat{\gamma}_{1}^{\dagger} - \widehat{\gamma}^{\dagger} + \int_{\partial \mat{\Omega}_{w, 0}} \sigma(\widehat{\mathbf{u}}^{\dagger}, - p^{\dagger}) \cdot \mathbf{n}_y \,\text{d}S_{w, 0} &= \mathbf{0}.
\end{eqnarray}
With that, we have the gradient $\dfrac{\partial \mathcal{L}}{\partial \widehat{\mathbf{q}}} \delta \widehat{\mathbf{q}}  = 0$.

\subsubsection{$\dfrac{\partial \mathcal{L}}{\partial \mathbf{Q}} \delta \mathbf{Q}$}
We use Gateaux derivative to calculate the gradient $\dfrac{\partial \mathcal{L}}{\partial \mathbf{Q}} \delta \mathbf{Q}$, arriving at:
\begin{multline*} \label{dif_state}
    \frac{\partial \mathcal{L}}{\partial \mathbf{Q}} \delta \mathbf{Q} = \frac{\partial \lambda}{\partial \mathbf{Q}} \delta \mathbf{Q} - \frac{\partial \lambda}{\partial \mathbf{Q}} \delta \mathbf{Q} \int_{\mat{\Omega}_0}  \widehat{\mathbf{q}}\cdot \widehat{\mathbf{q}}^{\dagger}  \text{d}\mathcal{V}_0 - \int_{\mat{\Omega}_0}   \left( \nabla \delta \mathbf{U} \cdot \widehat{\mathbf{u}} +  \nabla \widehat{\mathbf{u}} \cdot \delta \mathbf{U} \right) \cdot  \widehat{\mathbf{u}}^{\dagger}   \text{d}\mathcal{V}_0  + \\
     - \int_{\mat{\Omega}_0}  \left(\nabla \cdot\delta \mathbf{U}\right) \cdot P^{\dagger}  \text{d}\mathcal{V}_0 - \int_{\mat{\Omega}_0}   \left( \nabla \delta \mathbf{U}\cdot \mathbf{U}  +  \delta \mathbf{U} \cdot \nabla \mathbf{U} - \frac{1}{Re}\nabla^2 \delta \mathbf{U} + \nabla \delta P \right) \cdot \mathbf{U}^{\dagger}  \text{d}\mathcal{V}_0,
\end{multline*} 

If we use the following normalization, 
    \begin{equation}\label{normalization}
        \int_{\mat{\Omega}_0} \widehat{\mathbf{q}}^{\dagger}\cdot \widehat{\mathbf{q}} \,  \text{d}\mathcal{V}_0 = 1,  
    \end{equation}
and apply integral by parts and the divergence theorem, we obtain
\begin{multline*}
    \frac{\partial \mathcal{L}}{\partial \mathbf{Q}} \delta \mathbf{Q} = - \int_{\mat{\Omega}_0}  \left( \nabla \widehat{\mathbf{u}} \cdot \widehat{\mathbf{u}}^{\dagger} -  \nabla \widehat{\mathbf{u}}^{\dagger} \cdot \widehat{\mathbf{u}}  \right)\cdot \delta \mathbf{U}   \text{d}\mathcal{V}_0  - \int_{\mat{\Omega}_0}  \left(\nabla \cdot  \mathbf{U}^{\dagger}\right)  \cdot \delta P \text{d}\mathcal{V}_0 +\\
     - \int_{\mat{\Omega}_0}   \left(\nabla  \mathbf{U} \cdot \mathbf{U}^{\dagger} -  \nabla \mathbf{U}^{\dagger} \cdot \mathbf{U} - \frac{1}{Re}\nabla^2 \mathbf{U}^{\dagger} - \nabla P^{\dagger} \right) \cdot \delta \mathbf{U}  \text{d}\mathcal{V}_0   - \mathcal{B} , \nonumber
\end{multline*}
where the concomitant bilinear $\mathcal{B}$ is:
 \begin{multline*}
     \mathcal{B} =  \int_{\partial \Omega} \mathbf{n} \cdot
    \left\{ \hat{\mathbf{u}} \cdot ( \delta \mathbf{U} \widehat{\mathbf{u}}^{\dagger}) + \mathbf{U} \cdot ( \delta \mathbf{U} \mathbf{U}^{\dagger}) + \mathbf{U}^{\dagger}  P + P^{\dagger} \delta \mathbf{U} +  \right. \\
    + \left. Re^{-1} \left[ \delta \mathbf{U} \cdot \nabla \mathbf{U}^{\dagger} - \nabla \delta \mathbf{U} \cdot \mathbf{U}^{\dagger} \right] \right\} \text{d}S.  
\end{multline*}

We then set boundary conditions for this system such that $\mathcal{B} = 0$. To obtain these boundary conditions, we first consider the boundary conditions of the base flow system (\ref{NonLFSI}) and of the GEP system (\ref{PAG}), and impose the following  boundary conditions for the adjoint velocity:
\begin{itemize}
    \item $\mathbf{U}^{\dagger}  = \mathbf{0}$ at the inlet ($\partial \Omega_{i,0}$) and wall ($\partial \Omega_{w,0}$);
    \item $ (\hat{\mathbf{u}} \cdot \mathbf{n})\mathbf{u}^{\dagger} + (\mathbf{U} \cdot \mathbf{n})\mathbf{U}^{\dagger} + P^{\dagger} + Re^{-1} \nabla \mathbf{U}^{\dagger} = \mathbf{0}$ at the outlet ($\partial \Omega_{o,0}$).
\end{itemize}

Besides that, we solve the following adjoint system:
\begin{eqnarray}
    \nabla  \mathbf{U} \cdot \mathbf{U}^{\dagger} -  \nabla \mathbf{U}^{\dagger} \cdot \mathbf{U} - Re^{-1} \nabla^2 \mathbf{U}^{\dagger} - \nabla P^{\dagger} &=& -(\nabla  \widehat{\mathbf{u}} \cdot \widehat{\mathbf{u}}^{\dagger} -  \nabla \widehat{\mathbf{u}}^{\dagger}  \cdot  \widehat{\mathbf{u}}), \label{steadyforce_adj11}\\
    \nabla \cdot \mathbf{U}^{\dagger} &=& 0, \label{steadyforce_adj21}
\end{eqnarray}

With that, $\dfrac{\partial \mathcal{L}}{\partial \mathbf{Q}} \delta \mathbf{Q}=0$ is satisfied.

\subsection{Direct modes: Convergence analysis}
Numerical verifications of the direct modes were presented in \cite{Dolci2019}, in which comparisons with results from the previous works were shown. In this work, the convergence analysis is also introduced by monitoring the least stable eigenvalue of the direct mode with respect to the variation of the polynomial degree. Table~\ref{Pol_var} shows the convergence analysis results for the flow around a fixed cylinder at $\Rey = 46.6$ and for the flow around an elastically-mounted cylinder at $(\Rey, m^{\ast}, V_r) = (46.6, 20, 7)$. Table~\ref{Pol_var} shows the real and imaginary part of the least stable eigenvalue. The real part presents a major variation when the polynomial order changes. So one verifies that the variation from seventh polynomial order to eighth polynomial order is of $1.4\%$ for the flow around a fixed cylinder, and $0.007\%$ for the flow around an elastically-mounted cylinder. Therefore, the seventh polynomial order is considered enough to carry out the sensitivity analyses in this work.

\begin{table}
    \centering
    \begin{tabular}{ccc}
      \hline
      Polynomial order & Fixed cylinder & Mounted cylinder\\
      \hline
      5 & $2.3452e-04 + i7.3491e-01$  & $1.4177e-02 + i8.3617e-01$ \\
      6 & $-2.8129e-05 + i7.3598e-01$ & $1.4356e-02 + i8.3595e-01$ \\
      7 & $-2.3927e-05 + i7.3594e-01$ & $1.4439e-02 + i8.3563e-01$ \\
      8 & $-2.3588e-05 + i7.3593e-01$ & $1.4438e-02 + i8.3556e-01$ \\
      \hline
    \end{tabular}
    \caption{Convergence analysis of the least stable eigenvalue with respect to the polynomial degree.}
    \label{Pol_var}
\end{table}

\subsection{Verification of the adjoint modes} \label{adj_mode_ver}
In this work, the sensitivity was calculated using the adjoint modes. So, the adjoint mode calculation routines were verified by comparing the eigenvalues of the adjoint modes with the respective eigenvalues of the direct modes, shown in Table~\ref{CompI_eig_FSI}. The quantitative difference was computed using the expressions:
\[
    d_{\lambda_r} = \dfrac{\lambda_{r,d}- \lambda_{r,a}}{\lambda_{r,d}}, \quad
    d_{\lambda_i} = \dfrac{\lambda_{i,d}- \lambda_{i,a}}{\lambda_{i,d}},
\]
where $d_{\lambda_r}$ is the relative difference between the real part of the eigenvalues of the direct ($\lambda_{r,d}$) and adjoint modes ($\lambda_{r,a}$), $d_{\lambda_i}$ is the relative difference between the imaginary parts ($\lambda_{i,d}$ and $\lambda_{i,a}$) of these same eigenvalues. According to Table~\ref{CompI_eig_FSI}, the highest difference was 8\%. Although this difference could be considered a little high, we believe that the results of sensitivity using the adjoint and direct modes are satisfactory. Observe in Table~\ref{CompI_eig_FSI} that the difference of 8\% was also observed for the modes of a fixed cylinder. 
\begin{table}
\caption{Comparisons of least stable eigenvalue ($\lambda_1$) of direct and adjoint modes for flow around a fixed cylinder and flow around an elastically-mounted cylinder.}
    \centering
    \begin{tabular}{llccl}
    \hline
    \hline
    $(Re, m^{\ast}, \zeta)$  & $V_r$ & $\lambda_1$ of  $\widehat{\mathbf{u}}$ & $\lambda_1$ of $\mathbf{u}^{\ast}$ & ($d_{\lambda_r}$,$d_{\lambda_i}$)\% \\
    \hline
    \hline 
     \multirow{4}{*}{$(46.6, 20, 0)$} & Fixed cyl. & $-2.25\times10^{-5}+ i0.736$ & $-2.15\times10^{-5} + i0.736$& ($4, 0.004$) \\
     & $V_r = 6.3$ & $1.356 \times10^{-3} + i0.937$  & $1.373\times10^{-2} + i0.937$   & ($1.5, 0.0$) \\
     & $V_r = 8$ & $4.24 \times10^{-2} + i0.752$  & $4.39\times10^{-2} + i0.754$   & ($3.5, 0.2$) \\
     & $V_r = 9.5$ & $1.66\times10^{-2} + i0.717$  & $1.75\times10^{-2} + i0.717$   & ($5.4, 0.4$) \\
    \hline
      \hline 
     \multirow{3}{*}{$(46.6, 7, 0)$} & $V_r = 6.3$ & $2.75 \times10^{-2} + i0.837$  & $2.59\times10^{-2} + i0.836$   & ($5.8, 0.1$) \\
     & $V_r = 8$   & $6.27\times10^{-2} + i0.730$  & $6.07\times10^{-2} + i0.732$   & ($3, 0.3$) \\
     & $V_r = 9.5$ & $2.97\times10^{-2} + i0.710$  & $2.95 \times10^{-2}  + i0.712$   & ($0.7, 0.4$) \\
    \hline
    \end{tabular}
    \label{CompI_eig_FSI}
\end{table}

\bibliographystyle{arXiv}
\bibliography{references}


\end{document}